\documentclass[reprint,preprintnumbers,aps,prd,amsmath,amssymb,nobibnotes,nofootinbib,onecolumn,longbibliography]{revtex4-2}
\pdfoutput=1

\usepackage{rotating}
\usepackage{array}
\usepackage{amsmath}
\usepackage[normalem]{ulem}
\usepackage{slashed}

\usepackage{booktabs}
\usepackage[pdftex,table]{xcolor}
\usepackage{units}
\usepackage{xfrac}
\usepackage{mathtools}
\usepackage{empheq}
\usepackage[]{units}
\usepackage{multirow}
\usepackage{amssymb}
\usepackage{url}
\usepackage{comment}
\usepackage{physics}
\usepackage{color,soul}
\usepackage{bbm}
\usepackage[caption=false]{subfig}  %
\usepackage{adjustbox}
\usepackage[T1]{fontenc}
\usepackage{enumerate}
\usepackage{capt-of}

\usepackage{hyperref}
\definecolor{linkcolor}{HTML}{0b5394}
\hypersetup{
  colorlinks=true,
  citecolor=linkcolor,
  linkcolor=linkcolor,
  urlcolor=linkcolor
}

\newcommand{\omni}{\textsc{OmniLearned}~}

\begin{document}

\captionsetup{justification=justified, singlelinecheck=false}

\title{Pre-Training for Simulation-Based Science:\\ A Study on Jet Foundation Model Training Objectives}

\author{Ibrahim Elsharkawy$^{\&}$}
\email{ibrahim.elsharkawy@mail.utoronto.ca}
\affiliation{Department of Physics, University of Toronto and Vector Institute, Toronto, ON, Canada}
\affiliation{NERSC, Lawrence Berkeley National Laboratory, Berkeley, California, USA}

\author{Joschka Birk$^{\&}$}
\email{joschka.birk@uni-hamburg.de}
\affiliation{Institut f\"ur Experimentalphysik, Universit\"at Hamburg, 22761 Hamburg, Germany}

\author{Vinicius Mikuni}
\email{vmikuni@hepl.phys.nagoya-u.ac.jp}
\affiliation{Nagoya University, Kobayashi-Maskawa Institute, Aichi 464-8602, Japan}

\author{Wahid Bhimji}
\email{wbhimji@lbl.gov}
\affiliation{NERSC, Lawrence Berkeley National Laboratory, Berkeley, California, USA}

\author{Gregor Kasieczka}
\email{gregor.kasieczka@uni-hamburg.de}
\affiliation{Institut f\"ur Experimentalphysik, Universit\"at Hamburg, 22761 Hamburg, Germany}

\author{Benjamin Nachman}
\email{nachman@stanford.edu}
\affiliation{Department of Particle Physics and Astrophysics, Stanford University, Stanford, CA 94305, USA}
\affiliation{Fundamental Physics Directorate, SLAC National Accelerator Laboratory, Menlo Park, CA 94025, USA}

\begin{abstract}
Foundation models (FMs) trained on large datasets and fine-tuned on downstream tasks have emerged as a powerful paradigm in AI for science. 
Industrial FMs are typically trained using self-supervision with masking due to the lack of labels.  In many scientific domains, accurate simulations are plentiful and facilitate large, labeled datasets.  This opens up new possibilities for pre-training. 
We present a systematic comparison of pre-training methods using the \omni High Energy Physics FM framework. We test supervised classification, flow-matching generation, and self-supervised masked particle modeling. 
All models are pre-trained on the JetClass dataset and fine-tuned on two representative downstream tasks, top jet classification and JetNet conditional generation. Among other observations, for classification tasks, we find that pure classifier pre-training is optimal when downstream labels and model capacity are plentiful, but combining it with self-supervised masked particle modeling (MPM) is uniquely powerful in the low-finetuning label regime. Flow matching-based generative pre-training seems to provide little benefit for downstream classification, and interestingly, for downstream generation, we find that flow matching must be in the pre-training objective to see a significant finetuning advantage, hinting at the orthogonality of classification and generation tasks. That is, for a model to transfer to both generative and classification downstream tasks, it must be pre-trained on both.  %
This study provides a template for controlled scaling analysis of pre-training objectives for foundation models in simulation-based sciences. %
\end{abstract}

\maketitle
\let\thefootnote\relax\footnotetext{$^{\&}$\ Authors contributed equally.}
\vspace{-1em}
\tableofcontents

\section{Introduction}
\label{sec:intro}

The foundation model paradigm, pre-training a large model on a broad dataset and fine-tuning on specific downstream tasks, has recently been adopted for various AI for science applications~\cite{Bhimji:2025isp,Mikuni:2024qsr,Mikuni:2025tar,Birk:2024knn,Amram:2024fjg,Birk:2025fbs,Vigl:2024lat,Tani:2025osu,mccabe2025walruscrossdomainfoundationmodel,Parker_2024,morehead2026zatom1multimodalfoundationmodel,wood2026umafamilyuniversalmodels,Harris:2024sra,Golling:2024abg,Leigh:2024ked,Bardhan:2025icr}. Despite this progress, fundamental questions remain. Namely, which pre-training objective transfers best to which downstream task, and how strongly does this depend on the amount of pre-training data, downstream data, and model capacity? AI for science often has access to large simulated datasets, with labels available at scale. How does pre-training in a supervised fashion compare to self-supervised and generative objectives?

In this work, we attempt to address these questions using the \omni framework ~\cite{Bhimji:2025isp,Mikuni:2024qsr,Mikuni:2025tar}. \omni is a foundation model in high-energy physics (HEP) trained on jets, variable-number particle sprays that arise from the showering and subsequent hadronization of free partons in high-energy colliders.  Parton-shower and detector simulators such as \textsc{Pythia}, \textsc{Herwig}, \textsc{DELPHES}, and \textsc{Geant4}, and others, reproduce real collider data with a fidelity that is likely unrivaled among scientific domains, as validated by precision measurements at the LHC~\cite{Sjostrand:2014zea,Bellm:2015jjp,deFavereau:2013fsa,ATLAS:2020bbn,GEANT4:2002zbu}. Importantly, datasets at the $\sim10^9$ scale, including JetClass~\cite{Qu:2022mxj} and the billion-jet dataset released with \omni~\cite{Bhimji:2025isp}, are already publicly available, removing potential data-scaling limitations. Jets are also more compact than other data types targeted by foundation models, with $\mathcal{O}(100)$ particle constituents per example rather than the tens of thousands of pixels found in image/video data. This makes a full sweep across pre-training objectives, dataset sizes, and model capacities computationally tractable. Despite the compactness, jet data is complex and requires data and model scale for downstream tasks. Published scaling-law studies for jet classification and generation~\cite{Batson:2023ohn,Amram:2024fjg,ATLAS:2026zdb,Bhimji:2025isp,Vigl:2026ppx,Amram:2026zzv,Bahl:2026jvt} show smooth, predictable performance gains with both model size and pre-training-data up until at least $\sim10^9$ jets and parameters. Further, the downstream tasks we consider (jet classification and jet generation) are not toy benchmarks but useful to ongoing analyses at the Large Hadron Collider~(LHC) and beyond: jet tagging is an essential component of LHC physics analyses~\cite{review-of-heavy-flavor-jet} and generative models for jet constituents show promising potential for anomaly detection efforts~\cite{Buhmann:2023acn,Mikuni:2024qsr,Mikuni:2025tar,Bhimji:2025isp,Mikuni:2026ced}. 
Finally, jets have been used as a model for data in AI interpretability studies \cite {Bogorad:2026oxa,Faucett:2020vbu}, and the \omni model has shown cross-domain transfer to other physical domains~\cite{Elsharkawy:2026kwp,Mikuni:2025ocp,Krzmanc:2026fdw}. 

Given the setup, we pre-train a set of variants of \omni using the following objectives,
\begin{enumerate}
  \item \textbf{Jet classification}: supervised cross-entropy on jet
        type labels denoting the initial particle.
  \item \textbf{Jet generation}: flow-matching with a velocity-prediction
        objective.
  \item \textbf{Masked Particle Modeling (MPM)}: a self-supervised objective
         predicting masked jet constituents.
\end{enumerate}
We also train models with all pairwise combinations
(Classifier$+$Generator, Classifier$+$MPM, Generator$+$MPM) and all three. For selected
configurations, we also include \textit{perturbed} classification losses on noised or masked inputs, promoting robustness of the learned representations. We vary the three axes of pre-training dataset size, finetuning dataset size, and model size. Finetuning is performed on two downstream tasks: top-quark jet tagging (a classification
benchmark) and JetNet conditional generation (a generative benchmark). This study provides a template for controlled scaling analysis of pre-training objectives for jet foundation models.

The paper is organized as follows.
Section~\ref{sec:omnilearned} reviews the \omni architecture and presents the
architecture changes adopted in this work.
Section~\ref{sec:study-design} describes the study design, including
the pre-training configurations in Sec.~\ref{sec:pre-training-setup}
and the finetuning protocol in Sec.~\ref{sec:finetuning-setup}.
Section~\ref{sec:res} presents results on top tagging
(Sec.~\ref{sec:topres}) and JetNet generation
(Sec.~\ref{sec:jetnetres}), and conclusions are provided in Sec.~\ref{sec:conclusion}.

\begin{figure}[t]
    {\centering
    \includegraphics[width=.99\linewidth]{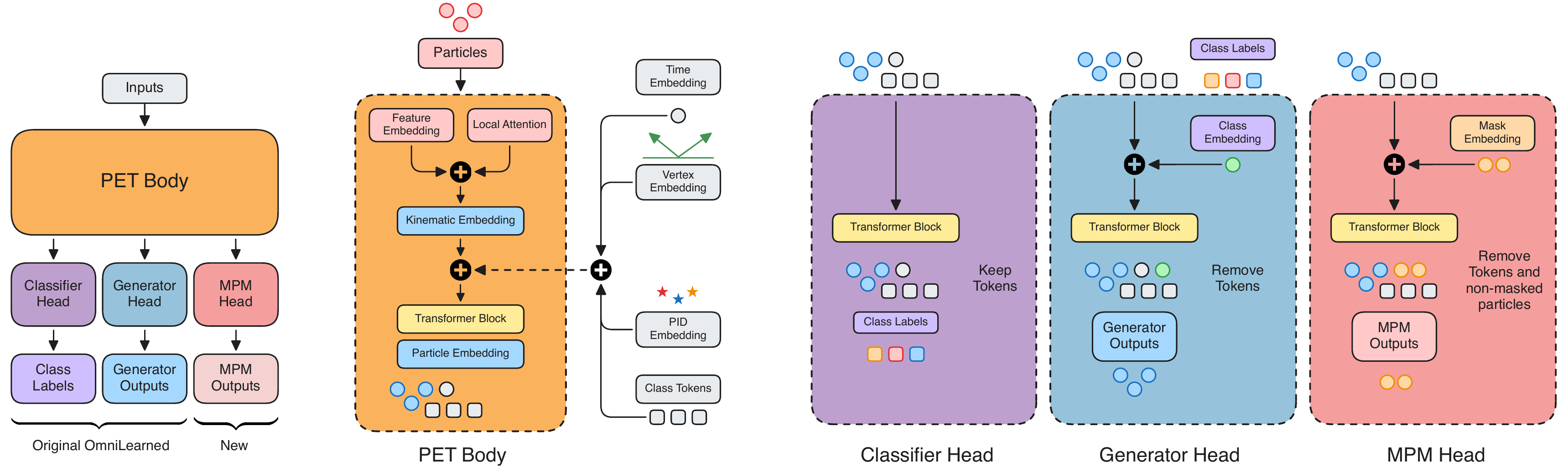}\par
    }
    \caption{
        Illustration of the model architecture used in this work.
        The generator and classifier components were already used in the
        original \omni{}\cite{Bhimji:2025isp} work. The additional
        MPM~\cite{Golling:2024abg,Leigh:2024ked} head is added to the \omni
        framework here.
    }
    \label{fig:omnilearned-arch}
\end{figure}

\section{\textsc{OmniLearned} Review}
\label{sec:omnilearned}

\omni\cite{Bhimji:2025isp} is a point-edge transformer (PET) foundation model for jet physics and an upgrade of the original
\textsc{OmniLearn}~\cite{Mikuni:2024qsr,Mikuni:2025tar}. \omni was pre-trained on $\sim 10^{9}$ jets drawn from JetClass, JetClass~2, ATLAS Top Tagging,
H1~DIS, Aspen Open Jets, and CMS QCD/BSM \cite{Qu:2022mxj,Li:2024htp,ATLAS:2024rua,H1:2021wkz,Amram:2024fjg,Bhimji:2025isp}. The model consists of
a shared body that produces per-particle and per-jet
representations, and task-specific heads that can be attached
or removed depending on the training objective. This design enables
multi-task pre-training where the body is shared across all objectives. A diagram of the model architecture used in this study can be found in Figure~\ref{fig:omnilearned-arch}.
The architecture follows the original \omni{} architecture, with the addition of another model head.
The original \omni framework was pre-trained on the joint classification +
generation objective~\cite{Mikuni:2024qsr,Mikuni:2025tar,Bhimji:2025isp}. 
To extend the comparison to fully self-supervised pre-training, we introduce a
masked particle modeling (MPM)~\cite{Golling:2024abg,Leigh:2024ked} head. The remaining architecture is unchanged.

Two defining aspects of \textsc{OmniLearned}'s PET are a local embedding block and a physics-informed attention bias, which are described in App. \ref{app:omnilearned}. Due to unique architecture choices and extensive pre-training, \omni has shown great performance and flexibility in jet-related tasks (jet classification, generation, anomaly detection, unfolding) and cross-domain point cloud tasks (neutrino physics, molecular dynamics, and cosmology)~\cite{Elsharkawy:2026kwp,Krzmanc:2026fdw,Mikuni:2025ocp}.

\paragraph{Input features}
Each jet is represented as a variable-length set $\{x_i\}_{i=1}^{N}$ of $< \sim150$ constituents. The input is the four-vector
$x_i = (\Delta\eta_i,\, \Delta\phi_i,\, \log p_{\mathrm{T},i},\, \log E_i)$,
with $\Delta\eta_i, \Delta\phi_i$ measured relative to the jet axis and $\log p_{\mathrm{T},i}$ computed relative to the beam axis. When available, particle identification (PID) is embedded through a learned lookup table, and charged-particle vertex information is encoded by a two-layer MLP. These optional embeddings are \textit{added} to the
kinematic embedding, and are set to zero when the
information is not present in a dataset. The flow-matching time
$t\sim\mathcal{U}(0,1)$, encoded with Fourier features followed by an
MLP, is appended to the point cloud as an additional token.

\paragraph{Classifier and generator head}
The classifier head takes the $N_{\mathrm{tok}}=4$ global tokens along with the input embeddings produced by the body, passes them through additional attention blocks. The global tokens are then flattened and passed through an MLP, and outputs $C$ class logits $\hat{y}$. 
The classifier is trained with standard cross-entropy loss against the true class label $y$,
\begin{equation}
    \label{eq:class-loss}
    \mathcal{L}_{\mathrm{class}}
    = -\sum_{c=1}^{C} y_{c}\log \hat{y}_{c}.
\end{equation}
The generator head is trained with flow
matching~\cite{lipman2023flowmatchinggenerativemodeling} for conditional jet generation. Given a clean jet $x$ and a time
$t\sim\mathcal{U}(0,1)$, the head receives the perturbed input $\varphi(t) = (1-t)x + t\epsilon$ with $\epsilon\sim\mathcal{N}(\mu,\sigma)$ and predicts the velocity field $\frac{d}{dt}\varphi(t) = \epsilon - x$. The generator is
trained with a mean squared error against the true velocity field,
\begin{equation}
    \label{eq:gen-loss}
    \mathcal{L}_{\mathrm{gen}}
    = \bigl\lVert \hat{v} - \frac{d}{dt}\varphi(t) \bigr\rVert^{2}.
\end{equation}

\paragraph{Masked Particle Modeling (MPM) head}
The MPM head follows a regression-based paradigm as introduced
in~\cite{Golling:2024abg,Leigh:2024ked}. 
A fixed fraction of the input particles (40\% in this case) is masked before
passing the point cloud to the body. The learned representation of the survived
(=not masked) particles is then used as context to reconstruct the feature
vector of the masked particles. 
Masked particles are re-introduced to the particle set after the body. This is
implemented with learnable mask tokens that encode partial information about the
$p_\mathrm{T}$ of the masked particles.
This head consists of two transformer blocks, followed by a linear layer. Only
the kinematic features of the masked particles are predicted by the MPM head,
following the same paradigm as in the generator head. The MPM loss is given by,
\begin{equation}
    \label{eq:MPM-Loss}
    \mathcal{L}_{\mathrm{MPM}} =
    \frac{1}{|\mathcal{M}|}
    \sum_{\mathcal{M}}
        \bigl\lVert \hat{x}_i - x_i \bigr\rVert,
\end{equation}
where $\mathcal{M}$ is the set of masked particle indices and $\hat{x}_i$ is the head's prediction.

\section{Study Design}
\label{sec:study-design}

This section describes how we use the \omni framework to study how the choice of pre-training objective influences downstream fine-tuning performance. The architecture is held fixed, only the set of active tasks
heads, their associated losses, and the size of the body are varied.
The first subsection (Sec.~\ref{sec:pre-training-setup}) describes the pre-training configurations, the second
(Sec.~\ref{sec:finetuning-setup}) describes the fine-tuning protocol used to evaluate them.

\subsection{Pre-training Setup}
\label{sec:pre-training-setup}

Our goal is to benchmark supervised, generative, and self-supervised pre-training
objectives and combinations of those objectives as a function of the downstream
task, fine-tuning dataset size, model size, and pre-training dataset size. 
Given the three model heads with their dedicated task, this yields three
pre-training objectives that can be activated independently or in
combination, giving seven pre-training modes. 
We pre-train each configuration at three model sizes and, for the smallest size
we additionally vary the number of pre-training examples to study how each
objective scales with data.

The pre-training is restricted to the publicly available JetClass dataset~\cite{Qu:2022mxj,JetClass:dataset} throughout the study.
JetClass contains jets across ten physically distinct classes generated with \textsc{MadGraph5}~\cite{Alwall:2014hca}, showered, and hadronized
with \textsc{Pythia} 8~\cite{Sjostrand:2014zea}, and passed through the
\textsc{Delphes} 3.4.3 detector simulation with the CMS card~\cite{deFavereau:2013fsa}. We use the standard 100M/5M/20M training-validation-test split. The use of a single labeled dataset ensures that all pre-training objectives have access to identical input data, the supervised classifier fit to the JetClass labels,  the generation objectives are conditioned on them, and the self-supervised objective ignores them (only trained on the per-particle kinematic and additional features).

\subsubsection{Pre-training Objectives}
\label{sec:pre-training-objectives}
\begin{figure}[t]
    \centering
    \vspace{1.0em}
    \includegraphics[width=0.7\linewidth]{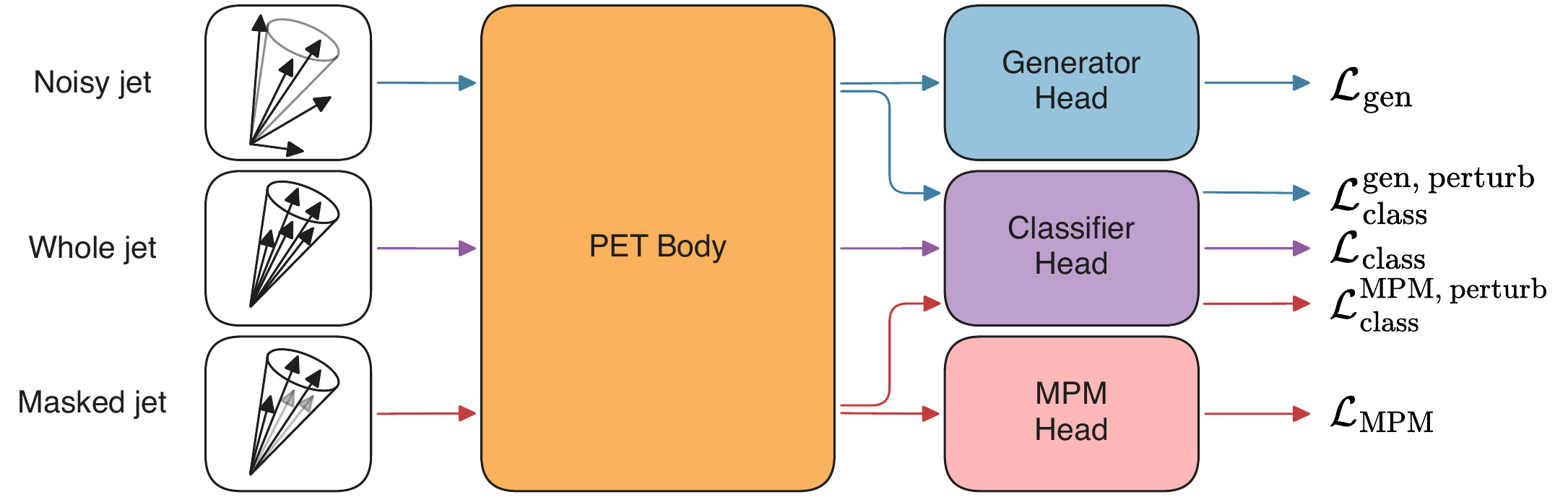}
    \vspace{0.5em}
    \caption{
        Illustration of the different jet augmentations for the different tasks
        and how the information flows through the model in the forward pass.
    }
    \label{fig:forward_pass_sketch}
\end{figure}

The total pre-training loss is a sum over the active head losses, and in the case of all heads being active,
\begin{equation}
    \label{eq:total-pre-training-loss}
    \mathcal{L} =
    \mathcal{L}_{\mathrm{class}}
    + \mathcal{L}_{\mathrm{gen}}
    + \mathcal{L}_{\mathrm{MPM}}
    + \alpha(t)^2\mathcal{L}_{\mathrm{class}}^{\mathrm{gen,~perturb}}
    + w_{\rm MPM}\mathcal{L}_{\mathrm{class}}^{\mathrm{MPM,~perturb}}, 
\end{equation}
where each term is included only when the corresponding head is
active in a given pre-training run. 
In addition to the three primary head losses, two perturbed-input classification losses appear in
Eq.~\ref{eq:total-pre-training-loss}. These terms are activated whenever the classifier head is active with the generator and/or the MPM
head. These terms originate from passing a perturbed body representation (noised for the flow-matching forward process, or masked for MPM) through the classifier head, and adding a cross-entropy loss of its prediction against the true label to the total loss. $\alpha(t)\equiv1-t$ is the interpolation schedule and $w_{\rm MPM}$ is set to $1-f_m$ where $f_m$ is the fraction of masked particles ($f_m=0.4$ in our study).
The main loss components of the individual tasks enter the total loss with equal weight. 
Investigating weighted combinations of the different tasks is left for future studies.
An illustration of the different forward passes is presented in Figure~\ref{fig:forward_pass_sketch}.

The seven pre-training configurations studied in this work are summarized in Table~\ref{tab:pre-training-objectives}, where the Classifier+Generator configuration corresponds to the original \omni pre-training objective~\cite{Bhimji:2025isp,Mikuni:2024qsr}.

\begin{table}[t]
    \centering
    \caption{The seven pre-training configurations studied in this
    work.}
    \label{tab:pre-training-objectives}
    \vspace{0.5em}
    \begin{tabular}{ll}
        \toprule
        Configuration & Pre-training loss \\
        \midrule
        Classifier
            & $\mathcal{L} = \mathcal{L}_{\mathrm{class}}$ \\
        Generator
            & $\mathcal{L} = \mathcal{L}_{\mathrm{gen}}$ \\
        MPM
            & $\mathcal{L} = \mathcal{L}_{\mathrm{MPM}}$ \\
        Classifier + Generator
            & $\mathcal{L} = \mathcal{L}_{\mathrm{class}}
                              + \mathcal{L}_{\mathrm{gen}}
                              + \alpha(t)^2\mathcal{L}_{\mathrm{class}}^{\mathrm{gen,~perturb}}$ \\
        Classifier + MPM
            & $\mathcal{L} = \mathcal{L}_{\mathrm{class}}
                              + \mathcal{L}_{\mathrm{MPM}}
                              + w_{\rm MPM}\mathcal{L}_{\mathrm{class}}^{\mathrm{MPM,~perturb}}$ \\
        Generator + MPM
            & $\mathcal{L} = \mathcal{L}_{\mathrm{gen}}
                              + \mathcal{L}_{\mathrm{MPM}}$ \\
        Classifier + Generator + MPM
            & $\mathcal{L} = \mathcal{L}_{\mathrm{class}}
                              + \mathcal{L}_{\mathrm{gen}}
                              + \mathcal{L}_{\mathrm{MPM}}
                              + \alpha(t)^2\mathcal{L}_{\mathrm{class}}^{\mathrm{gen,~perturb}}
                              + w_{\rm mpm}\mathcal{L}_{\mathrm{class}}^{\mathrm{mpm,~perturb}}$ \\
        \bottomrule
    \end{tabular}
\end{table}

\subsubsection{Model Sizes and Training Hyperparameters}
\label{sec:sizes-and-scale}
We train each of the seven configurations at three model sizes,
summarized in Table~\ref{tab:sizes}. The local-attention block, the input embedding of Eq.~\ref{eq:omni-embed}, the interaction bias of Eq.~\ref{eq:omni-interaction}, and the head architectures are held fixed across sizes. The body depth, the embedding
dimension, and the number of attention heads varies with model size.

\begin{table}[t]
    \centering
    \caption{Model configurations used in this study.
    $N_{\mathrm{body}}$ is the number of transformer blocks in the
    body, $d_{\mathrm{base}}$ is the embedding dimension, and
    $N_{\mathrm{heads}}$ is the number of attention heads.
    $N_{\mathrm{nodes}}$ is the number of NERSC A100x4 compute nodes used during pre-training.}
    \label{tab:sizes}
    \begin{tabular}{lcccccc}
        \toprule
        & $N_{\mathrm{body}}$ & $d_{\mathrm{base}}$
        & $N_{\mathrm{heads}}$ & Params
        & $N_{\mathrm{nodes}}$ & Local Batch Size \\
        \midrule
        Micro  & 3  & 32  & 4  & ${\sim}100$k & 4  & 256 \\
        Small  & 8  & 128 & 8  & ${\sim}7$M   & 8  & 128 \\
        Medium & 12 & 512 & 16 & ${\sim}51$M  & 32 & 32  \\
        \bottomrule
    \end{tabular}
\end{table}

Because the micro model is small enough to train to convergence rapidly, for the Micro model only, we vary the number of pre-training jets $N_{\mathrm{pre}}$ over a log sweep from $N_{\mathrm{pre}} = 10^{5}$ to the full $10^{8}$ JetClass training
set.
\paragraph{Pre-training hyperparameters and schedule}
\label{sec:pre-training-hparams}
Pre-training proceeded on NERSC Perlmutter A100x4 nodes~\cite{nersc_perlmutter}. All pre-training runs use a fixed global batch size of 8192 jets, kept constant across model sizes. The medium models, and \textit{all} models in
which the MPM head is active (regardless of size), are trained with the
Lookahead optimizer~\cite{NEURIPS2019_90fd4f88} with RAdam~\cite{Liu2020On} as the inner optimizer, referred to as \textsc{Ranger}~\cite{wright2019ranger,Buss_RangerLite_2025}. All remaining configurations are trained with the \textsc{Lion}~\cite{chen2023lion} optimizer matching the original \omni~\cite{Bhimji:2025isp}. We use a cosine learning-rate schedule with a linear warmup phase. All remaining hyperparameters were tuned independently for each model size, pre-training objective pair to maximize that configuration's fine-tuning performance. Training was terminated when either the validation loss plateaued or began to
overfit, or downstream finetuning performance ceased to
improve. All hyperparameters not mentioned but used in this study are laid out in App. \ref{app:hyper}.

\subsection{Finetuning Setup}
\label{sec:finetuning-setup}

Each pre-trained body is evaluated on two downstream tasks: a standard top vs. QCD jet classification task, and a standard jet generation task. For every pre-trained checkpoint we initialize the model from the pre-training weights and train on the downstream
task, including all weights of pre-training task heads whose shapes are
compatible with the downstream task head. For both tasks, we additionally train models from scratch as a reference. 

\subsubsection{Top Jet Classification}
\label{sec:top-tagging-finetune}
For top-jet classification we use the publicly available Top Quark
Tagging Reference Dataset~\cite{Kasieczka:2019dbj, toptag:dataset},
the same benchmark used to evaluate the original
\omni~\cite{Bhimji:2025isp,Mikuni:2024qsr}. The
dataset contains top jets (produced via a weak decay $t\rightarrow W^+ b\rightarrow q\bar{q}' b$ and QCD (produced via one free parton) generated with \textsc{Pythia} 8 and passed through the \textsc{Delphes} fast detector simulation, and is split into 1.2M training, 400k validation, and 400k test jets. All final results reported in this work are computed on the test set.

We initialize each finetuning run from one of the pre-trained
checkpoints described in Sec.~\ref{sec:pre-training-setup}. All body
and embedding weights are loaded, and the pre-trained classifier head
is also loaded when available. The classifier
output layer is reinitialized to match the binary output of the task, and unused task heads (generator and MPM) are discarded. 

Finetuning uses the Lion optimizer~\cite{chen2023lion} with a
learning rate and weight decay chosen separately for each model
size to maximize validation performance. The from-scratch baseline randomly initializes the entire model and is trained with a separately tuned learning rate and weight decay. For finetuning we use a linear warmup followed by a cosine decay. Early stopping on the validation loss is applied throughout. To map out the effect of increasing the finetuning dataset size and efficiency of each pre-training objective, we finetune over a logarithmic sweep of the number of available jets, $N_{\mathrm{ft}} \in \{10^{3}, 10^{4}, 10^{5}, 1.2
\times 10^{6}\}$, where the last point uses the entire training set.

For each pre-training configuration, model size, $N_{\mathrm{ft}}$
combination, finetuning is performed \textit{three times} with different random seeds for the data subset selection and weight initialization of fresh layers, and we report the mean and standard deviation across the three runs. After finetuning, each model is evaluated on the held-out test set. We report standard top-tagging metrics as a function of the pre-training configuration, the model size, and the number of finetuning examples $N_{\mathrm{ft}}$.

\subsubsection{Jet Generation}

The generative downstream task is evaluated on the public JetNet dataset~\cite{Kansal:2021cqp,kansal_2022_6975118}.
This dataset contains jets originating from gluons, light quarks, $W$-bosons,
$Z$-bosons and top quarks produced in proton-proton collisions at a center-of-mass energy of
13\,TeV and carry a transverse momentum of $p_\mathrm{T}^\mathrm{jet} \approx 1\,\text{TeV}$.
The used dataset split corresponds to 550k jets for model training, 60k jets for
validation, and 260k jets for testing. 
The maximum number of available particles stored in the dataset (150) is used.
Only kinematic features of the jet constituents are available in the JetNet dataset
and the approximation $m \approx 0$ is used to calculate the particle's energy
for the input features.
While the generative pre-training component is conditioned only on the jet type,
we further add the jet mass and the jet transverse momentum as conditioning
features. 
The jet type information enters the model after the backbone, whereas the jet mass
and the jet $p_\mathrm{T}$ information is fed to the body in the form of an additional
particle.
In our study here, we use the jet-level values from the test
dataset as conditioning features during generation. In an actual application of this
generative model those would be generated by another model for jet-level features.

We evaluate the generative downstream task for three downstream dataset sizes
$N_\mathrm{JetNet} \in \{ 10^4, 10^5, 5.5 \times 10^5 \}$.
When initialized from a pre-trained checkpoint, the weights of the pre-trained
body are loaded. If the pre-training includes the generative task they corresponding
weights of the generator head are loaded as well.
As in the top tagging downstream task, the last layer of the generator head is
re-initialized even when the remaining weights are loaded from a pre-trained
checkpoint. 
In addition to that, the jet type
embedding at the beginning of the generator head is always re-initialized as
well, as the jet type mapping differs between pre-training and downstream task.
The trainings on the generative task are performed with the Lion~\cite{chen2023lion} optimizer,
training each configuration for 70k training steps with a global batch size of 4096, 
corresponding to a maximum of 520 epochs when the largest dataset size of
$5.5\times 10^5$ is used.
A cosine learning rate schedule with linear warmup is used.
The final epoch is used for evaluation, unless overtraining occurs in which case
the checkpoint corresponding to the lowest smoothed validation loss is used.
The limited training duration of 70k training steps on the generative task
is chosen as this was found to be a suitable training duration for the small
model size.
Initial tests of longer trainings with the medium model size showed that
further training can improve the generative performance of those models. 
Yet, we stick with a maximum of 70k training steps for all generative trainings
due to the computational constraints imposed by the large number of
configurations studied in this work.

\section{Results}
\label{sec:res}
Below, we summarize our findings for the two downstream tasks (top-jet classification and jet generation). Other metrics not reported in this section can be found in App. \ref{app:MoreTopRes} and App. \ref{app:othergenresults}.
\subsection{Top Tagging}
\label{sec:topres}
\begin{figure}[t]
    \centering
    \subfloat{
        \includegraphics[width=0.8\linewidth]{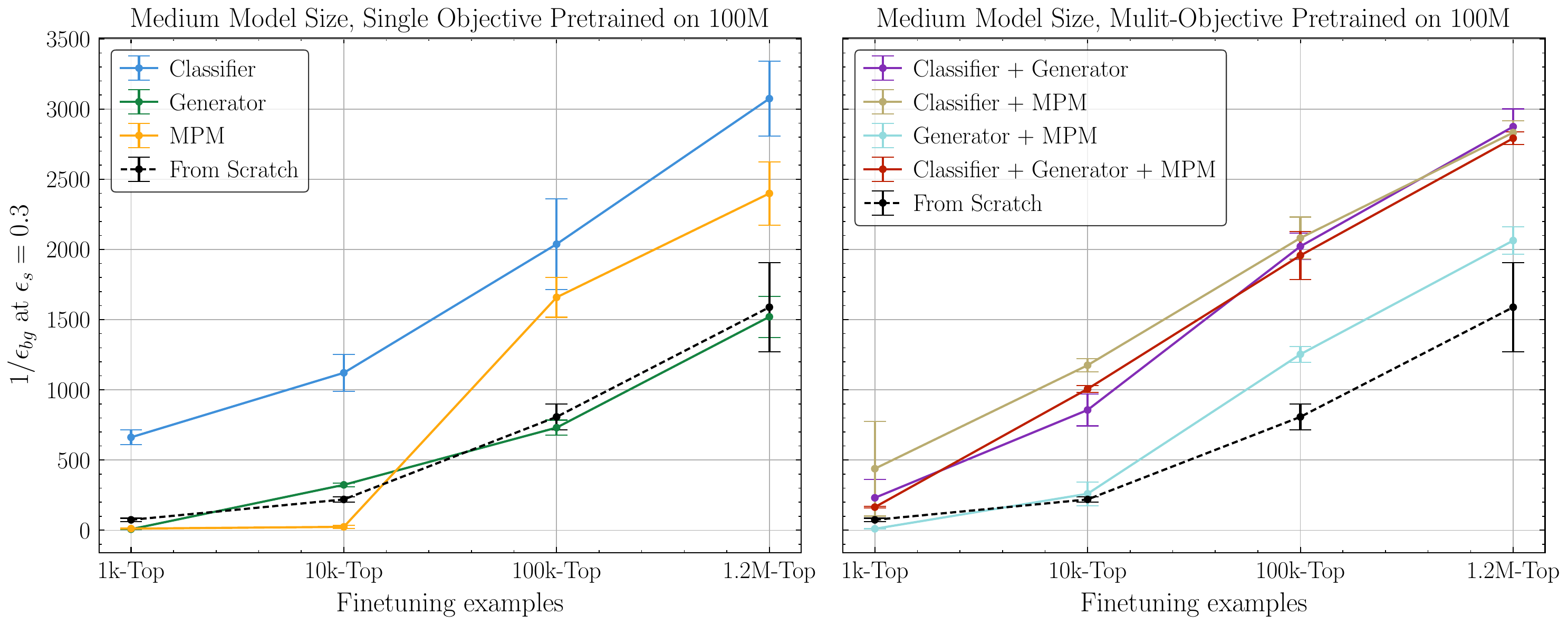}
    }\\{\footnotesize (a) Medium model size}\\
    \subfloat{
        \includegraphics[width=0.8\linewidth]{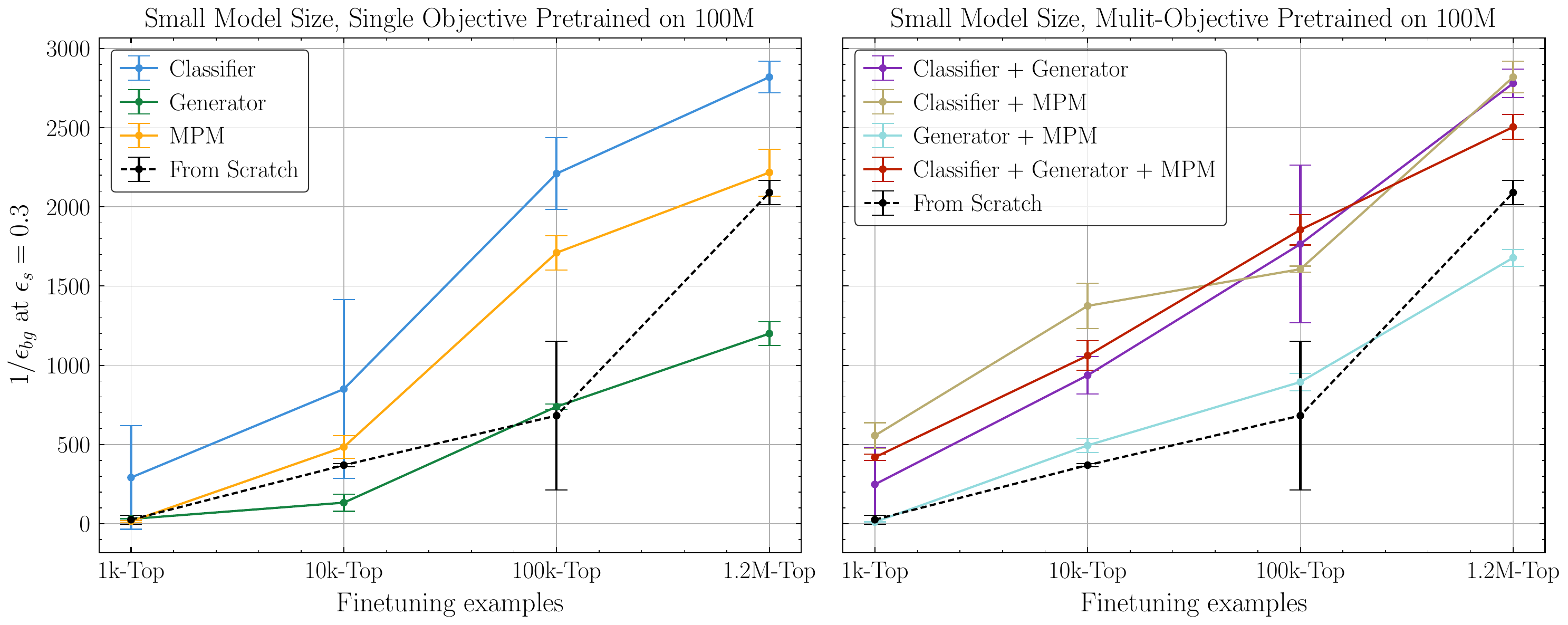}
    }\\{\footnotesize (b) Small model size}\\
    \subfloat{
        \includegraphics[width=0.8\linewidth]{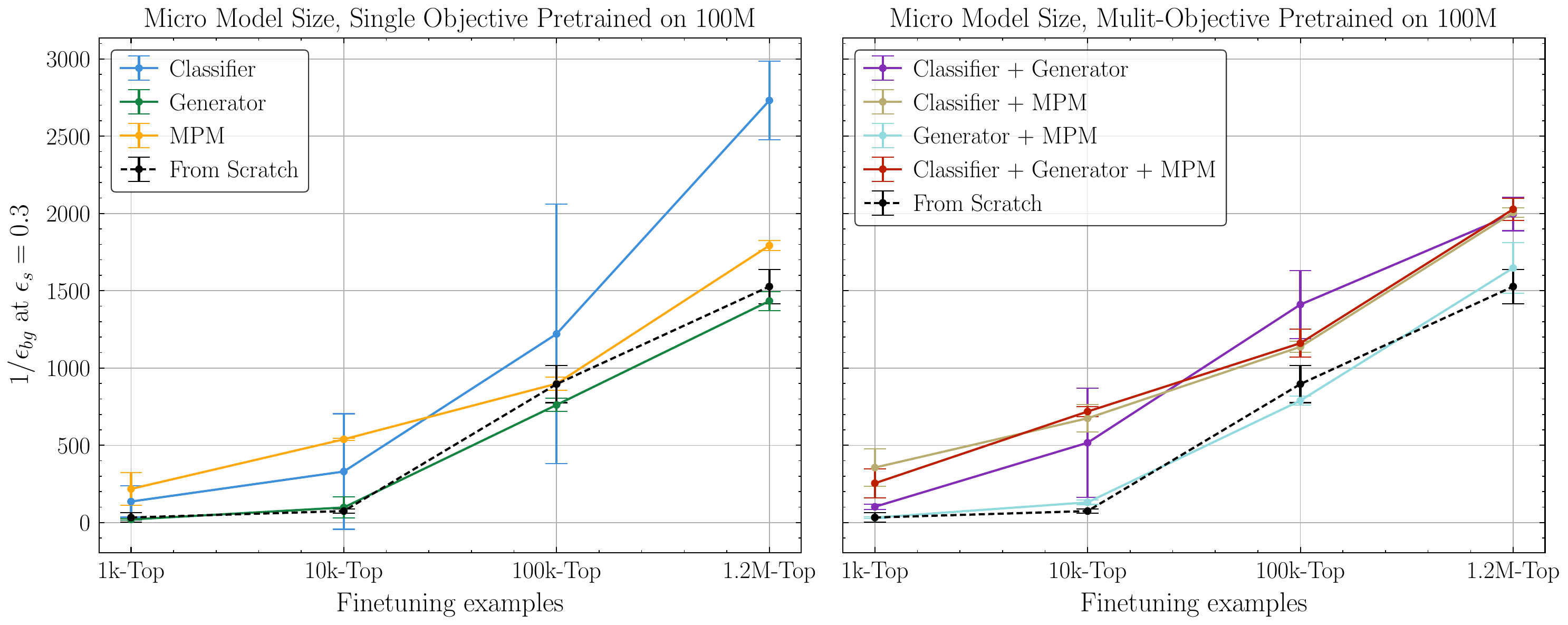}
    }\\{\footnotesize (c) Micro model size}\\
    \caption{
        Top-tagging background rejection at $\epsilon_s=0.3$ vs. top-tagging
        dataset size for different model sizes (rows), pre-trained on 100M JetClass
        jets. Single pre-training losses are shown on the left while combinations of multiple losses are shown on the right.
    }
    \label{fig:micro-small-medium-doublet}
\end{figure}

We run the finetuning setup described in Sec.~\ref{sec:top-tagging-finetune} and
report the standard top-tagging metric background rejection $1/\epsilon_{\rm
bg}$ at a signal efficiency of $\epsilon_s = 0.3$ used in the
literature~\cite{Kasieczka:2019dbj, Bhimji:2025isp,Mikuni:2024qsr}. We plot the
number of finetuning examples (the number of training jets taken from the top tagging
dataset) vs. performance for various model
sizes in Figs.~\ref{fig:micro-small-medium-doublet}\,a-c.
In these, we see that at $N_{\mathrm{ft}} = N_{\max}\equiv1.2\times10^6$, pure classifier pre-training is the single best configuration at every model size when using the best pre-train validation loss checkpoint for fine-tuning. As we will see in Sec \ref{sec:finetuning-traj}, this does not necessarily hold when looking at the entire pretraining trajectory.

The low-data regime is qualitatively different. To summarize each pre-training configuration, model size,
$N_{\mathrm{top}}$ combination with a single number in the low data regime, we define the data-efficiency score,
\begin{equation}
    \label{eq:des}
    \mathrm{DES}(N_{\mathrm{top}}) \;=\;
        \frac{\mathrm{Perf}_{\mathrm{pre-trained}}(N_{\mathrm{top}})}
             {\mathrm{Perf}_{\mathrm{scratch}}(N_{\max})}
    \;\times\;
        \frac{N_{\max}}{N_{\mathrm{top}}},
\end{equation}
where $\mathrm{Perf}$ is again background rejection at fixed signal efficiency $\epsilon_s = 0.3$,
$N_{\mathrm{top}}$ is the number of finetuning examples used during
finetuning, and $N_{\max} = 1.2\times 10^{6}$ is the full
top-tagging training set. A value
$\mathrm{DES}(N_{\mathrm{top}}) \gg 1$ indicates that pre-training provides an advantage compared to a model trained from scratch.

\begin{figure}[t]
    \centering
    \includegraphics[width=0.9\linewidth]{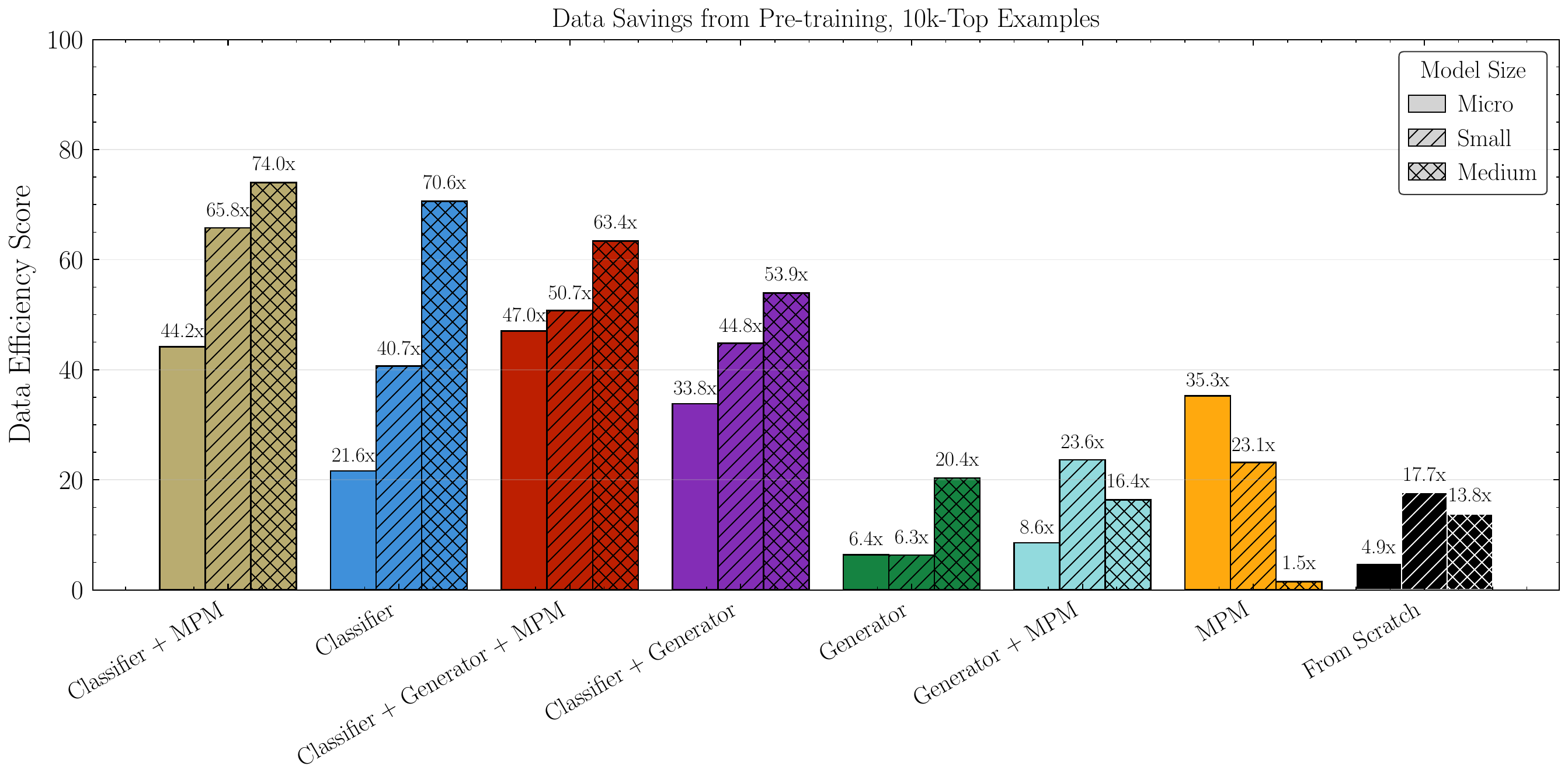}
    \caption{
        Data-efficiency score $\mathrm{DES}(N_{\mathrm{top}})$
        of Eq.~\eqref{eq:des} at $N_{\mathrm{top}}=10^{4}$ across the
        seven pre-training configurations and all three model sizes.
    }
    \label{fig:combined-dataeff}
\end{figure}

At $N_{\mathrm{top}}=10^{4}$ shown in Fig.~\ref{fig:combined-dataeff},
Classifier+MPM is the top configuration at \textit{every} model
size. Pure classifier pre-training is a clear second at Small and Medium but drops to fifth at Micro, where it is beaten by both Classifier+MPM and the fully self-supervised standalone MPM head. The
Classifier+Generator+MPM configuration tracks Classifier+MPM
closely across all three sizes, indicating that the gain over
classifier-only comes from the masked-particle term rather than from
generation.

\begin{figure}[t]
    \centering
    \subfloat{
        \includegraphics[width=0.49\linewidth]{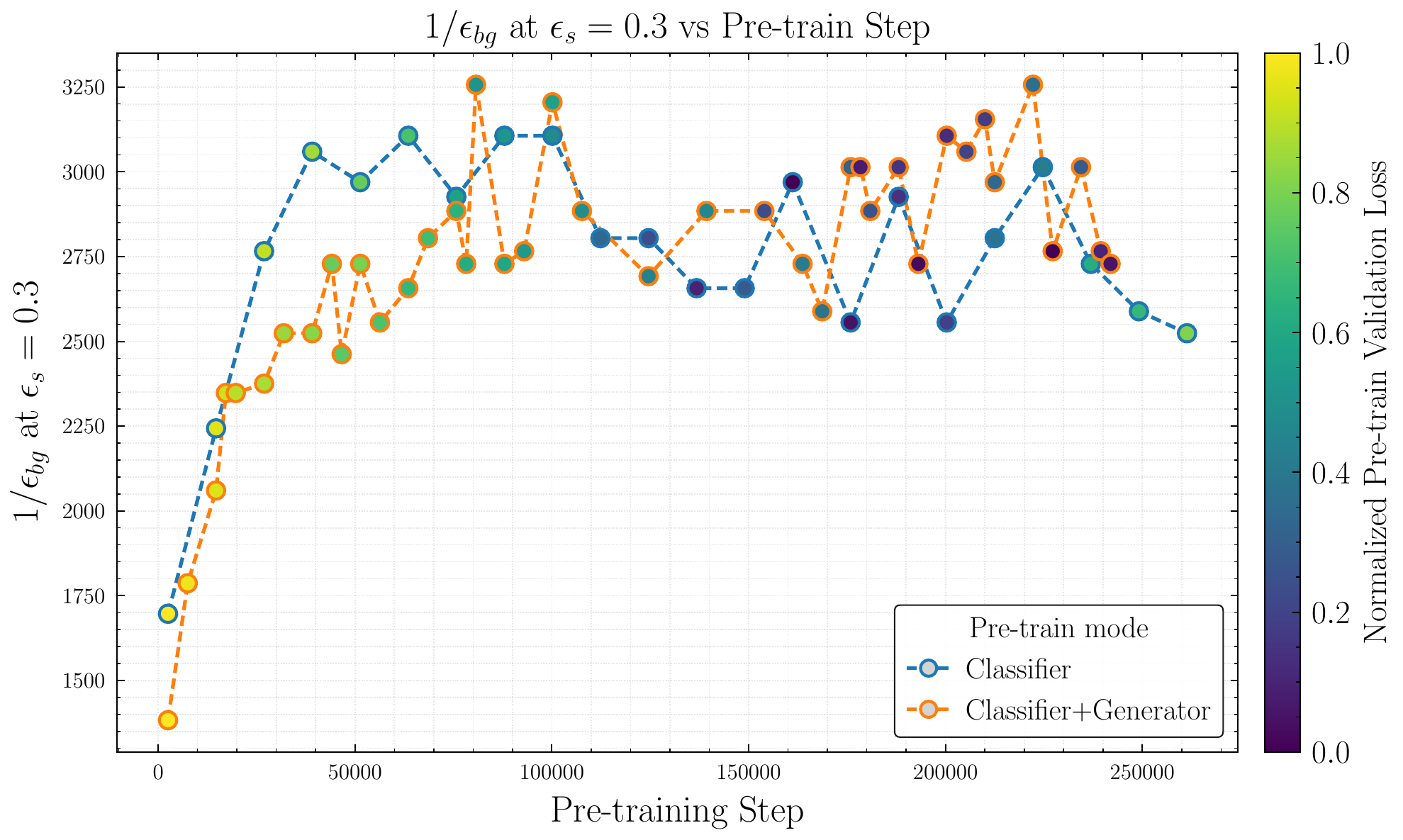}
    }
    \subfloat{
        \includegraphics[width=0.49\linewidth]{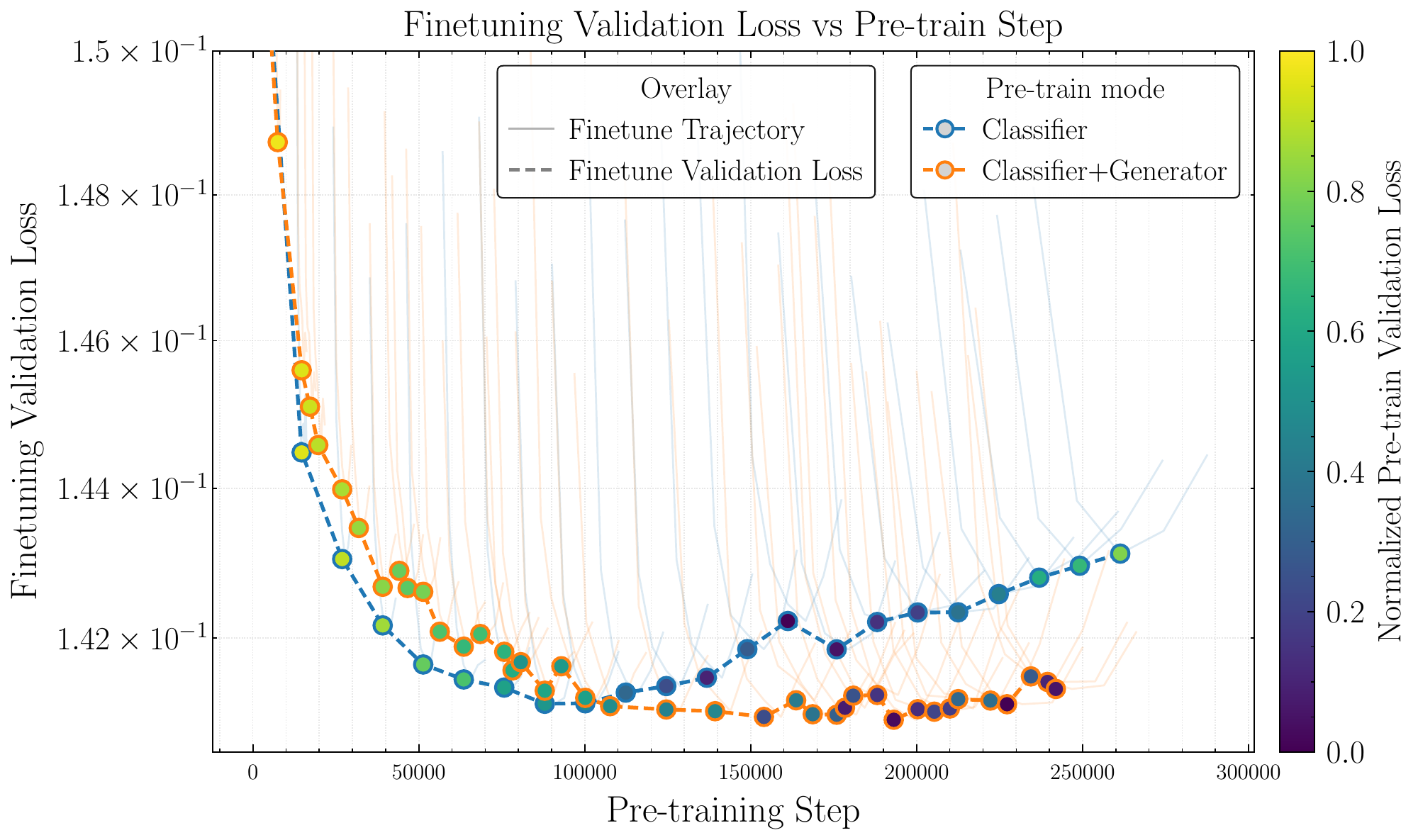}
    }
    \caption{
        Downstream metrics vs. pre-training step for the Classifier and
        Classifier+Generator configurations at the Medium model size, fine-tuned
        on the full $N_{\mathrm{ft}} = N_{\max}$ top-tagging set. Left gives
        background rejection $1/\epsilon_{\mathrm{bg}}$ at $\epsilon_s = 0.3$. Right
        gives finetuning validation loss, with overlaid finetuning trajectories per
        checkpoint. Here, only one finetuning with one seed is performed. Marker color encodes the normalized pre-training validation loss. Results in Fig \ref{fig:micro-small-medium-doublet} correspond to finetuning checkpoint with the lowest \textit{pre-training validation loss}. Raw pre-training and finetuning metrics as a function of pre-train step can be found in App. \ref{app:MoreTopRes} in Table \ref{tab:medium-classifier-stepsweep} and Table \ref{tab:medium-classgen-stepsweep}.
    }
    \label{fig:finetune-trajectory}

    \centering
    \captionof{table}{Pearson correlation $r$ between pre-training validation
    loss and two downstream metrics across the
    sequence of pre-training checkpoints shown in
    Fig.~\ref{fig:finetune-trajectory}. $N$ is the number of
    checkpoints saved along the trajectory, plotted in Figure \ref{fig:finetune-trajectory}.}
    \label{tab:pre-train-finetune-correlation}
    \vspace{0.8em}
    \begin{tabular}{lccc}
        \toprule
        Configuration
            & $N$
            & $r(\mathcal{L}_{\mathrm{pre-train}}, \mathcal{L}_{\mathrm{FT}})$
            & $r(\mathcal{L}_{\mathrm{pre-train}}, 1/\epsilon_{\mathrm{bg}})$ \\
        \midrule
        Classifier
            & 22
            & $+0.962$
            & $-0.751$ \\
        Classifier + Generator
            & 40
            & $+0.896$
            & $-0.765$ \\
        \bottomrule
    \end{tabular}
\end{figure}

Two additional patterns stand out. 
First, all classifier-containing multi-objective configurations converge to
within a few percent of each other at $N_{\mathrm{ft}} = N_{\max}$ within
uncertainties, which is not the case for single objective pre-training.
Second, pure generator pre-training
is the weakest performing configuration throughout, at every
$N_{\mathrm{top}}$, model size pair it either matches or
underperforms the from-scratch baseline. Generator+MPM
inherits this pattern in the low-data regime, suggesting that
flow-matching alone fails to teach the body's representation in a
way that transfers usefully to classification, and that combining it
with MPM only dilutes the masked-particle signal. 
The pattern of purely generative pre-training being outperformed by MPM
pre-training when fine-tuned on top tagging was previously seen
in~\cite{Birk:2025fbs}, where the generative task is implemented with next token
prediction.

Taken together, these observations suggest that when both the model and the labeled downstream sample are large, supervised classifier pre-training performs the best for classification downstream tasks. When either is limited, combining classifier pre-training with the self-supervised MPM head yields the largest improvement.

\subsubsection{Finetuning trajectory over pre-training steps}
\label{sec:finetuning-traj}
To probe how finetuning performance evolves as a function of pre-training steps, we finetune a sequence of
pre-training checkpoints for the Classifier and Classifier+Generator
configurations at the Medium model size. We see that more pre-training is not always better, both configurations reach a peak in downstream performance well
before pre-training converges, after which top-tagging $1/\epsilon_{\mathrm{bg}}$ either plateaus or decreases (Fig.~\ref{fig:finetune-trajectory}). Adding Generation seems to regularize this effect. Pearson
coefficients quantifying the correlation between the two metrics and pre-training validation loss are given in Table~\ref{tab:pre-train-finetune-correlation} and show strong but not perfect correlation.

\subsubsection{Model size scaling}
\begin{figure}[t]
    \centering
    \includegraphics[width=.85\linewidth]{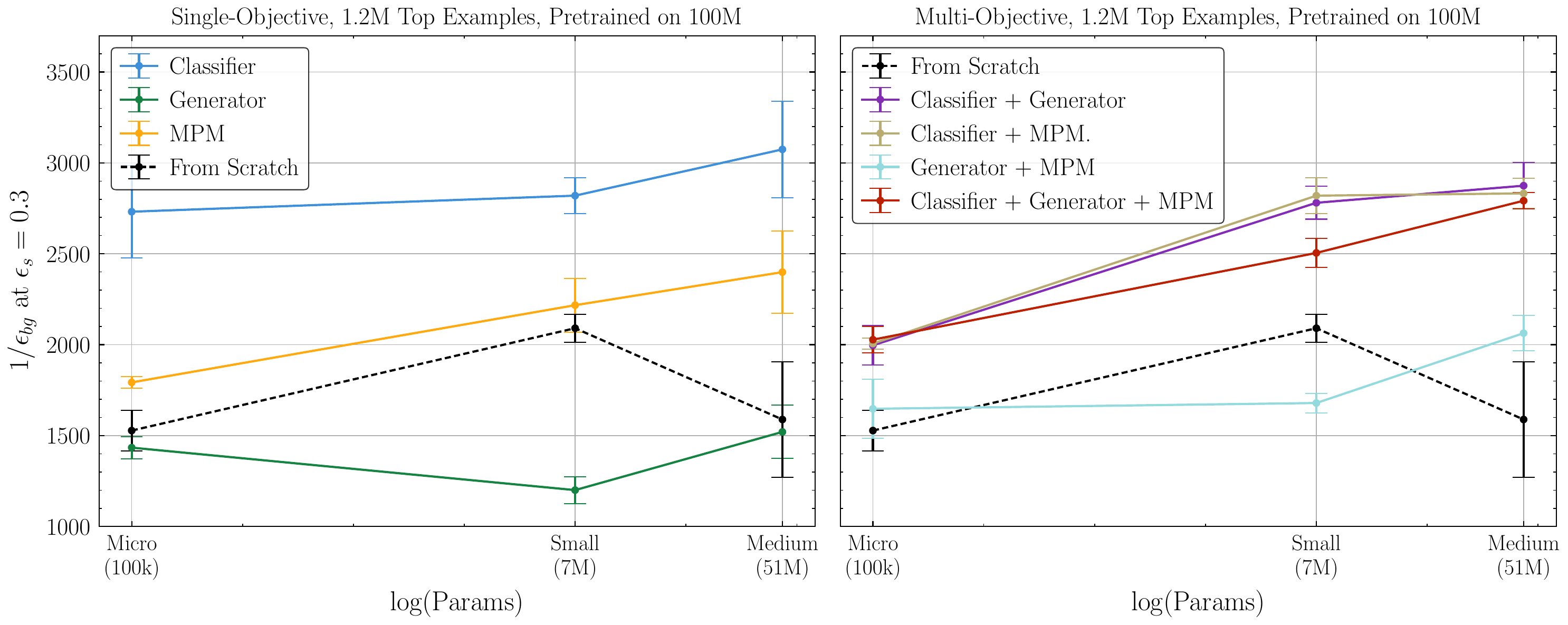}
    \caption{Top-tagging background rejection at $\epsilon_s = 0.3$
    vs model size for the seven pre-training configurations at
    $N_{\mathrm{ft}} = N_{\max}$.}
    \label{fig:model-size-scaling}
\end{figure}

Figure~\ref{fig:model-size-scaling} shows the full top tagging data
($N_{\mathrm{ft}} = N_{\max}$) background rejection as a function
of model size, in $\log(N_{\mathrm{params}})$, for all seven
pre-training configurations and the from-scratch baseline. Every
pre-training configuration improves with size, but to very different
degrees, and Table~\ref{tab:size-scaling-fits} quantifies this with
a linear fit of $1/\epsilon_{\mathrm{bg}}$ vs
$\log(N_{\mathrm{params}})$. The from-scratch baseline
and the pure-Generator configuration are essentially flat, confirming
that flow-matching pre-training alone fails to deliver usable
scaling. Generator+MPM shows similar behavior. On the other hand, the three classifier-containing multi-objective configurations have the
steepest slopes. Interestingly, they converge to within
$\sim$1\% of each other at the Medium scale, indicating that once capacity is sufficient, the choice of \textit{which} self-supervised head accompanies the classifier loss may have little effect on the final performance. 

\begin{table}[t]
    \centering
    \caption{Linear fits of background rejection
    $1/\epsilon_{\mathrm{bg}}$ at $\epsilon_s = 0.3$ vs
    $\log(N_{\mathrm{params}})$. Slope $m$ gives the performance gain per parameter count, $R^2$ measures the linearity of the trend. Bold represents the largest slope (best scaling). }
    \label{tab:size-scaling-fits}
    \vspace{1em}
    \begin{tabular}{lcc}
        \toprule
        Configuration & $m$ & $R^2$ \\
        \midrule
        Classifier                       & $114.2$ & $0.787$ \\
        Generator                        & $\phantom{00}6.9$ & $0.003$ \\
        MPM                              & \textbf{225.0} & $1.000$ \\
        From Scratch                     & $67.2$ & $0.090$\\
        \midrule
        Classifier + Generator           & \textbf{340.2} & $0.953$ \\
        Classifier + MPM                 & $326.6$ & $0.910$ \\
        Generator + MPM                  & $132.1$ & $0.623$ \\
        Classifier + Generator + MPM     & $278.6$ & $0.996$ \\
        \bottomrule
    \end{tabular}
\end{table}

Standalone MPM is the most striking single-objective result, its slope is the largest among single-objective configurations and nearly $2\times$ that of the pure Classifier, suggesting that the masked-particle objective alone, with no supervised signal, scales more efficiently with capacity than the supervised classifier loss, even though the supervised loss yields higher absolute performance at every model size studied. We leave scaling models to larger sizes for future work, where it would be interesting to see if the MPM, and  Classifier+Generator+MPM continue to scale beyond other configurations.

\subsubsection{Scan over pre-training dataset size.}
\begin{figure}[t]
    \centering
    \includegraphics[width=.8\linewidth]{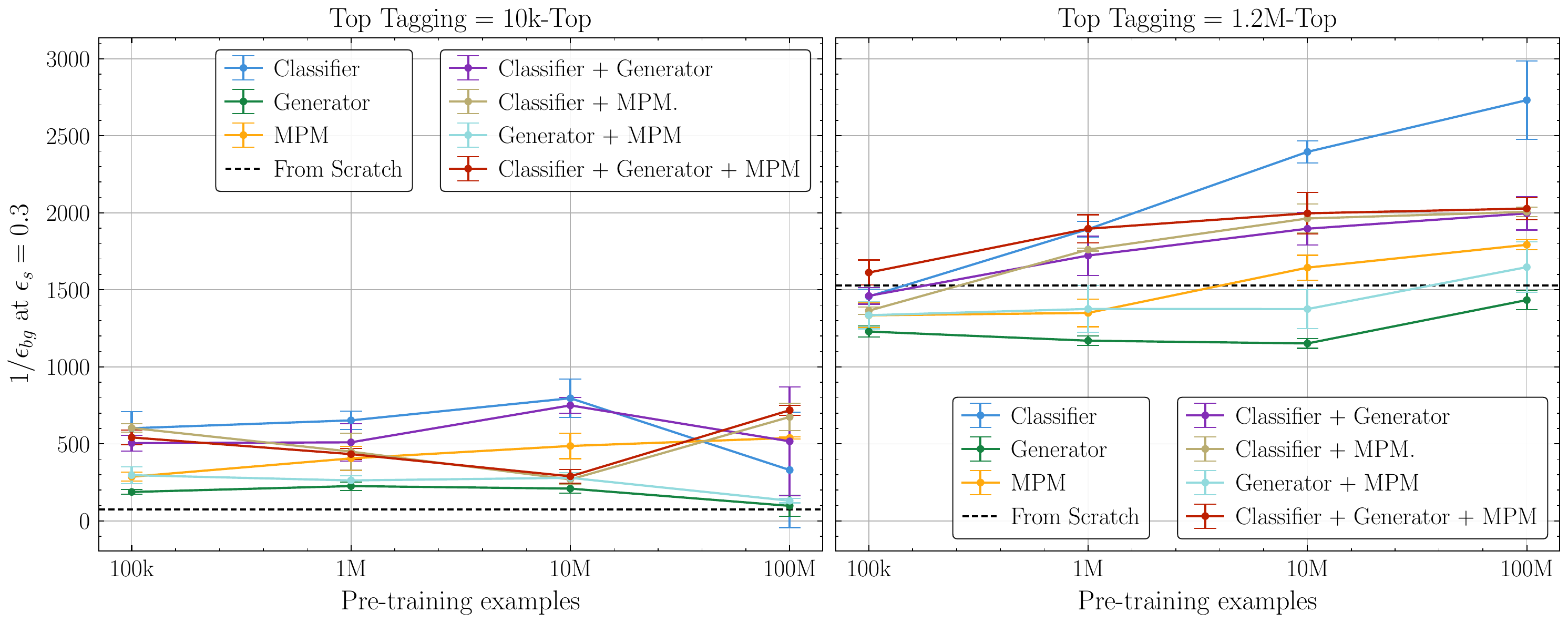}
    \caption{Background rejection $1/\epsilon_{\mathrm{bg}}$ at
    $\epsilon_s = 0.3$ vs pre-training dataset size
    $N_{\mathrm{pre}}$ for the micro model under all seven
    pre-training configurations.
    $N_{\mathrm{ft}} = 10^{4}$ is shown on the left.
    $N_{\mathrm{ft}} = N_{\max} = 1.2\times10^{6}$ is shown on the right. The from-scratch baseline (dashed) does not depend on $N_{\mathrm{pre}}$.}
    \label{fig:pre-train-scan}
\end{figure}

To probe how each pre-training objective responds to the amount of JetClass data, we sweep the pre-training sample size
$N_{\mathrm{pre}} \in \{10^{5}, 10^{6}, 10^{7}, 10^{8}\}$ for
the Micro model under all seven pre-training configurations and
finetune at two top-tagging dataset sizes,
$N_{\mathrm{top}}=10^{4}$ and $N_{\mathrm{ft}} = N_{\max}$
(Fig.~\ref{fig:pre-train-scan}). Table~\ref{tab:pre-train-scan-fits}
gives the slope $m$ and $R^2$ of a linear fit of
$1/\epsilon_{\mathrm{bg}}$ vs $\log(N_{\mathrm{pre}})$ for each configuration and each $N_{\mathrm{top}}$.

\begin{table}[t]
    \centering
    \caption{Linear fits of $1/\epsilon_{\mathrm{bg}}$ at
    $\epsilon_s=0.3$ vs $\log(N_{\mathrm{pre}})$ for the
    Micro model, at two fine-tuning sample sizes. $m$ is the slope and
    $R^2$ measures the linearity of the trend. Bold represents best scaling}
    \label{tab:pre-train-scan-fits}
    \vspace{1em}
    \begin{tabular}{lcccc}
        \toprule
        & \multicolumn{2}{c}{$N_{\mathrm{top}}=10^{4}$}
        & \multicolumn{2}{c}{$N_{\mathrm{top}}=N_{\max}$} \\
        \cmidrule(lr){2-3} \cmidrule(lr){4-5}
        Configuration & $m$ & $R^2$ & $m$ & $R^2$ \\
        \midrule
        Classifier
            & $-67.0$ & $0.197$ & \textbf{432.5} & $0.994$ \\
        Generator
            & $-28.7$ & $0.419$ & $\phantom{0}59.5$ & $0.353$ \\
        MPM
            & $\phantom{-}$\textbf{83.7} & $0.969$ & $166.8$ & $0.912$ \\
        \midrule
        Classifier + Generator
            & $\phantom{-}27.5$ & $0.088$ & $178.1$ & $0.960$ \\
        Classifier + MPM
            & $\phantom{-0}3.4$ & $0.001$ & \textbf{212.7}& $0.878$ \\
        Generator + MPM
            & $-48.1$ & $0.673$ & $\phantom{0}93.5$ & $0.705$ \\
        Classifier + Generator + MPM
            & $\phantom{-}38.6$ & $0.076$ & $134.6$ & $0.844$ \\
        \midrule
        Average (over pre-train modes)
            & $\phantom{-0}1.4$ & $0.346$ & \textbf{182.5} & $0.807$ \\
        \bottomrule
    \end{tabular}
\end{table}

At the high $N_{\mathrm{top}}$ (shown in the right panel of
Fig.~\ref{fig:pre-train-scan}), every configuration improves
substantially with $N_{\mathrm{pre}}$, with an average slope across
pre-training modes of $\bar{m} = 182.5$ and an average
$\bar{R}^2 = 0.807$. Pure Classifier pre-training dominates this
regime, its slope is over $2\times$ the next-best
configuration and yields a near-perfect linear scaling, in clear contrast to the
much shallower scaling of the Generator-only and Generator+MPM configurations.
In short, when downstream labels are abundant, more pre-training data gives
better performance, and the uplift is the most for the purely supervised
objective.

The low-$N_{\mathrm{top}}$ regime is quite different. At
$N_{\mathrm{top}}=10^{4}$ (left panel of
Fig.~\ref{fig:pre-train-scan}), the average pre-training-mode
slope collapses to $\bar{m} = 1.4$ with
$\bar{R}^2 = 0.346$, and several configurations have \textit{negative}
slopes. Interestingly, the only configuration that shows strong positive
scaling with low downstream data is standalone MPM. 
These results may imply (depending on how strongly conditioned this result is on the micro model size) that the value of additional pre-training data is highly dependent on the size of the downstream dataset. 
In the high-finetuning data regime, more pre-training is a near-universal good and supervised classifier pre-training gains the most. In the low-finetuning data regime on the other hand, the choice of objective dominates over the amount of pre-training data, and most configurations show essentially no improvement as $N_{\mathrm{pre}}$ grows.

\begin{table}[ht]
    \centering
    \caption{
        Performance of the different generative models averaged over all jet types.
        Each model is trained for 520 epochs on the full dataset
        ($N_\mathrm{JetNet}=550\mathrm{k}$).
        The shown values represent the Wasserstein distance $W_1$ between the generated
        jets and jets from the JetNet dataset.
    }
    \label{tab:jetnet_metrics_full_dataset_last_epoch}
    
\vspace{1.5em} (a) Medium \\[0.5em]

\begin{tabular}{lcccc}
\toprule
 & $\tau_{21}$ & $\tau_{32}$ & Jet energy & Jet mass \\
 & $\times 10^{-2}$ & $\times 10^{-2}$ & (GeV) & (GeV) \\
\midrule
From Scratch & $6.7 \pm 1.1$ & $4.6 \pm 1.2$ & $8.8 \pm 1.5$ & $1.7 \pm 0.4$ \\
Generator & $1.7 \pm 0.3$ & $1.8 \pm 0.4$ & $9.5 \pm 2.2$ & $1.4 \pm 0.3$ \\
Classifier + Generator + MPM & $\mathbf{1.3 \pm 0.2}$ & $\mathbf{1.7 \pm 0.4}$ & $8.8 \pm 1.9$ & $\mathbf{1.2 \pm 0.2}$ \\
Classifier + Generator & $1.9 \pm 0.4$ & $2.1 \pm 0.4$ & $9.1 \pm 1.1$ & $1.7 \pm 0.4$ \\
Generator + MPM & $1.6 \pm 0.2$ & $1.9 \pm 0.4$ & $\mathbf{8.8 \pm 1.9}$ & $1.2 \pm 0.3$ \\
Classifier & $9.1 \pm 1.8$ & $6.5 \pm 1.6$ & $9.3 \pm 2.3$ & $2.9 \pm 0.8$ \\
MPM & $7.9 \pm 1.4$ & $5.7 \pm 1.5$ & $10.9 \pm 2.3$ & $1.8 \pm 0.4$ \\
Classifier + MPM & $9.4 \pm 2.0$ & $6.8 \pm 1.8$ & $9.8 \pm 2.0$ & $3.1 \pm 0.9$ \\
\bottomrule
\end{tabular}

\vspace{1.5em} (b) Small \\[0.5em]

\begin{tabular}{lcccc}
\toprule
 & $\tau_{21}$ & $\tau_{32}$ & Jet energy & Jet mass \\
 & $\times 10^{-2}$ & $\times 10^{-2}$ & (GeV) & (GeV) \\
\midrule
From Scratch & $2.0 \pm 0.3$ & $2.1 \pm 0.5$ & $8.0 \pm 1.2$ & $0.9 \pm 0.1$ \\
Generator & $1.9 \pm 0.3$ & $2.0 \pm 0.4$ & $9.0 \pm 2.3$ & $1.2 \pm 0.3$ \\
Classifier + Generator + MPM & $1.2 \pm 0.1$ & $1.4 \pm 0.4$ & $8.1 \pm 1.4$ & $0.8 \pm 0.2$ \\
Classifier + Generator & $\mathbf{1.1 \pm 0.2}$ & $\mathbf{1.4 \pm 0.3}$ & $\mathbf{7.6 \pm 1.2}$ & $\mathbf{0.7 \pm 0.1}$ \\
Generator + MPM & $1.2 \pm 0.2$ & $1.4 \pm 0.3$ & $8.6 \pm 1.3$ & $0.9 \pm 0.1$ \\
Classifier & $3.6 \pm 0.5$ & $3.3 \pm 0.6$ & $9.2 \pm 1.5$ & $1.1 \pm 0.2$ \\
MPM & $3.0 \pm 0.4$ & $2.5 \pm 0.6$ & $9.2 \pm 2.2$ & $1.0 \pm 0.2$ \\
Classifier + MPM & $4.8 \pm 0.8$ & $4.0 \pm 0.8$ & $8.5 \pm 1.4$ & $1.2 \pm 0.2$ \\
\bottomrule
\end{tabular}

\vspace{1.5em} (c) Micro \\[0.5em]

\begin{tabular}{lcccc}
\toprule
 & $\tau_{21}$ & $\tau_{32}$ & Jet energy & Jet mass \\
 & $\times 10^{-2}$ & $\times 10^{-2}$ & (GeV) & (GeV) \\
\midrule
From Scratch & $10.9 \pm 2.4$ & $6.7 \pm 1.5$ & $10.1 \pm 1.6$ & $\mathbf{4.8 \pm 1.6}$ \\
Generator & $10.4 \pm 3.4$ & $\mathbf{5.5 \pm 1.2}$ & $11.6 \pm 3.1$ & $10.9 \pm 3.5$ \\
Classifier + Generator + MPM & $13.4 \pm 3.0$ & $8.8 \pm 0.7$ & $12.9 \pm 3.2$ & $12.8 \pm 2.5$ \\
Classifier + Generator & $\mathbf{10.4 \pm 2.6}$ & $7.0 \pm 1.2$ & $12.8 \pm 1.8$ & $4.9 \pm 1.9$ \\
Generator + MPM & $13.0 \pm 3.0$ & $7.9 \pm 0.7$ & $11.4 \pm 2.7$ & $10.3 \pm 2.9$ \\
Classifier & $15.3 \pm 3.1$ & $6.7 \pm 1.5$ & $11.4 \pm 1.9$ & $8.4 \pm 2.3$ \\
MPM & $13.4 \pm 2.6$ & $7.0 \pm 1.5$ & $\mathbf{9.9 \pm 2.2}$ & $7.0 \pm 2.2$ \\
Classifier + MPM & $11.8 \pm 2.6$ & $6.1 \pm 1.4$ & $11.3 \pm 1.8$ & $5.8 \pm 2.0$ \\
\bottomrule
\end{tabular}

\vspace{1em}
\end{table}

\subsection{JetNet Generation}
\label{sec:jetnetres}

The performance of the generative models is evaluated both on particle-level and
jet-level observables based on generated samples of 50k jets per type. 
Uncertainties are estimated with bootstrapping by calculating each metric five
times, drawing batches of 10k jets from the
generated sample and the JetNet test set with replacement. 
The Wasserstein distance $W_1$ is used as a metric to describe how well
the generated distribution matches the target (=JetNet) distribution.
The evaluated particle-level features are the features directly generated by
the model, i.e. $\Delta\eta_i$, $\Delta\phi_i$, $\log p_\mathrm{T,i}$ and $\log E_i$.
The jet-level features quantify if inter-particle correlations are modeled
accurately. To this end, we calculate the jet mass, the subjettiness~\cite{Thaler:2010tr}
ratios $\tau_{21} = \tau_2/\tau_1$ and $\tau_{32} = \tau_3/\tau_2$ as well as the jet
pseudorapidity and jet energy from the particle features.
The jet energy is the scalar sum of the constituent energy whereas other
jet-level features combine multiple particle-level features in a
non-trivial way.
The jet mass and the subjettiness ratios are calculated from the four-momenta
based on $\Delta\phi_i$, $\Delta\eta_i$ and $\log p_\mathrm{T,i}$ values
(assuming massless particle). 
The absolute value of the jet pseudorapidity $\eta_\mathrm{jet}$ is obtained by
averaging\footnote{
    We only consider particles with $E_i \geq p_{\mathrm{T},i}$ in this average,
    thus removing unphysical particles. The fraction of unphysical particles with
    $E_i < p_{\mathrm{T},i}$ is at the order of $\mathcal{O}(1\,\%)$.
} over the per-particle estimate 
$\eta_{\mathrm{jet},i} \approx - \Delta\eta_i + \cosh^{-1}\left(E_i/p_{\mathrm{T},i}\right)$.
All metrics are first calculated for each jet type individually and then averaged
to reflect the overall performance across all five jet types.
The performance of the models trained on the whole JetNet dataset is shown in
Table~\ref{tab:jetnet_metrics_full_dataset_last_epoch}, where a selection
of metrics is presented. 
Further metrics can be found in the Appendix in
Table~\ref{tab:jetnet_metrics_full_dataset_last_epoch_additional_metrics}.

\begin{figure}[t]
    \centering
    \subfloat{
        \includegraphics[width=0.99\linewidth]{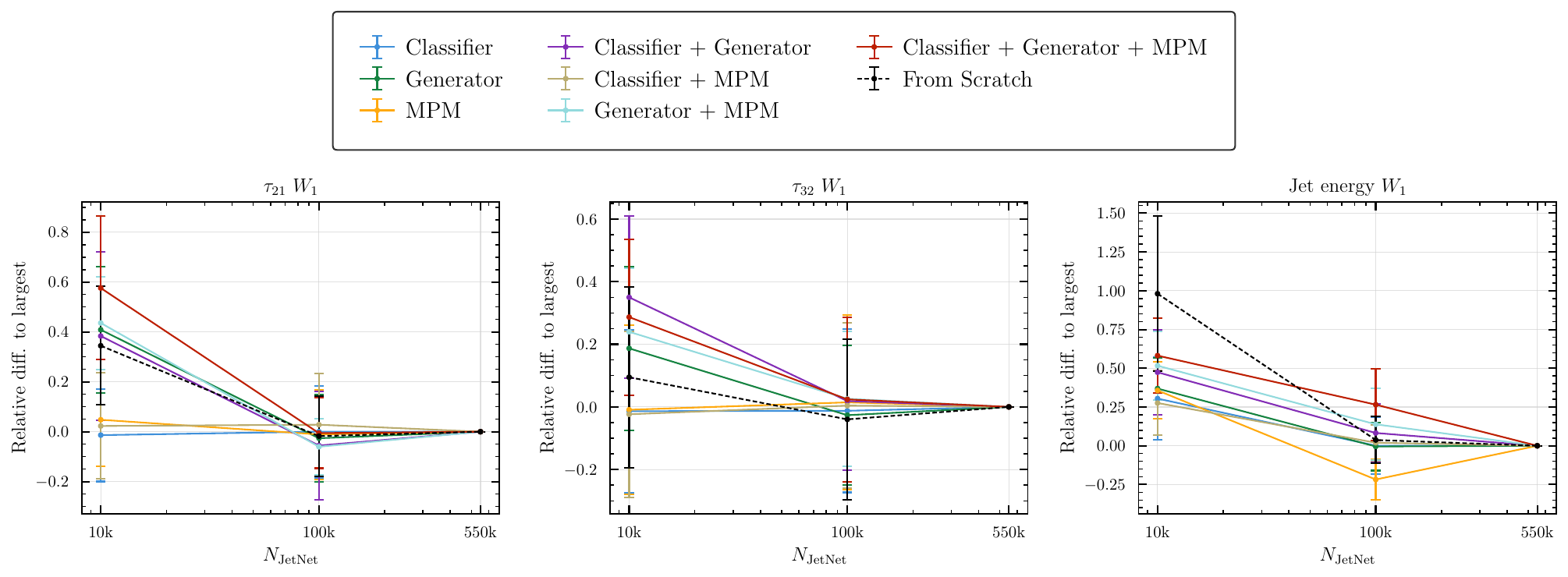}
    }\\{\footnotesize (a) Medium model size}\\\vspace{0.5em}
    \subfloat{
        \includegraphics[width=0.99\linewidth]{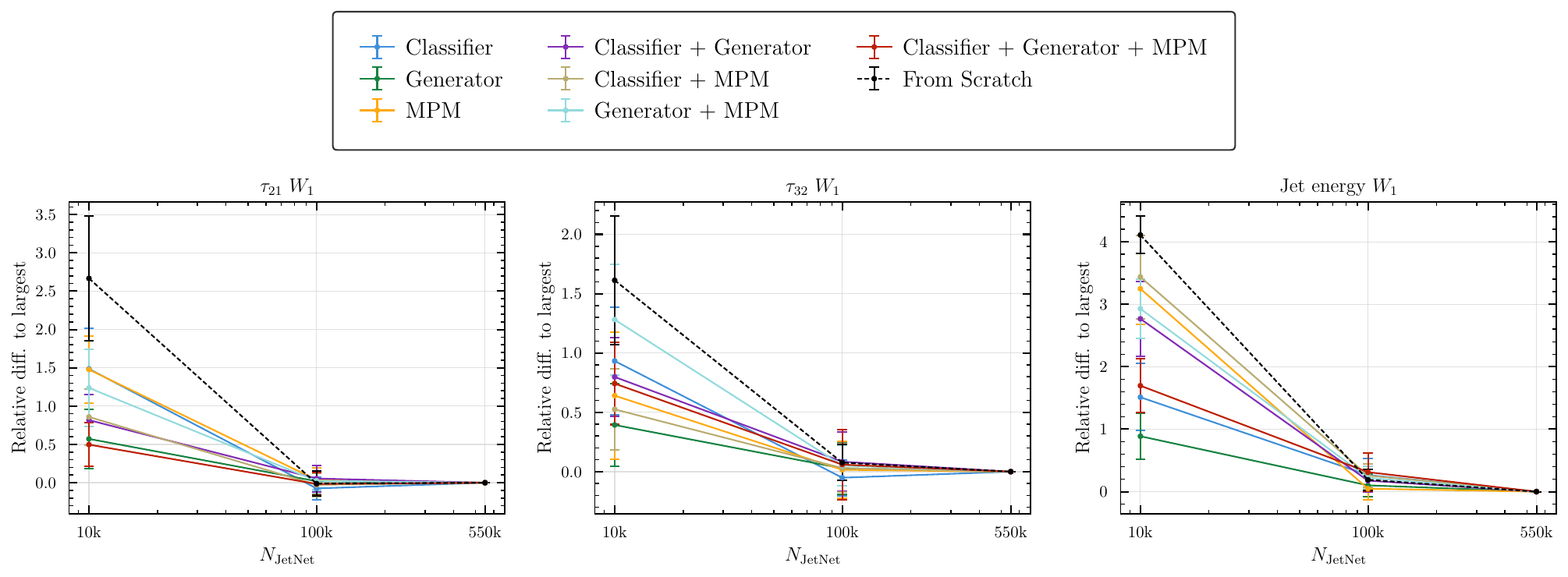}
    }\\{\footnotesize (b) Small model size}\\
    \caption{
        Relative performance difference of the generative model as a function of the training
        dataset size compared to the corresponding model obtained with the full
        dataset size ($N_\mathrm{JetNet}=550\mathrm{k}$) for the medium model
        (top) and the small model (bottom).
        The relative performance difference is given by 
        $\left(W_1(N_\mathrm{JetNet}) - W_1(550\mathrm{k})\right)/W_1(550\mathrm{k})$
        for the corresponding metric.
    }
    \label{fig:jetnet_metrics_vs_dataset_size}
\end{figure}

\paragraph{From scratch baselines} 
The small model size provides the best from scratch baseline across the different
model sizes. The performance of the micro models is by a margin lower than the
corresponding small or medium counterparts, potentially due to the limited model
size that is simply not sufficient for this generative task.
The medium model is significantly better than the micro model, but also falls
short in performance compared to the from scratch baseline obtained with the small
model size.

\paragraph{Different pre-training strategies}
Small and medium models show the best performance when pre-trained 
solely or partially on the generative task. 
The combination of all three pre-training tasks (Classifier+Generator+MPM)
leads to the best downstream generative performance for the medium model while
the Classifier+Generator pre-training strategy is the best-performing for the small model.
Most notably, all methods that include the generative component during
pre-training outperform the from scratch baseline. 
This is not true for models that did not include the generative task during
pre-training.
Those models show worse performance than the from scratch baseline, which
already indicates that those pre-training strategies are not well-aligned with the
generative task.
Micro models show worse than from scratch performance on most metrics across all
pre-training methods. Yet, the effect seen for small and medium models is inversed
here: models pre-trained on the generative task show lower performance than the other
pre-training methods. This might be due to the limited model capacity, and the generative
model (which was pre-trained to generate JetClass-like jets) not being able to adjust to
the new dataset. Due to these limitations, the micro model is omitted in the
subsequent evaluation.

\begin{figure}[t]
    \centering
    \subfloat{
        \includegraphics[width=0.98\linewidth]{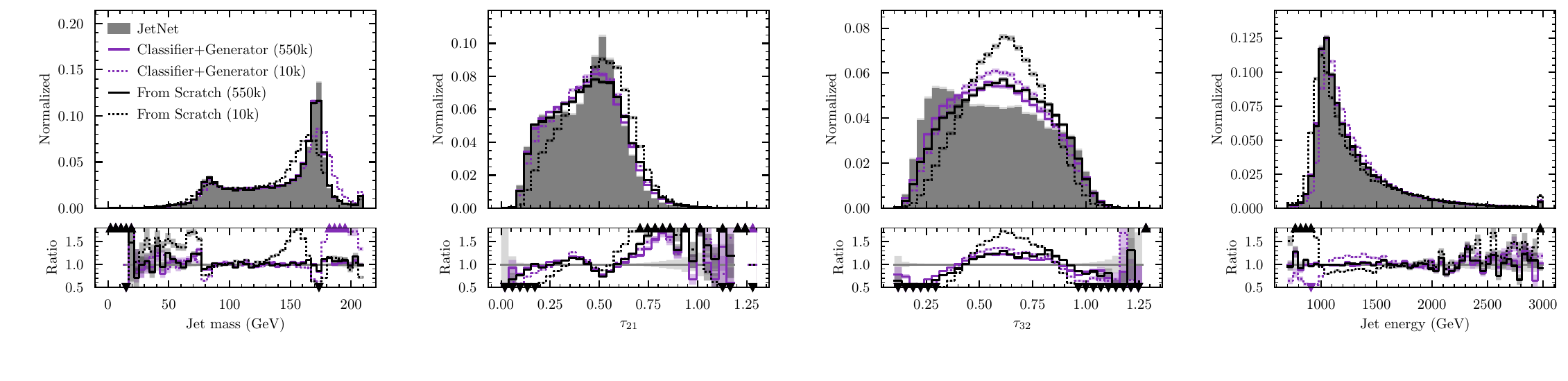}
    }\\\vspace{-1.5em}
    \subfloat{
        \includegraphics[width=0.98\linewidth]{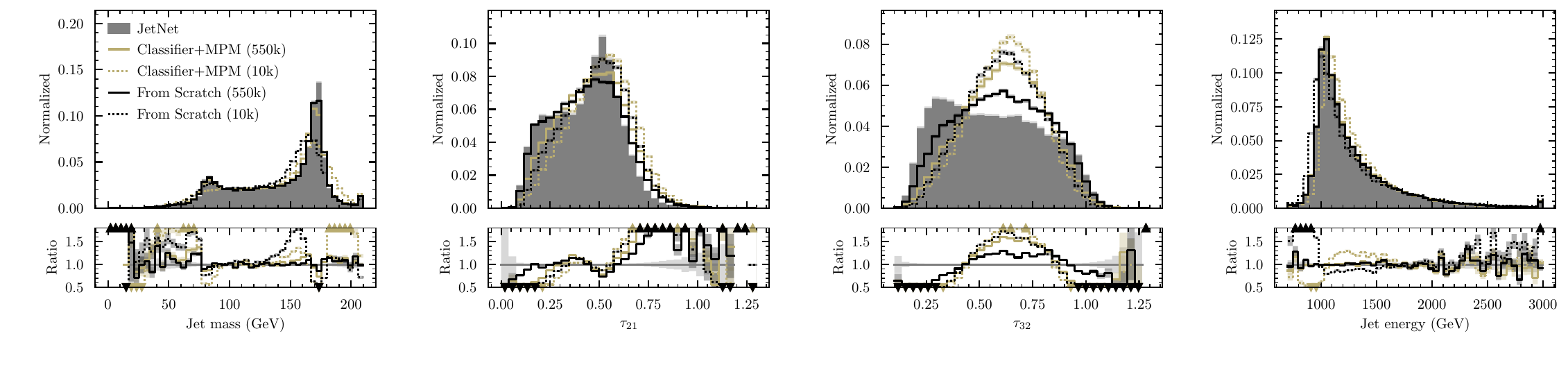}
    }\\
    \vspace{-0.5cm}
    \caption{
        Jet-level distributions of generated \textbf{top jets} obtained from the small model
        with two different pre-training strategies: Classifier+Generator (top panel)
        and Classifier+MPM (bottom panel) compared with the from scratch baseline.
        Two downstream dataset sizes are shown in dotted ($N_\mathrm{JetNet}=10\mathrm{k}$)
        and solid ($N_\mathrm{JetNet}=550\mathrm{k}$) lines.
        The gray histogram shows jets from the JetNet dataset.
    }
    \label{fig:jetnet_distributions_vs_dataset_size}
\end{figure}

\begin{figure}[t]
    \centering
    \subfloat{
        \includegraphics[width=0.98\linewidth]{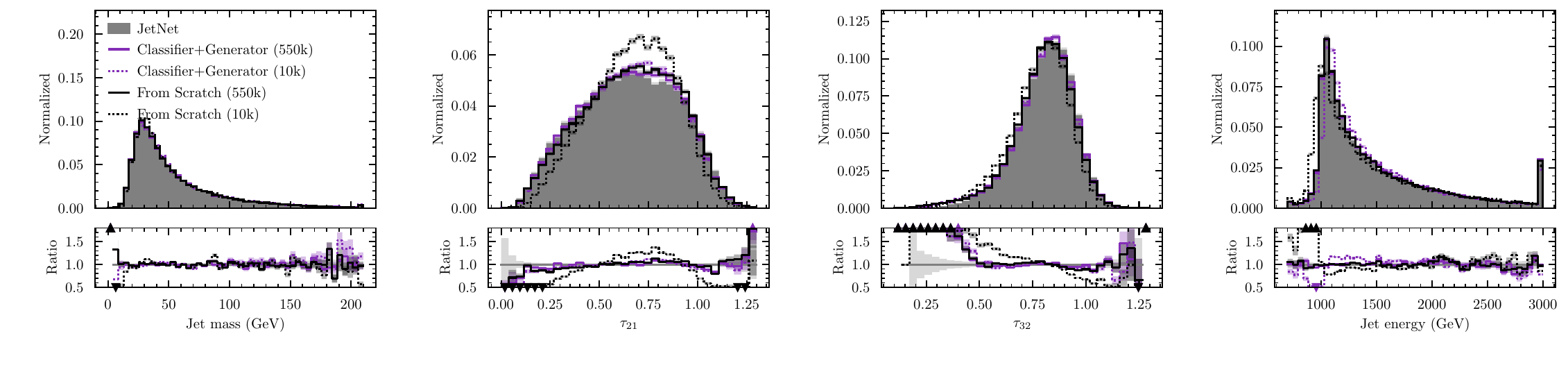}
    }\\\vspace{-1.5em}
    \subfloat{
        \includegraphics[width=0.98\linewidth]{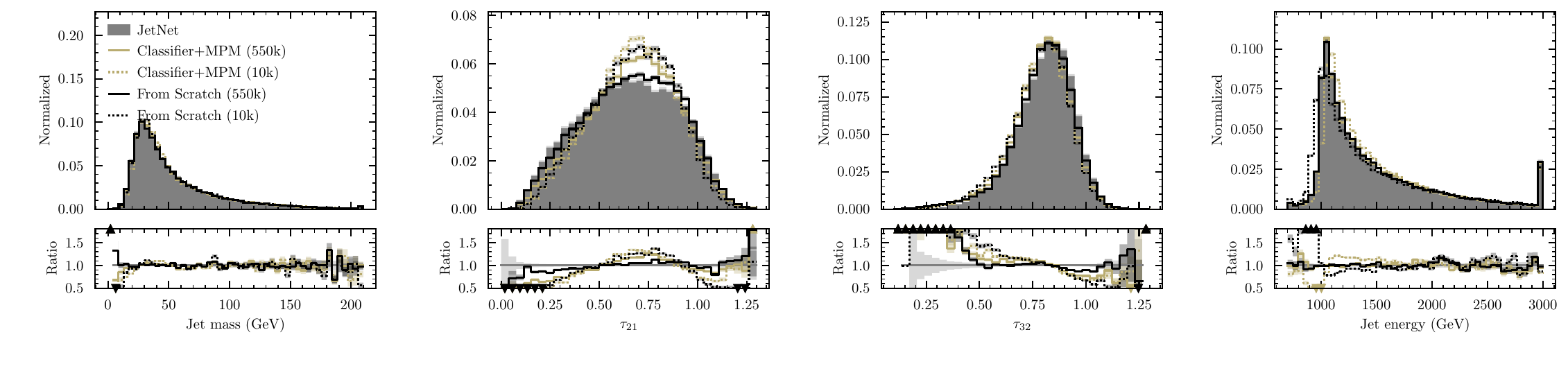}
    }\\
    \vspace{-0.5cm}
    \caption{
        Jet-level distributions of generated \textbf{light quark jets} obtained from the small model
        with two different pre-training strategies: Classifier+Generator (top panel)
        and Classifier+MPM (bottom panel) compared with the from scratch baseline.
        Two downstream dataset sizes are shown in dotted ($N_\mathrm{JetNet}=10\mathrm{k}$)
        and solid ($N_\mathrm{JetNet}=550\mathrm{k}$) lines.
        The gray histogram shows jets from the JetNet dataset.
    }
    \label{fig:jetnet_distributions_vs_dataset_size_light_quark}
\end{figure}

\begin{figure}[t]
    \centering
    \subfloat{
        \includegraphics[width=0.99\linewidth]{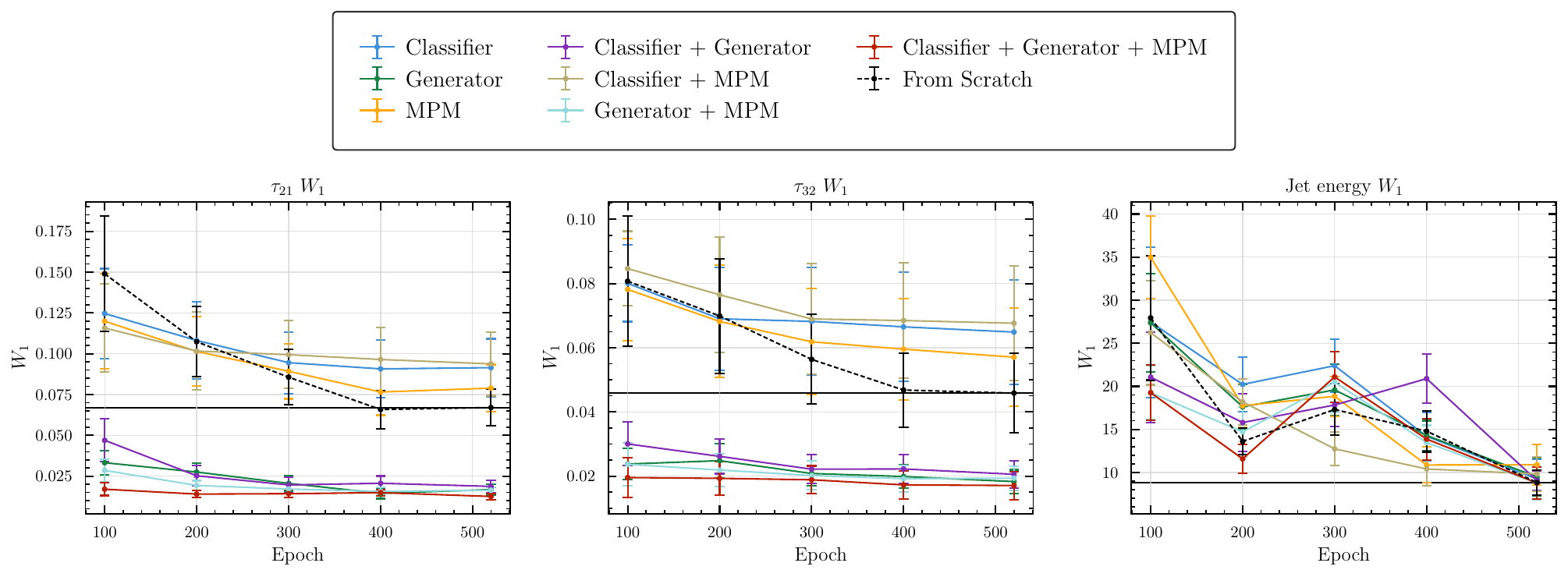}
    }\\{\footnotesize (a) Medium model size}\\\vspace{0.5em}
    \subfloat{
        \includegraphics[width=0.99\linewidth]{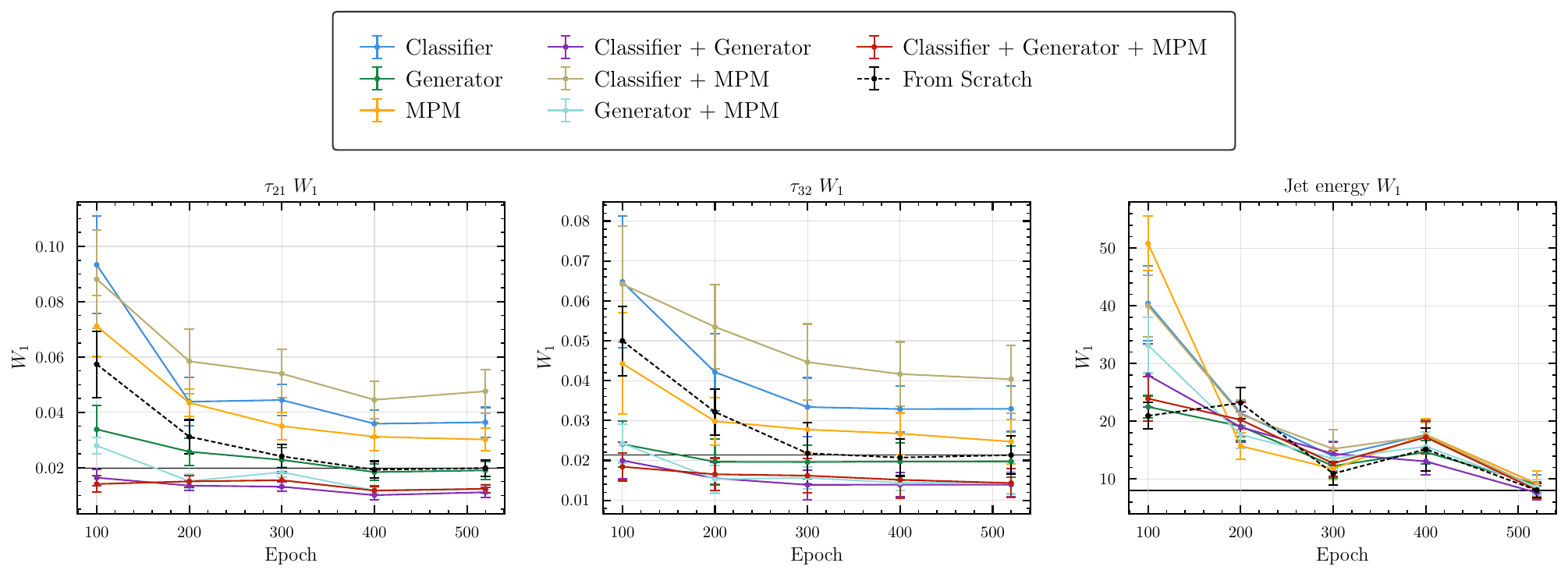}
    }\\{\footnotesize (b) Small model size}\\\vspace{0.5em}
    \caption{
        Performance of the generative models as a function of the training epoch 
        using a training dataset size $N_\mathrm{JetNet} = 550\mathrm{k}$ for
        (a) the medium and (b) the small model size.
        The black horizontal line marks the value corresponding to the final epoch 
        of the from scratch baseline.
    }
    \label{fig:jetnet_metrics_vs_epoch}
\end{figure}
\begin{table}[h]
    \centering
    \caption{
        Improvement $\Delta W_1(f)$ of the Wasserstein distance metric for
        different features $f$ between epoch 100 and epoch 520.
    }
    \label{tab:jetnet_metrics_improvement_epoch_100_to_520}

\vspace{0.5em} (a) Medium model size \\[0.5em]

\begin{tabular}{lcccc}
\toprule
 & $\Delta W_1(\tau_{21})$ & $\Delta W_1(\tau_{32})$ & $\Delta W_1(\mathrm{Jet~energy})$ & $\Delta W_1(\mathrm{Jet~mass})$ \\
 & $\times 10^{-2}$ & $\times 10^{-2}$ & (GeV) & (GeV) \\
\midrule
From Scratch & $8.2 \pm 3.7$ & $3.5 \pm 2.4$ & $19.1 \pm 7.4$ & $6.5 \pm 1.8$ \\
Generator & $1.6 \pm 0.8$ & $0.5 \pm 0.6$ & $17.9 \pm 6.1$ & $\mathbf{2.2 \pm 1.2}$ \\
Classifier + Generator + MPM & $\mathbf{0.4 \pm 0.4}$ & $\mathbf{0.2 \pm 0.8}$ & $\mathbf{10.5 \pm 3.7}$ & $4.2 \pm 1.3$ \\
Classifier + Generator & $2.8 \pm 1.4$ & $0.9 \pm 0.8$ & $12.0 \pm 5.4$ & $6.2 \pm 2.1$ \\
Generator + MPM & $1.2 \pm 0.7$ & $0.4 \pm 0.8$ & $10.5 \pm 3.6$ & $3.2 \pm 1.1$ \\
Classifier & $3.3 \pm 3.3$ & $1.5 \pm 2.0$ & $18.2 \pm 9.0$ & $8.9 \pm 1.4$ \\
MPM & $4.1 \pm 3.3$ & $2.1 \pm 2.2$ & $24.1 \pm 5.3$ & $5.2 \pm 1.8$ \\
Classifier + MPM & $2.2 \pm 3.3$ & $1.7 \pm 2.1$ & $16.4 \pm 6.4$ & $9.0 \pm 2.1$ \\
\bottomrule
\end{tabular}

\vspace{1.5em} (b) Small model size \\[0.5em]

\begin{tabular}{lcccc}
\toprule
 & $\Delta W_1(\tau_{21})$ & $\Delta W_1(\tau_{32})$ & $\Delta W_1(\mathrm{Jet~energy})$ & $\Delta W_1(\mathrm{Jet~mass})$ \\
 & $\times 10^{-2}$ & $\times 10^{-2}$ & (GeV) & (GeV) \\
\midrule
From Scratch & $3.8 \pm 1.2$ & $2.9 \pm 1.0$ & $\mathbf{12.9 \pm 2.6}$ & $2.8 \pm 0.4$ \\
Generator & $1.5 \pm 0.9$ & $0.4 \pm 0.7$ & $13.5 \pm 3.0$ & $1.7 \pm 0.9$ \\
Classifier + Generator + MPM & $\mathbf{0.2 \pm 0.3}$ & $\mathbf{0.4 \pm 0.5}$ & $15.8 \pm 4.1$ & $\mathbf{0.6 \pm 0.3}$ \\
Classifier + Generator & $0.5 \pm 0.3$ & $0.6 \pm 0.5$ & $20.4 \pm 5.6$ & $0.8 \pm 0.3$ \\
Generator + MPM & $1.6 \pm 0.3$ & $1.0 \pm 0.6$ & $24.6 \pm 5.0$ & $1.4 \pm 0.4$ \\
Classifier & $5.7 \pm 1.8$ & $3.2 \pm 1.7$ & $31.2 \pm 6.7$ & $2.8 \pm 1.0$ \\
MPM & $4.1 \pm 1.2$ & $2.0 \pm 1.4$ & $41.6 \pm 5.2$ & $2.3 \pm 0.8$ \\
Classifier + MPM & $4.1 \pm 2.0$ & $2.4 \pm 1.7$ & $31.5 \pm 5.5$ & $2.6 \pm 1.0$ \\
\bottomrule
\end{tabular}

\end{table}

\subsubsection{Scan over dataset size}

To assess the effect of downstream dataset size in the generative task, the
performance difference of models trained on either 10k or 100k JetNet jets
is compared to the values shown in Table~\ref{tab:jetnet_metrics_full_dataset_last_epoch}.
The corresponding relative performance difference among different downstream
dataset sizes is shown in Fig.~\ref{fig:jetnet_metrics_vs_dataset_size},
illustrating how close the different models get to the best performance (at
$N_\mathrm{JetNet}=550\mathrm{k}$ with the same pre-training strategy).
The Figure shows the corresponding metrics for small and medium model sizes and all
seven pre-training methods. 
All models benefit from being exposed to a larger training sample during downstream
generative training. 
The extend of performance improvement varies for the different metrics and different model sizes.
The most notable change is seen when increasing the dataset size from 10k to 100k jets.
Only minor improvements are seens when further increasing the dataset size to
$N_\mathrm{JetNet}=550\mathrm{k}$.

For medium-sized models with generative pre-training, the Wasserstein distance
calculated on the subjettiness ratio decreases by around 20\% to 60\% when going
from $N_\mathrm{JetNet} = 10\mathrm{k}$ to $N_\mathrm{JetNet} = 550\mathrm{k}$.
The medium size from scratch baseline also shows improved performance for
larger $N_\mathrm{JetNet}$, for example with a 100\% improvement in the jet energy $W_1$ when 
going from $N_\mathrm{JetNet} = 10\mathrm{k}$ to $N_\mathrm{JetNet} = 550\mathrm{k}$.
No clear performance gain is seen when varying the training dataset size for pre-trained models
that were not pre-trained on the generative task. This however does not mean that those models
perform well for all downstream dataset sizes. Instead, those models show bad agreement
with the target distributions no matter how large the training dataset they were exposed to
during downstream training on the JetNet dataset.

For small models we observe even larger performance gains with larger $N_\mathrm{JetNet}$.
A massive improvement is seen in the jet energy modeling, ranging from 100\% (Generator-only) to  400\% (From Scratch) when comparing the smallest and largest dataset size.
In the subjettiness ratio Wasserstein distances, the from scratch model shows
improvements of 260\% in $\tau_{21}$ and 160\% in $\tau_{32}$, showing that the
small from scratch baseline benefits the most from an increase in the dataset
size.
To put these numbers into context, distributions of top jets obtained from the methods with
the highest (Classifier+Generator) and lowest (Classifier+MPM) performance on
the small model size are shown in
Fig.~\ref{fig:jetnet_distributions_vs_dataset_size} for 10k and 550k training
jets, compared to the corresponding from scratch baselines. 
The from scratch baseline shows a clear improvement with more training data,
leading to much better agreement with the target distribution.
However, the difference is visually much less dominant for the pre-trained models.
While both methods benefit from additional training data during fine-tuning, the overall 
performance difference between the different pre-training methods is clearly visible.
The model pre-trained with the Classifier+Generator method also improves when the dataset
size is increased, but the improvements are visually less dominant. Even more important, the
Classifier+Generator model fine-tuned with $N_\mathrm{JetNet} = 10\mathrm{k}$ jets
clearly outperforms the corresponding from scratch baseline.
This is not true for the model pre-trained with the Classifier+MPM method, which
does not show a benefit over the from scratch model at either of the downstream
dataset sizes.
It should be noted that top jets are the most challenging jets to model within
the JetNet dataset, and were chosen for Fig.~\ref{fig:jetnet_distributions_vs_dataset_size}
as the mentioned trends are most visible there.
Other jet types show the same overall trends, with the distribution
of the generated jets showing better agreement with the target distribution. This can be seen in Fig.~\ref{fig:jetnet_distributions_vs_dataset_size_light_quark} for light quark jets with more jet types found in appendix plots Fig.~\ref{fig:jetnet_distributions_vs_dataset_size_class_gen_W-Z-light-gluon}
and Fig.~\ref{fig:jetnet_distributions_vs_dataset_size_class_mpm_W-Z-light-gluon}.

\begin{figure}[h]
    \centering
    \subfloat{
        \includegraphics[width=0.98\linewidth]{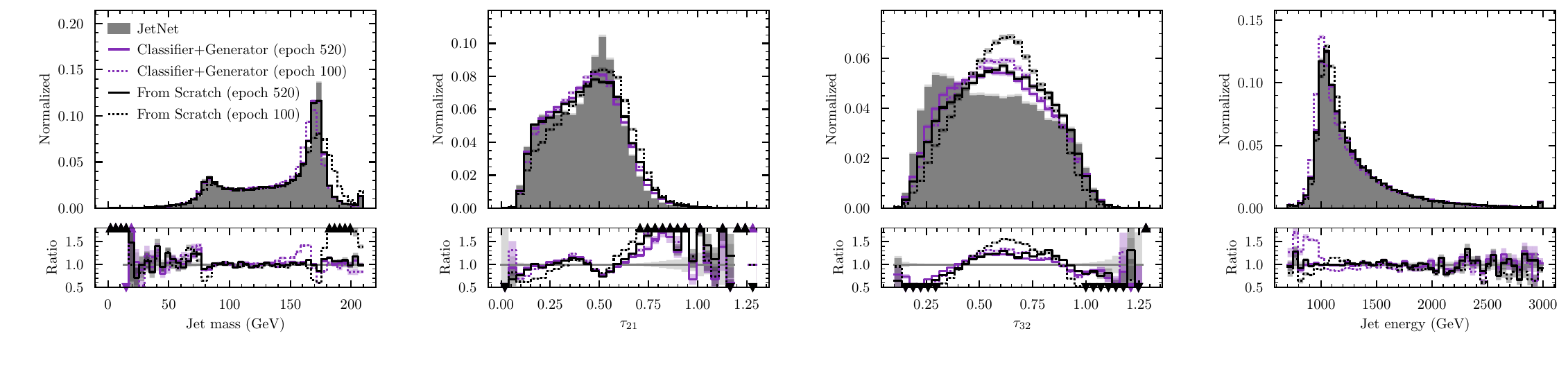}
    }\\\vspace{-1.5em}
    \subfloat{
        \includegraphics[width=0.98\linewidth]{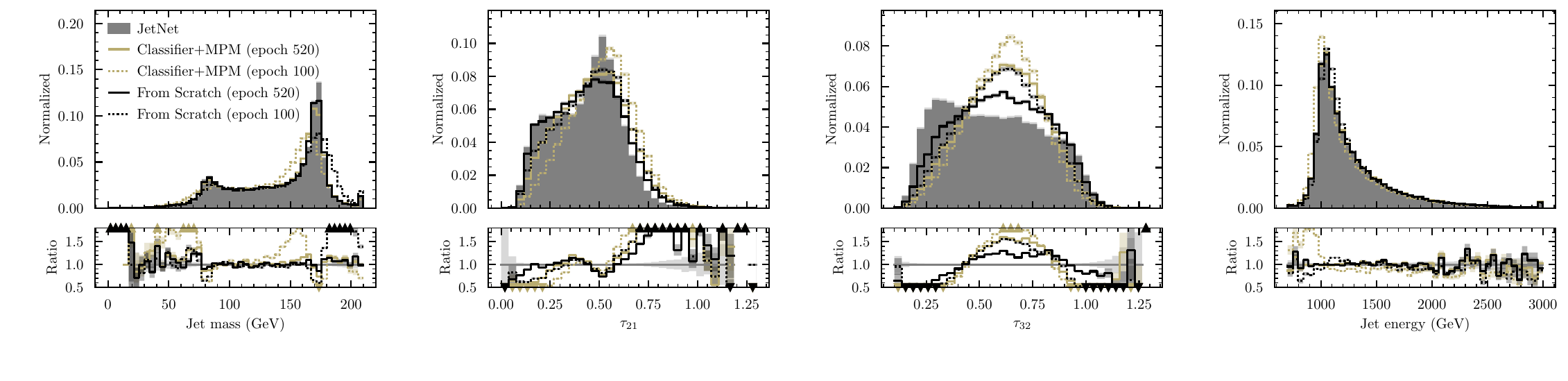}
    }\\
    \vspace{-0.7cm}
    \caption{
        Jet-level distributions of generated top jets obtained from the small model
        with two different pre-training strategies: Classifier+Generator (top panel)
        and Classifier+MPM (bottom panel) compared with the from scratch baseline.
        Two evaluated epochs are shown in dotted (epoch 100)
        and solid (epoch 520) lines. The gray histogram shows jets from the JetNet dataset.
    }
    \label{fig:jetnet_distributions_vs_epoch}
\end{figure}

\subsubsection{Performance vs. training duration}

To investigate how the pre-training affects the convergence speed of the
downstream training, we also evaluate the models trained on the whole JetNet
dataset as a function of the training epoch.
The corresponding curves of the previously introduced metrics are shown in
Fig.~\ref{fig:jetnet_metrics_vs_epoch}. Further plots with the extended
set of metrics are shown in Fig.~\ref{fig:jetnet_metrics_vs_epoch_medium_extended} 
and Fig.~\ref{fig:jetnet_metrics_vs_epoch_small_extended} in App.~\ref{app:othergenresults}.
Some features,  e.g. the jet energy shown in Fig.~\ref{fig:jetnet_metrics_vs_epoch}
do not show a clear improvement in convergence speed when pre-trained with either of the
investigated methods.
Other features such as the subjettiness ratios, however, show a clear trend.
Those show again the clear overall performance gain obtained from generative
pre-training (single-objective or combined with classification or MPM), with
those models performing better than the from scratch baseline at any of the
evaluated checkpoints. Even more, the generative pre-trained models already
outperform the final from-scratch performance after 100 training epochs.
For small models this effect is less pronounced, but still visible. Generative
pre-trained models with that model size surpass the final from scratch performance
in $\tau_{21}$ and $\tau_{32}$ already after 100 or 200 epochs, except for the 
generator-only pre-trained model that takes 400 epochs to surpass the final from-scratch
performance.
Another notable effect seen in the subjettiness ratios is the overall improvement
from epoch 100 to epoch 520, with the corresponding values shown in
Table~\ref{tab:jetnet_metrics_improvement_epoch_100_to_520}.
Most generative pre-trained models only show modest improvements between
epoch 100 and epoch 520, indicating that even short fine-tunings of those
models can yield performance close to the performance obtained with longer
downstream training.
The remaining pre-training methods, together with the from scratch baseline,
show larger performance gains when trained for longer, revealing that
the corresponding pre-trainings yield a model state that needs significantly
more computationally expensive fine-tuning.
The respective distributions of top jets are shown in Fig.~\ref{fig:jetnet_distributions_vs_epoch}
for the Classifier+Generator, the Classifier+MPM and the from scratch model,
where this effect is also visually noticeable, underlining yet again the
importance of including the generative task during pre-training if downstream
generation is of interest.
As mentioned earlier in the comparison of different downstream dataset sizes,
the overall quality of the generated jets is much better for the other jet
types, with the corresponding plots shown in the appendix in
Fig.~\ref{fig:jetnet_distributions_vs_epoch_class_gen_W-Z-light-gluon} and 
Fig.~\ref{fig:jetnet_distributions_vs_epoch_class_mpm_W-Z-light-gluon}.

\section{Conclusion}
\label{sec:conclusion}

We present a  study of pre-training objectives
for jet foundation models, using the \omni framework. Holding the
architecture fixed and varying only the active task heads, we
benchmark seven pre-training configurations across three axes,
pre-training dataset size, finetuning dataset size, and model size.
All models are pre-trained on JetClass and fine-tuned on two
representative downstream tasks, top-quark jet tagging, and JetNet
conditional generation. We aim to disentangle the
contributions of supervised, generative, and self-supervised
signals to the downstream transfer, and to map out how that picture
depends on the amount of data and the capacity available at both
stages, given we have access to labels at scale.

The top-tagging results show that no single pre-training objective is
universally optimal. When labeled datasets for pre-training and model capacity are available, plain
supervised classifier pre-training is the best configuration at every
model size and scales near-linearly with the amount of pre-training
data when performing downstream classification. The ranking inverts in the low-label regime, at $10^{4}$ top
labels, Classifier+MPM is the top-ranked configuration at every model
size, and at the smallest model size, standalone MPM outperforms pure
classifier pre-training despite using no labels at all. Standalone MPM
is also the most scale-efficient single-objective configuration with
respect to model capacity, with a slope roughly $2\times$ that of the
pure classifier even though the classifier pre-training is still the most performant. 
We further found that pre-training loss does not reliably track downstream
performance. Instead, downstream performance peaks well before pre-training
converges. 
The value of additional pre-training data may depend sharply on the
downstream sample size, ranging from near-linear gains when downstream labels
are abundant to essentially no gain (or, for the pure classifier, mild
degradation) when downstream labels are scarce. This effect needs to be confirmed at larger model sizes.
Finally, we notice that standalone flow-matching-based generative pre-training does not lead
to a performance improvement in downstream classification. This indicates that
the flow matching task is not well aligned with the classification task, and that the
learned representation can not be easily transferred to the classification task.
Yet, the classification performance can be enhanced by combining flow matching
with either MPM in a self-supervised setting or with classification if labels
are available during pre-training. 
This result, i.e. that purely generative pre-training can yield sub-optimal
pre-trained model components, but can be improved in a self-supervised way by being
combined with MPM is similar to the findings in~\cite{Birk:2025fbs}, where this
combination was studied in the form of next token prediction (generative) and
masked token prediction (MPM).

Our studies on the generative downstream task based on the JetNet dataset
show the importance of aligning the pre-training task(s) with the downstream task. While the performance obtained with different pre-training strategies varies among the different feature distributions investigated here, we notice that
neither the Classifier nor the MPM task help for downstream generation quality, unless combined with the generative task during pre-training.
All models that were pre-trained at least partially on the generative task show a performance gain when fine-tuned to jet generation on another dataset. Not only do these fine-tuned models outperform the corresponding from-scratch baseline, but the gap between the performance early on and at the end of the downstream training is much smaller for jet substructure features like the subjettiness ratios, often surpassing the results obtained from the model trained from scratch with 5 times fewer training epochs.

Taken together, these results suggest that the optimal pre-training objective is not a property of the model alone but of the joint configuration of model
capacity, pre-training data, and downstream data availability and
compatibility. The combination of multiple objectives, while not necessarily always the best performing for each individual task, provides a balance in benefits from both downstream tasks studied in this work. Additional pre-training targets avoid over-specialization from the supervised pre-training and promote robustness and improved transferability during downstream tasks.

Three directions for future work follow naturally from this study. First, our
finetuning-trajectory results show that pre-training loss is an
unreliable proxy for downstream performance, with downstream
background rejection peaking well before pre-training converges. Understanding this effect, its scaling, and mitigating it by introducing pre-training modifications (including multi-objective pre-training) would be an interesting next step. It may be necessary to add additional downstream tasks, especially those with less overlap with the pre-training tasks.

Second, we have benchmarked seven configurations at three model
sizes spanning $\sim 100$k to $\sim 51$M parameters, which we believe is
sufficient to reveal pre-training ordering, but does not yet probe the regime
where the largest HEP foundation models operate ($\sim 10^{9}$ parameters and
pre-training on billions of tokens~\cite{Bhimji:2025isp}). 
Three configurations are of
particular interest at this scale, pure Classifier, which dominates
when capacity and labels are plentiful and seem to have the cleanest
data scaling, standalone MPM, which exhibits the steepest
parameter-scaling and data scaling (at low downstream dataset sizes) slope among single-objective configurations
despite lower absolute performance (which may change with scale), and Classifier+Generator+MPM, which combines the strengths of all three signals. Scaling these three configurations to the size of the published \textsc{OmniLearned}-l model and to the full ${\sim} 10^{9}$-jet pre-training would reveal the continuation of these trends. 

Finally, understanding the origin of the seeming orthogonality between generative and classification tasks would be quite interesting.  Classification focuses on the differences between datasets while generation focuses on the bulk of the datasets - while a shared latent space may be effective at solving both types of problems, it need not be.  The difference in model representation between the two pre-training tasks may be a good probe to answer this and broader interpretability questions.

\section*{Code Availability}

The code used for this study can be found at \url{https://github.com/ibrahimEls/PretrainingForScience}. 

\section*{Acknowledgments}

This research used resources of the National Energy Research Scientific Computing Center, a DOE Office of Science User Facility supported by the Office of Science of the U.S. Department of Energy under Contract No. DE-AC02-05CH11231 using NERSC award HEP-ERCAP0035546.
I.E. is supported in part by the Connaught International Scholarship at University of Toronto and Natural Sciences and Engineering Research Council of Canada (NSERC) Canada Graduate Research Scholarship. I.E thanks Yonatan Kahn, Andrew Larkoski, and David Curtin for valuable conversations on various aspects of parton showers. I.E. thanks Marat Freytsis and Anthropic for access to Claude Code, which assisted with aspects of orchestrating training models used in this work. 
V.M. is supported by JST EXPERT-J, Japan Grant Number JPMJEX2509.
J.B. and G.K. are supported by the DFG under the German Excellence Initiative -- EXC 2121  Quantum Universe – 390833306.
J.B. is supported by a scholarship of the German Academic Exchange Service~(DAAD).
J.B. also acknowledges support via the Hamburg VISTA/VISOR --- Virtual Initiative for Science \& Technology in AI --- network.
B.N. is supported by the Department of Energy (DOE), Office of Science under contract DE-AC02-76SF00515.

\bibliography{bib}
\bibliographystyle{apsrev4-2}

\appendix
\section{\omni Modules}
\label{app:omnilearned}
Below is the description of the defining aspects of the \omni foundation model. A more detailed description of the model can be found in the original \omni and \textsc{OmniLearn} papers. \cite{Bhimji:2025isp,Mikuni:2024qsr,Mikuni:2025tar}
\paragraph{Local Embedding}
A local block captures short-range correlations between nearby particles
before the main body. For each particle $i$, we identify its $k=10$ nearest neighbors in $(\eta, \phi)$ space and compute, for each pair $(x_i,x_j)$, the pairwise features
\begin{equation}
    f(x_i,x_j) =
    \bigl[
        x_i-x_j,~
        \log m(x_i,x_j),~
        \log \Delta_R(x_i,x_j),~
        \log k_T(x_i,x_j)
    \bigr],
    \label{eq:omni-interaction}
\end{equation}
where $m(x_i,x_j)$ is the invariant mass of the pairwise four-vector sum, $\Delta_R$ is the distance in rapidity-azimuth space, and $k_T$ is the pairwise transverse momentum standard in jet analysis~\cite{Dreyer:2018nbf}. The features
$f(x_i,x_j)$ are passed through an MLP and then a transformer block
that attends only to the neighbors of each particle. The resulting local representation is added to the per-particle embedding $h_i$. The total embedding $h_i$ is given by,
\begin{equation}
    h_i =  \mathrm{MLP}_{\mathrm{kin}}(x_i)
           + \mathrm{EMB}_{\mathrm{PID}}(PID)
           + \mathrm{MLP}_{\mathrm{add}}(v_i) + \mathrm{ATTN}_\mathrm{local}(f(x_i,x_j)),
    \qquad x_i = (\Delta\eta_i,\Delta\phi_i,\log p_{T,i},\log E_i).
    \label{eq:omni-embed}
\end{equation}
where $x_i$ is the input jet constituent feature vector, of said jet constituent, and $v_i$ is additional vertex information for that constituent.
\paragraph{Physics-informed attention bias}
The same pairwise features $f(x_i,x_j)$ from
Eq.~\ref{eq:omni-interaction}, now evaluated over \textit{all} particle
pairs rather than only $k$-neighbors, are reused in the main body of
the model. Each pair is passed through a shared MLP that outputs one
scalar per attention head $B_{ij}$. 
These scalars are added as an additive bias to the corresponding self-attention
logits in every body transformer block as originally introduced in~\cite{Qu:2022mxj}.

\paragraph{Global attention}
The main body consists of $N_{\mathrm{body}}$ transformer blocks
operating on the full set of particle embeddings, with the interaction
bias added to every attention matrix. The model carries $N_{\mathrm{tok}}=4$
learnable global tokens~\cite{darcet2024vision} that are concatenated to the
particle set from the very first embedding block, so that they attend to all
particles throughout the body and aggregate jet-level information.
\section{Hyperparameters}
\label{app:hyper}

This appendix describes hyperparameters not stated in the main text.
Model architecture (depth, embedding width, attention-head count,
parameter count, and per-node local batch size) is given in
Tab.~\ref{tab:sizes}.

\subsection{Pre-training Hyperparameters}
\label{sec:app-pretrain-hparams}

Tab. \ref{tab:pretrain-size} lists the learning rate and weight decay
used for each size, optimizer, and pair.  As described in the main text,
configurations whose loss function includes the regression MPM head, as well as
all Medium runs, are trained with Ranger. The remaining configurations
use Lion.  Settings that are common to all pre-training run are listed in
Tab. \ref{tab:pretrain-common}.

\begin{table}[ht]
    \centering
    \caption{Size and optimizer specific pre-training hyperparameters.
    Lion is used unless the configuration is Medium or contains
    an active MPM head, in which case Ranger is used.}
    \label{tab:pretrain-size}
    \vspace{0.8em}
    \begin{tabular}{lcc}
        \toprule
        & Lion LR / WD & Ranger LR / WD \\
        \midrule
        Micro  & $1\cdot 10^{-3}$ / $0.01$ & $1\cdot 10^{-3}$ / $0.01$  \\
        Small  & $1\cdot 10^{-4}$ / $0.30$ & $1\cdot 10^{-3}$ / $0.01$  \\
        Medium & N/A                      & $5\cdot 10^{-4}$ / $0.01$  \\
        \bottomrule
    \end{tabular}
\end{table}

\begin{table}[ht]
    \centering
    \caption{Pre-training hyperparameters held fixed across every run.
    The optimizer hyperparameters listed apply to whichever optimizer is in
    use for a given configuration Lion or Ranger.}
    \label{tab:pretrain-common}
    \vspace{0.8em}
    \begin{tabular}{l r}
        \toprule
        Hyperparameter & Value \\
        \midrule
        Global batch size              & $8192$ \\
        Learning-rate schedule         & Cosine with linear warmup \\
        Warmup steps                   & $1000$ \\
        Lion betas $(\beta_1, \beta_2)$ & $(0.95,\,0.99)$ \\
        Ranger betas $(\beta_1, \beta_2)$ & $(0.95,\,0.999)$ \\
        Feature dropout                & $0.1$ \\
        Gradient accumulation          & $2$ \\
        Number of summary tokens       & $4$ \\
        Precision                      & FP32 \\
        \bottomrule
    \end{tabular}
\end{table}

\subsection{Top Tagging Hyperparameters}
\label{sec:app-toptag-hparams}

All top-tagging finetuning use Lion with the betas of
Tab. \ref{tab:pretrain-common}, a cosine learning-rate schedule with linear
warmup, a per-GPU batch size of $64$, $4$ nodes (global batch size $1024$),
and a fixed number of steps corresponding to a model-dependent epoch budget of 1.2 million top-tagging jets. When a pre-trained body is loaded the classification head receives a $5\times$ multiplier on the base learning rate.  Per-size values are summarized in Tab.~\ref{tab:toptag-size} and fixed settings are in Tab.~\ref{tab:toptag-common}.

\begin{table}[ht]
    \centering
    \caption{Top-tagging finetuning hyperparameters by model size.
    ``fs'' rows use a randomly initialized body.}
    \label{tab:toptag-size}
    \vspace{0.8em}
    \begin{tabular}{lcccc}
        \toprule
        & Learning rate & Weight decay & Warmup steps & Epoch budget \\
        \midrule
        Micro     & $5\cdot 10^{-5}$ & $0.1$ & $1028$ & $30$ \\
        Small     & $5\cdot 10^{-6}$ & $0.1$ & $1028$ & $30$ \\
        Medium    & $1\cdot 10^{-6}$ & $5.0$ & $0$    & $30$ \\
        Micro fs  & $1\cdot 10^{-4}$ & $0.5$ & $0$    & $30$ \\
        Small fs  & $5\cdot 10^{-4}$ & $0.5$ & $0$    & $15$ \\
        Medium fs & $1\cdot 10^{-5}$ & $0.5$ & $0$    & $30$ \\
        \bottomrule
    \end{tabular}
\end{table}

\begin{table}[ht]
    \centering
    \caption{Top-tagging hyperparameters held fixed across all sizes.}
    \label{tab:toptag-common}
    \vspace{0.5em}
    \begin{tabular}{l r}
        \toprule
        Hyperparameter & Value \\
        \midrule
        Optimizer                  & Lion \\
        Optimizer $(\beta_1, \beta_2)$ & $(0.95,\,0.99)$ \\
        Per-GPU batch size         & $64$ \\
        Number of nodes     & $4$ \\
        Global batch size          & $1024$ \\
        Gradient accumulation      & $1$ \\
        Head LR multiplier         & $5\times$ (pre-trained); $1\times$ (fs) \\
        Learning-rate schedule     & Cosine with linear warmup \\
        Precision                  & FP32 \\
        \bottomrule
    \end{tabular}
\end{table}

\subsection{Generation Hyperparameters}
\label{sec:app-gen-hparams}

Hyperparameters for the JetNet generative downstream task are given in
Tab. \ref{tab:genhparams}. 
Pre-trained weights are fine-tuned with a learning rate five times smaller
than the randomly initialized weights. The maximum training duration of
70k steps is kept fixed for all model sizes. This leads to a sufficient training
duration for the small models, though initial studies showed that medium-sized
models can benefit from longer trainings. 
However, due to computational constraints, the same maximum training step limit
is kept for the results presented for the medium size.

\begin{table}[ht]
    \centering
    \caption{
        Hyperparameters used for the generative downstream task.
        The stated learning rate corresponds to the learning rate used for
        untrained parameters. For pre-trained parameters a factor five smaller
        learning rate is used.
    }
    \label{tab:genhparams}
    \vspace{0.5em}
    \begin{tabular}{l|r}
        Hyperparameter & Value \\
        \midrule
        Optimizer & Lion \\
        Optimizer $(\beta_1, \beta_2)$ & $(0.95,\,0.99)$ \\
        Weight decay & 0.01 \\
        Global batch size & 4096 \\
        Max. training steps & 70\,000 \\
        Learning rate & (micro) $2\cdot 10^{-4}$ \\
                      & (small) $1\cdot 10^{-4}$ \\
                      & (medium) $5\cdot 10^{-6}$ \\
        Learning-rate schedule     & Cosine with linear warmup \\
        Precision                  & FP32 \\
    \end{tabular}
\end{table}

\section{More Top Jet Classification Results}
\label{app:MoreTopRes}
\begin{figure}[h]
    \centering
    \includegraphics[width=0.8\linewidth]{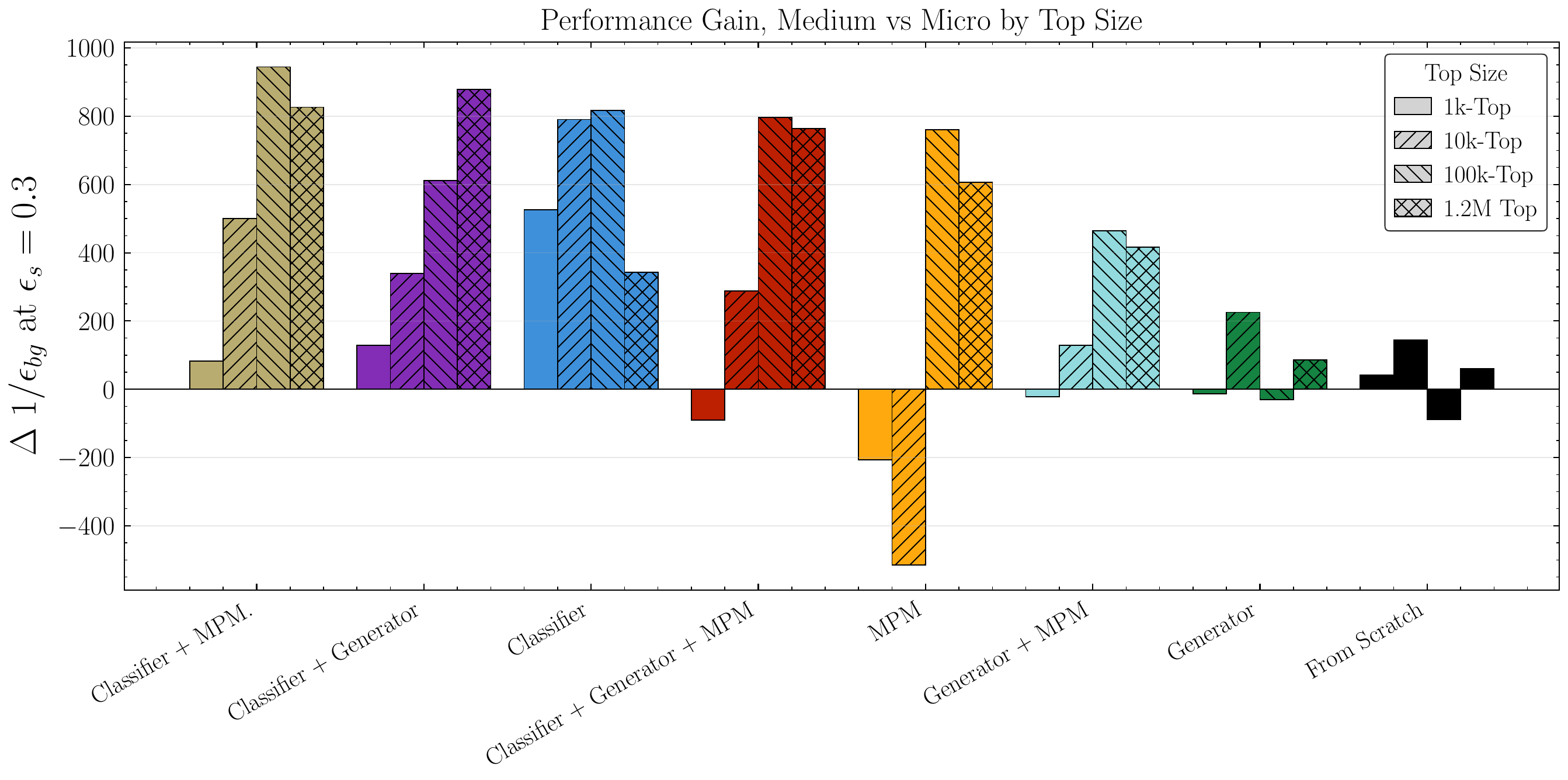}
    \caption{
        Top-Tagging $1/\epsilon_{\mathrm{bg}}$ performance gain from the micro
        model to the medium model for all $N_{ft}$.
    }
    \label{fig:eff-model-size-gain}
\end{figure}

\begin{figure}[h]
    \centering
    \includegraphics[width=0.8\linewidth]{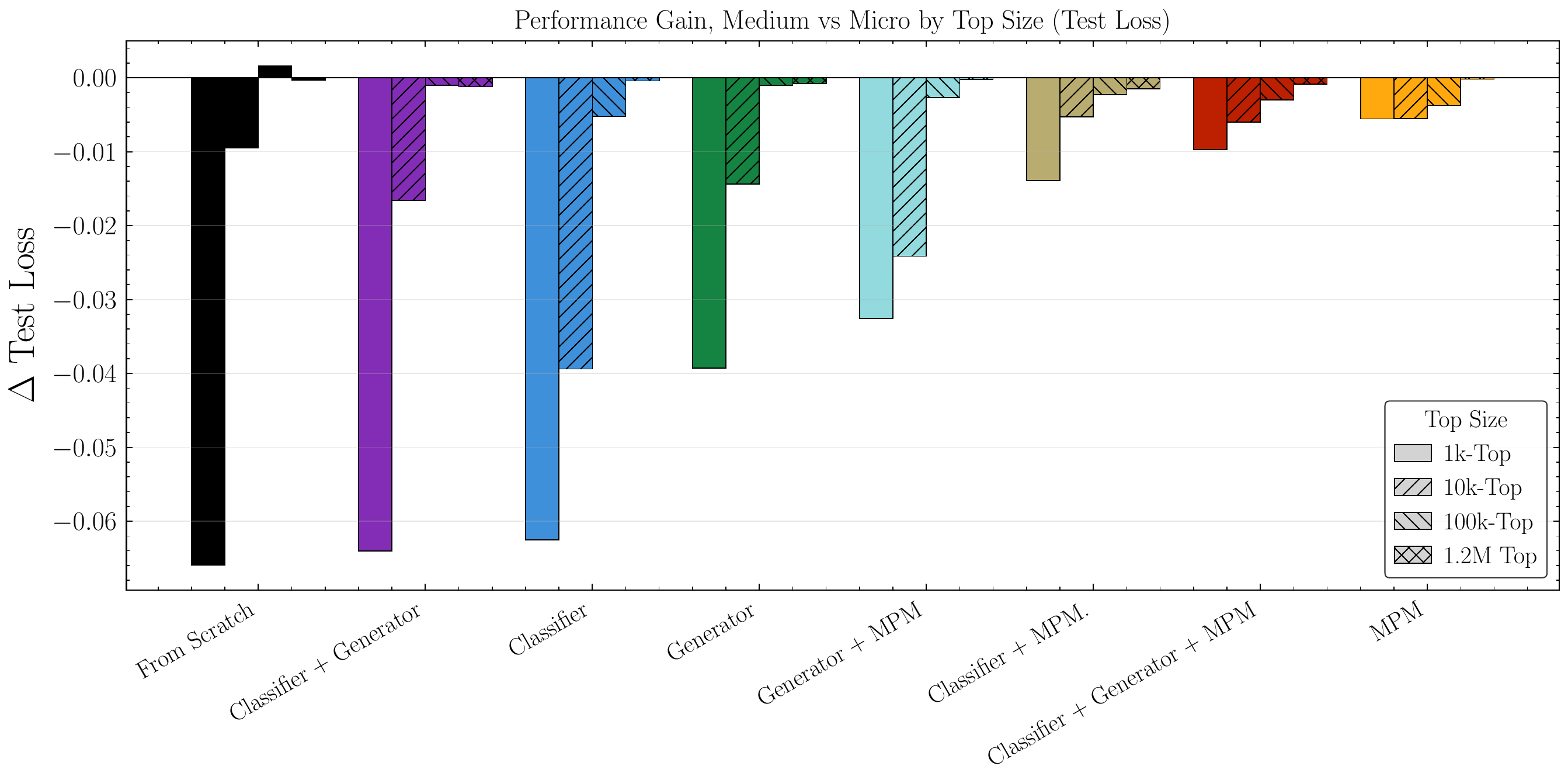}
    \caption{Top-Tagging Loss performance gain from the micro model to the medium model for all $N_{ft}$}
    \label{fig:loss-model-size-gain}
\end{figure}

\begin{figure}[h]
    \centering
    \subfloat{
        \includegraphics[width=0.8\linewidth]{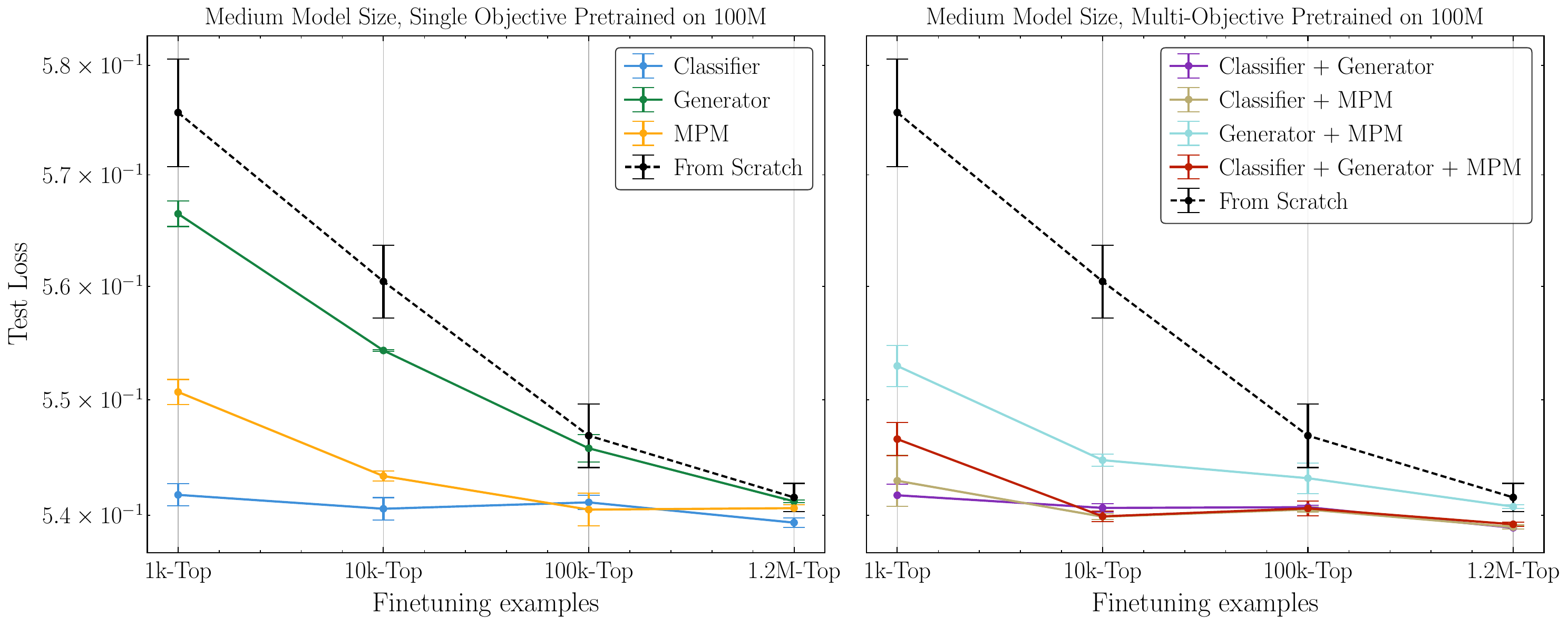}
    }\\{(a) Medium model size}\\
    \subfloat{
        \includegraphics[width=0.8\linewidth]{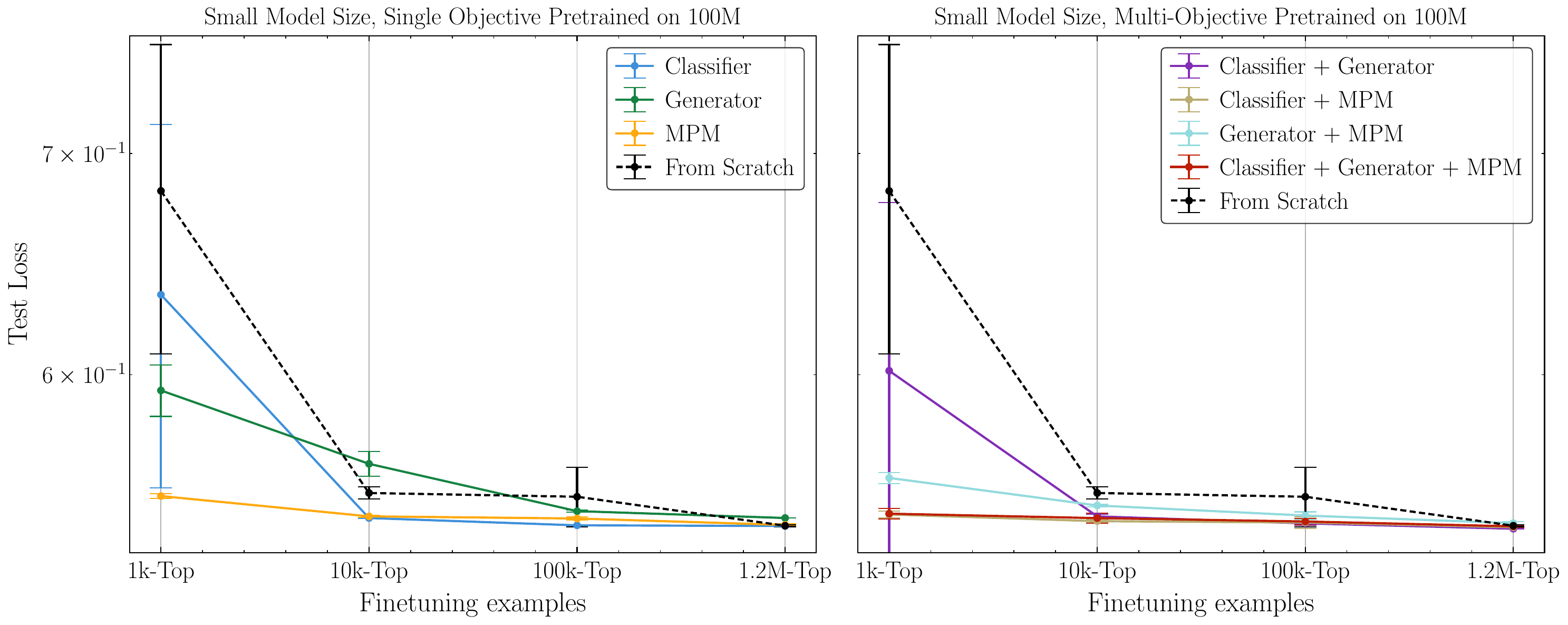}
    }\\{(b) Small model size}\\
    \subfloat{
        \includegraphics[width=0.8\linewidth]{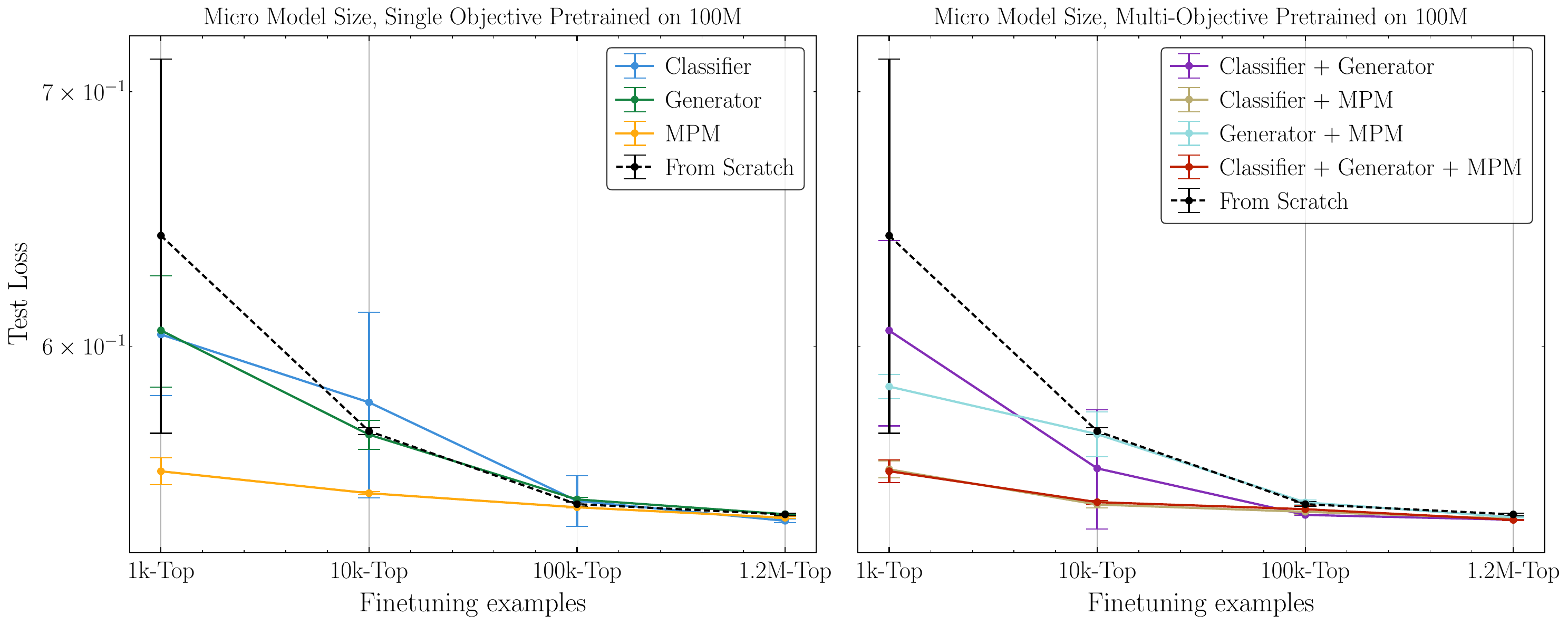}
    }\\{(c) Micro model size}\\
    \caption{
        Loss  vs top-tagging dataset size for different model sizes, pre-trained
        on 100M JetClass jets.
    }
    \label{fig:loss-medium-small-micro-doublet}
\end{figure}

\begin{figure}[h]
    \centering
    \includegraphics[width=0.8\linewidth]{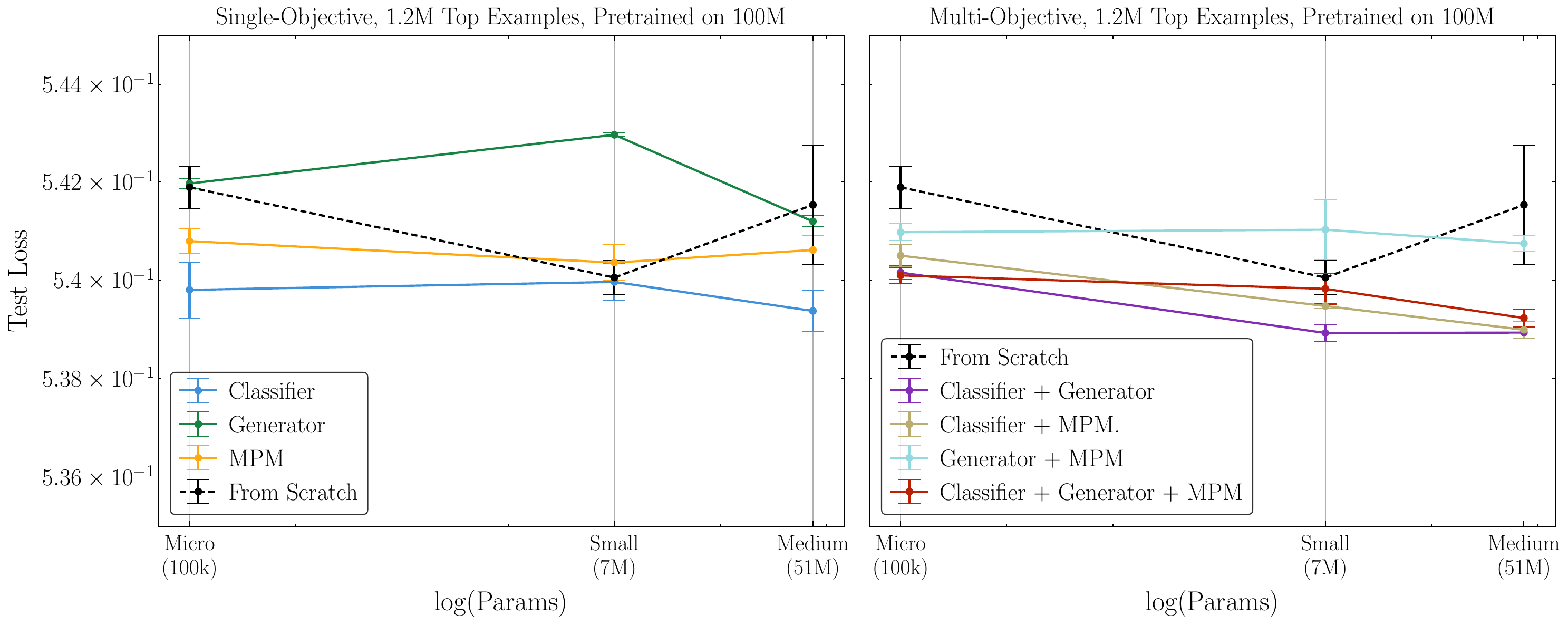}
    \caption{Top-tagging test loss
    vs model size for the seven pre-training configurations at
    $N_{\mathrm{ft}} = N_{\max}$.}
    \label{fig:loss-model-size-scaling}
\end{figure}

\begin{figure}[t]
    \centering
    \includegraphics[width=0.8\linewidth]{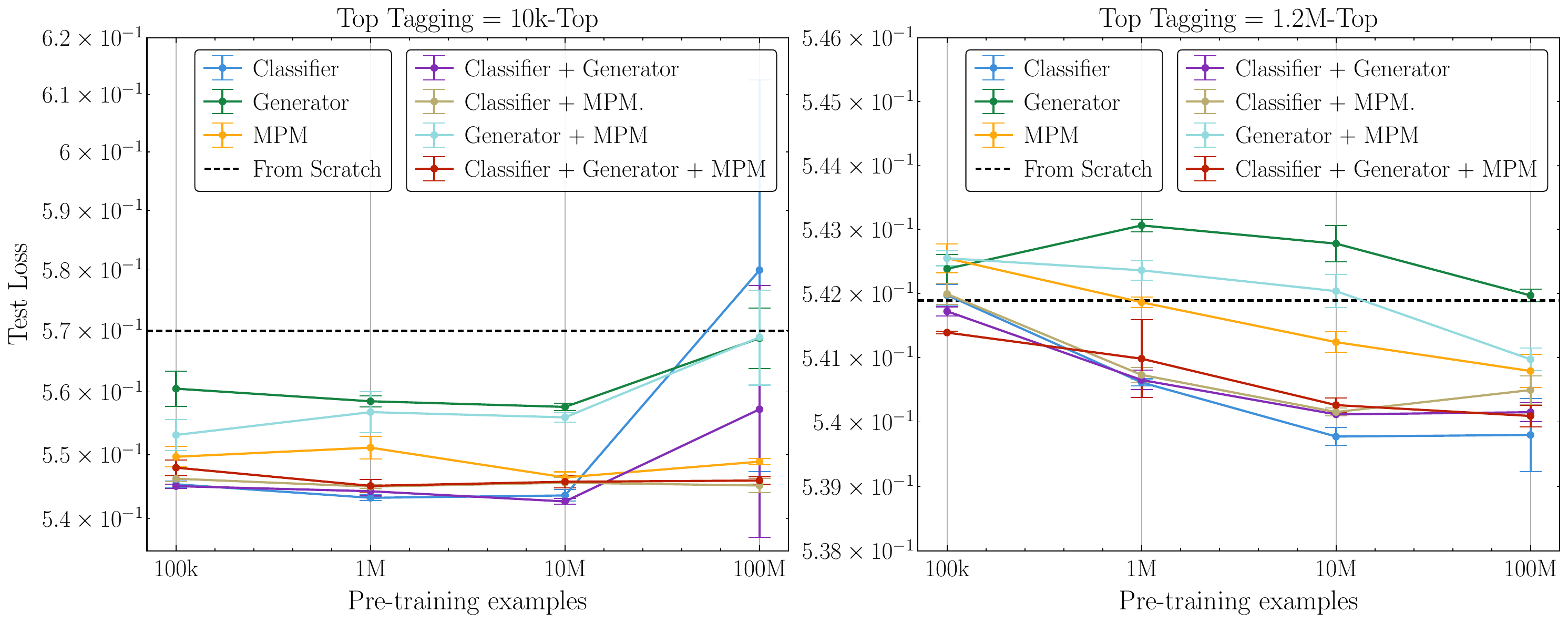}
    \caption{Loss vs pre-training dataset size
    $N_{\mathrm{pre}}$ for the Micro model under all seven
    pre-training configurations.
    $N_{\mathrm{ft}} = 10^{4}$ is shown on the left.
    $N_{\mathrm{ft}} = N_{\max} = 1.2\times10^{6}$ is shown on the right. The from-scratch baseline (dashed) does not depend on $N_{\mathrm{pre}}$.}
    \label{fig:loss-pre-train-scan}
\end{figure}

This section shows similar plots and tables as Section~\ref{sec:topres} but with the test loss (binary cross-entropy) metric as a function of dataset/pre-training mode/model size/pre-training size. Figures~\ref{fig:loss-medium-small-micro-doublet}\,a-c show the loss scaling with respect to top-tagging size for the various pre-training modes for all three model sizes. Figure~\ref{fig:loss-model-size-scaling} depicts the test-loss scaling with respect to model size. The linear fits for these scaling are found in Table~\ref{tab:size-scaling-fits-loss}. Figure \ref{fig:eff-model-size-gain} and Figure \ref{fig:loss-model-size-gain} shows the delta performance for all $N_{ft}$ gained by scaling the model from micro to medium. Figure~\ref{fig:loss-pre-train-scan} shows test loss scaling with respect to pre-training size. Table~\ref{tab:pre-train-scan-fits-loss} gives the scaling with respect to pre-training size. Finally, Table \ref{tab:medium-classifier-stepsweep} and Table \ref{tab:medium-classgen-stepsweep} contain raw pre-training loss and finetuning loss along the trajectory corresponding to Figure \ref{fig:finetune-trajectory}.


\begin{table}[h]
    \centering
    \caption{Linear fits of test loss vs $\log(N_{\mathrm{params}})$.
    Slope $m$ gives the loss reduction per decade of parameters,
    $R^2$ measures the linearity of the trend. Bold represents the most
    negative slope (best scaling).}
    \label{tab:size-scaling-fits-loss}
    \begin{tabular}{lcc}
        \toprule
        Configuration & $m, (10^{-4})$ & $R^2$ \\
        \midrule
        Classifier                       & $-1.20$ & $0.294$ \\
        Generator                        & $-1.56$ & $0.059$ \\
        MPM                              & $-0.94$ & $0.345$ \\
        From Scratch                     & $-$\textbf{2.69} & $0.146$ \\
        \midrule
        Classifier + Generator           & $-4.86$ & $0.899$ \\
        Classifier + MPM                 & $-$\textbf{5.58} & $1.000$ \\
        Generator + MPM                  & $-0.68$ & $0.386$ \\
        Classifier + Generator + MPM     & $-2.94$ & $0.837$ \\
        \bottomrule
    \end{tabular}
\end{table}

\begin{table}[h]
    \centering
    \caption{Linear fits of validation loss vs $\log(N_{\mathrm{pre}})$ for the
    Micro model, at two fine-tuning sample sizes. $m$ is the slope (in units of $10^{-3}$)
    and $R^2$ measures the linearity of the trend. Bold represents best scaling
    (most negative slope).}
    \label{tab:pre-train-scan-fits-loss}
    \begin{tabular}{lcccc}
        \toprule
        & \multicolumn{2}{c}{$N_{\mathrm{top}}=10^{4}$}
        & \multicolumn{2}{c}{$N_{\mathrm{top}}=N_{\max}$} \\
        \cmidrule(lr){2-3} \cmidrule(lr){4-5}
        Configuration & $m,(10^{-3})$ & $R^2$ & $m,(10^{-3})$ & $R^2$ \\
        \midrule
        Classifier
            & $\phantom{-0}10.42$ & $0.560$ & $-$\textbf{0.74} & $0.848$ \\
        Generator
            & $\phantom{-00}2.37$ & $0.363$ & $-0.15$ & $0.172$ \\
        MPM
            & $-$\textbf{0.70} & $0.212$ & $-0.59$ & $0.992$ \\
        \midrule
        Classifier + Generator
            & $\phantom{-00}3.49$ & $0.454$ & $-0.52$ & $0.819$ \\
        Classifier + MPM
            & $\phantom{-0}{-}0.26$ & $0.363$ & $-0.51$ & $0.666$ \\
        Generator + MPM
            & $\phantom{-00}4.64$ & $0.738$ & $-0.51$ & $0.859$ \\
        Classifier + Generator + MPM
            & $\phantom{-0}{-}0.54$ & $0.321$ & $-0.46$ & $0.954$ \\
        \midrule
        Average (over pre-train modes)
            & $\phantom{-00}2.78$ & $0.430$ & $-$\textbf{0.50} & $0.759$ \\
        \bottomrule
    \end{tabular}
\end{table}

\begin{table*}[h]
    \centering
    \small
    \setlength{\tabcolsep}{4pt}
    \caption{Small model, top tagging mean test loss, and mean background rejection
    $1/\epsilon_{\mathrm{bg}}$ at $\epsilon_s=0.3$ for all pre-training modes
    across fine-tuning sample sizes $N_{\mathrm{ft}}\in\{10^3,10^4,10^5,10^6\}$.
    Bold marks the best value per column.}
    \label{tab:small-allmodes}
    \begin{tabular}{lcccccccc}
        \toprule
        & \multicolumn{2}{c}{$N_{\mathrm{ft}}=10^{3}$}
        & \multicolumn{2}{c}{$N_{\mathrm{ft}}=10^{4}$}
        & \multicolumn{2}{c}{$N_{\mathrm{ft}}=10^{5}$}
        & \multicolumn{2}{c}{$N_{\mathrm{ft}}=10^{6}$} \\
        \cmidrule(lr){2-3}\cmidrule(lr){4-5}\cmidrule(lr){6-7}\cmidrule(lr){8-9}
        Configuration
            & Loss & $1/\epsilon_{\mathrm{bg}}$
            & Loss & $1/\epsilon_{\mathrm{bg}}$
            & Loss & $1/\epsilon_{\mathrm{bg}}$
            & Loss & $1/\epsilon_{\mathrm{bg}}$ \\
        \midrule
        Classifier
            & $0.6344$ & $291$
            & $0.5430$ & $850$
            & \textbf{0.5402} & \textbf{2210}
            & $0.5400$ & \textbf{2820} \\
        Generator
            & $0.5935$ & $30$
            & $0.5639$ & $132$
            & $0.5456$ & $739$
            & $0.5430$ & $1200$ \\
        MPM
            & $0.5513$ & $15$
            & $0.5436$ & $484$
            & $0.5428$ & $1711$
            & $0.5404$ & $2217$ \\
        From Scratch
            & $0.6820$ & $25$
            & $0.5525$ & $369$
            & $0.5511$ & $682$
            & $0.5400$ & $2091$ \\
        \midrule
        Classifier + Generator
            & $0.6016$ & $248$
            & $0.5436$ & $937$
            & $0.5409$ & $1766$
            & \textbf{0.5389} & $2781$ \\
        Classifier + MPM
            & \textbf{0.5443} & \textbf{556}
            & \textbf{0.5418} & \textbf{1375}
            & $0.5413$ & $1607$
            & $0.5395$ & \textbf{2820} \\
        Generator + MPM
            & $0.5583$ & $10$
            & $0.5477$ & $494$
            & $0.5439$ & $894$
            & $0.5410$ & $1679$ \\
        Classifier + Generator + MPM
            & $0.5446$ & $419$
            & $0.5429$ & $1061$
            & $0.5416$ & $1856$
            & $0.5398$ & $2505$ \\
        \bottomrule
    \end{tabular}
\end{table*}
\begin{table*}[h]
    \centering
    \small
    \setlength{\tabcolsep}{4pt}
    \caption{Medium model, top tagging mean test loss and mean background rejection 
    $1/\epsilon_{\mathrm{bg}}$ at $\epsilon_s=0.3$ for all pre-training modes
    across fine-tuning sample sizes $N_{\mathrm{ft}}\in\{10^3,10^4,10^5,10^6\}$.
    Bold marks the best value per column.}
    \label{tab:medium-allmodes}
    \begin{tabular}{lcccccccc}
        \toprule
        & \multicolumn{2}{c}{$N_{\mathrm{ft}}=10^{3}$}
        & \multicolumn{2}{c}{$N_{\mathrm{ft}}=10^{4}$}
        & \multicolumn{2}{c}{$N_{\mathrm{ft}}=10^{5}$}
        & \multicolumn{2}{c}{$N_{\mathrm{ft}}=10^{6}$} \\
        \cmidrule(lr){2-3}\cmidrule(lr){4-5}\cmidrule(lr){6-7}\cmidrule(lr){8-9}
        Configuration
            & Loss & $1/\epsilon_{\mathrm{bg}}$
            & Loss & $1/\epsilon_{\mathrm{bg}}$
            & Loss & $1/\epsilon_{\mathrm{bg}}$
            & Loss & $1/\epsilon_{\mathrm{bg}}$ \\
        \midrule
        Classifier
            & $0.5418$ & \textbf{662}
            & $0.5406$ & $1121$
            & $0.5411$ & $2038$
            & $0.5394$ & \textbf{3074} \\
        Generator
            & $0.5665$ & $\phantom{00}7$
            & $0.5543$ & $323$
            & $0.5458$ & $731$
            & $0.5412$ & $1520$ \\
        MPM
            & $0.5507$ & $12$
            & $0.5434$ & $\phantom{00}24$
            & \textbf{0.5405} & $1659$
            & $0.5406$ & $2399$ \\
        From Scratch
            & $0.5757$ & $74$
            & $0.5604$ & $219$
            & $0.5469$ & $807$
            & $0.5415$ & $1589$ \\
        \midrule
        Classifier + Generator
            & \textbf{0.5417} & $231$
            & $0.5406$ & $857$
            & $0.5407$ & $2022$
            & \textbf{0.5389} & $2874$ \\
        Classifier + MPM
            & $0.5430$ & $438$
            & \textbf{0.5399} & \textbf{1175}
            & \textbf{0.5405} & \textbf{2082}
            & $0.5390$ & $2832$ \\
        Generator + MPM
            & $0.5530$ & $10$
            & $0.5447$ & $260$
            & $0.5432$ & $1253$
            & $0.5407$ & $2063$ \\
        Classifier + Generator + MPM
            & $0.5466$ & $165$
            & $0.5399$ & $1007$
            & $0.5406$ & $1957$
            & $0.5392$ & $2792$ \\
        \bottomrule
    \end{tabular}
\end{table*}

\begin{table*}[h]
    \centering
    \small
    \setlength{\tabcolsep}{4pt}
    \renewcommand{\arraystretch}{1.05}
    \caption{Micro model, top tagging mean test loss and mean background rejection
    $1/\epsilon_{\mathrm{bg}}$ at $\epsilon_s=0.3$, for every pre-training objective and
    pre-training dataset size $N_{\mathrm{pt}}$, across fine-tuning sample sizes
    $N_{\mathrm{ft}}\in\{10^3,10^4,10^5,10^6\}$. Configuration blocks are separated by
    double rules; $N_{\mathrm{pt}}=10^{8}$ denotes the full ($100$M) pre-training set, and
    From Scratch ($N_{\mathrm{pt}}=\text{--}$) uses no pre-training. Bold marks the best
    value in each column across all rows.}
    \label{tab:micro-allmodes-allpt}
    \begin{tabular}{llcccccccc}
        \toprule
        & & \multicolumn{2}{c}{$N_{\mathrm{ft}}=10^{3}$}
        & \multicolumn{2}{c}{$N_{\mathrm{ft}}=10^{4}$}
        & \multicolumn{2}{c}{$N_{\mathrm{ft}}=10^{5}$}
        & \multicolumn{2}{c}{$N_{\mathrm{ft}}=10^{6}$} \\
        \cmidrule(lr){3-4}\cmidrule(lr){5-6}\cmidrule(lr){7-8}\cmidrule(lr){9-10}
        Configuration & $N_{\mathrm{pt}}$
            & Loss & $1/\epsilon_{\mathrm{bg}}$
            & Loss & $1/\epsilon_{\mathrm{bg}}$
            & Loss & $1/\epsilon_{\mathrm{bg}}$
            & Loss & $1/\epsilon_{\mathrm{bg}}$ \\
        \midrule
        \multirow{4}{*}{Classifier} & $10^{5}$ & $0.5554$ & $357$ & $0.5453$ & $602$ & $0.5439$ & $1038$ & $0.5420$ & $1457$ \\
         & $10^{6}$ & $0.5535$ & $324$ & $0.5432$ & $653$ & $0.5422$ & $1135$ & $0.5406$ & $1894$ \\
         & $10^{7}$ & $0.5861$ & $131$ & $0.5436$ & \textbf{796} & $0.5420$ & \textbf{1742} & \textbf{0.5398} & $2396$ \\
         & $10^{8}$ & $0.6043$ & $136$ & $0.5800$ & $331$ & $0.5463$ & $1221$ & \textbf{0.5398} & \textbf{2731} \\
        \hline\hline
        \multirow{4}{*}{Generator} & $10^{5}$ & $0.5647$ & $50$ & $0.5605$ & $188$ & $0.5446$ & $747$ & $0.5424$ & $1229$ \\
         & $10^{6}$ & $0.5684$ & $26$ & $0.5585$ & $226$ & $0.5467$ & $633$ & $0.5431$ & $1170$ \\
         & $10^{7}$ & $0.5694$ & $11$ & $0.5576$ & $210$ & $0.5463$ & $646$ & $0.5428$ & $1152$ \\
         & $10^{8}$ & $0.6058$ & $20$ & $0.5687$ & $98$ & $0.5468$ & $762$ & $0.5420$ & $1434$ \\
        \hline\hline
        \multirow{4}{*}{MPM} & $10^{5}$ & $0.6787$ & $19$ & $0.5497$ & $287$ & $0.5455$ & $814$ & $0.5425$ & $1335$ \\
         & $10^{6}$ & $0.5754$ & $45$ & $0.5511$ & $406$ & $0.5447$ & $766$ & $0.5419$ & $1350$ \\
         & $10^{7}$ & $0.5679$ & $91$ & $0.5464$ & $487$ & $0.5448$ & $886$ & $0.5412$ & $1644$ \\
         & $10^{8}$ & $0.5562$ & $218$ & $0.5489$ & $539$ & $0.5442$ & $899$ & $0.5408$ & $1793$ \\
        \hline\hline
        From Scratch & $\text{--}$ & $0.6416$ & $33$ & $0.5699$ & $75$ & $0.5453$ & $896$ & $0.5419$ & $1528$ \\
        \hline\hline
        \multirow{4}{*}{Classifier + Generator} & $10^{5}$ & $0.5555$ & $27$ & $0.5450$ & $505$ & $0.5431$ & $960$ & $0.5417$ & $1461$ \\
         & $10^{6}$ & \textbf{0.5481} & $178$ & $0.5443$ & $510$ & $0.5421$ & $1128$ & $0.5407$ & $1723$ \\
         & $10^{7}$ & $0.5530$ & $14$ & \textbf{0.5427} & $750$ & \textbf{0.5417} & $1342$ & $0.5401$ & $1897$ \\
         & $10^{8}$ & $0.6057$ & $102$ & $0.5572$ & $517$ & \textbf{0.5417} & $1411$ & $0.5402$ & $1996$ \\
        \hline\hline
        \multirow{4}{*}{Classifier + MPM} & $10^{5}$ & $0.5620$ & $211$ & $0.5462$ & $603$ & $0.5444$ & $939$ & $0.5420$ & $1365$ \\
         & $10^{6}$ & $0.5531$ & \textbf{437} & $0.5450$ & $450$ & $0.5423$ & $1077$ & $0.5407$ & $1761$ \\
         & $10^{7}$ & $0.5535$ & $311$ & $0.5456$ & $269$ & $0.5424$ & $1188$ & $0.5402$ & $1963$ \\
         & $10^{8}$ & $0.5569$ & $356$ & $0.5451$ & $675$ & $0.5428$ & $1137$ & $0.5405$ & $2006$ \\
        \hline\hline
        \multirow{4}{*}{Generator + MPM} & $10^{5}$ & $0.6366$ & $44$ & $0.5531$ & $297$ & $0.5451$ & $822$ & $0.5425$ & $1335$ \\
         & $10^{6}$ & $0.6009$ & $76$ & $0.5568$ & $263$ & $0.5450$ & $788$ & $0.5424$ & $1376$ \\
         & $10^{7}$ & $0.5650$ & $62$ & $0.5559$ & $280$ & $0.5454$ & $760$ & $0.5420$ & $1375$ \\
         & $10^{8}$ & $0.5855$ & $32$ & $0.5689$ & $131$ & $0.5458$ & $789$ & $0.5410$ & $1647$ \\
         \hline\hline
        \multirow{4}{*}{Classifier + Generator + MPM} & $10^{5}$ & $0.5517$ & $117$ & $0.5480$ & $542$ & $0.5446$ & $899$ & $0.5414$ & $1612$ \\
         & $10^{6}$ & $0.5531$ & $17$ & $0.5451$ & $434$ & $0.5429$ & $1058$ & $0.5410$ & $1897$ \\
         & $10^{7}$ & $0.5549$ & $14$ & $0.5457$ & $290$ & $0.5427$ & $1033$ & $0.5403$ & $1997$ \\
         & $10^{8}$ & $0.5563$ & $255$ & $0.5459$ & $719$ & $0.5436$ & $1161$ & $0.5401$ & $2028$ \\
        \bottomrule
    \end{tabular}
\end{table*}

\begin{table}[h]
    \centering
    \small
    \setlength{\tabcolsep}{6pt}
    \caption{Medium model, classifier pre-training, downstream top tagging as a function of
    pre-training checkpoint along the pre-training trajectory. Each row is one pre-training
    checkpoint fine-tuned on the full \texttt{1.2M-Top} set. Bold marks the best value in each metric column.}
    \label{tab:medium-classifier-stepsweep}
    \begin{tabular}{rrrr}
        \toprule
        Pre-train step & Pre-train val loss & Finetune val loss & $1/\epsilon_{\mathrm{bg}}\,(\epsilon_s{=}0.3)$ \\
        \midrule
        $2500$ & $0.7666$ & $0.1524$ & $1697$ \\
        $14707$ & $0.4501$ & $0.1445$ & $2243$ \\
        $26914$ & $0.4299$ & $0.1431$ & $2766$ \\
        $39121$ & $0.4198$ & $0.1422$ & $3059$ \\
        $51328$ & $0.4160$ & $0.1417$ & $2969$ \\
        $63535$ & $0.4139$ & $0.1415$ & \textbf{3106} \\
        $75742$ & $0.4119$ & $0.1413$ & $2926$ \\
        $87949$ & $0.4102$ & \textbf{0.1411} & \textbf{3106} \\
        $100156$ & $0.4097$ & \textbf{0.1411} & \textbf{3106} \\
        $112363$ & $0.4082$ & $0.1413$ & $2804$ \\
        $124570$ & $0.4062$ & $0.1414$ & $2804$ \\
        $136777$ & $0.4039$ & $0.1415$ & $2657$ \\
        $148984$ & $0.4070$ & $0.1419$ & $2657$ \\
        $161191$ & \textbf{0.4033} & $0.1422$ & $2969$ \\
        $175898$ & $0.4036$ & $0.1419$ & $2556$ \\
        $188105$ & $0.4053$ & $0.1422$ & $2926$ \\
        $200312$ & $0.4057$ & $0.1423$ & $2556$ \\
        $212519$ & $0.4089$ & $0.1423$ & $2804$ \\
        $224726$ & $0.4095$ & $0.1426$ & $3014$ \\
        $236933$ & $0.4126$ & $0.1428$ & $2729$ \\
        $249140$ & $0.4134$ & $0.1430$ & $2589$ \\
        $261347$ & $0.4185$ & $0.1431$ & $2524$ \\
        \bottomrule
    \end{tabular}
\end{table}

\begin{table}[t]
    \centering
    \small
    \setlength{\tabcolsep}{6pt}
    \caption{Medium model, classifier+generation pre-training, downstream top tagging as a function of
    pre-training checkpoint along the pre-training trajectory. Each row is one pre-training
    checkpoint fine-tuned on the full \texttt{1.2M-Top} set. Bold marks the best value in each metric column.}
    \label{tab:medium-classgen-stepsweep}
    \begin{tabular}{rrrr}
        \toprule
        Pre-train step & Pre-train val loss & Finetune val loss & $1/\epsilon_{\mathrm{bg}}\,(\epsilon_s{=}0.3)$ \\
        \midrule
        $2500$ & $4.1142$ & $0.1523$ & $1383$ \\
        $7500$ & $3.4389$ & $0.1487$ & $1787$ \\
        $14707$ & $3.3342$ & $0.1456$ & $2060$ \\
        $17207$ & $3.3287$ & $0.1451$ & $2348$ \\
        $19707$ & $3.3203$ & $0.1446$ & $2348$ \\
        $26914$ & $3.2925$ & $0.1440$ & $2375$ \\
        $31914$ & $3.2649$ & $0.1435$ & $2524$ \\
        $39121$ & $3.2578$ & $0.1427$ & $2524$ \\
        $44121$ & $3.2540$ & $0.1429$ & $2729$ \\
        $46621$ & $3.2477$ & $0.1427$ & $2462$ \\
        $51328$ & $3.2561$ & $0.1426$ & $2729$ \\
        $56328$ & $3.2391$ & $0.1421$ & $2556$ \\
        $63535$ & $3.2368$ & $0.1419$ & $2657$ \\
        $68535$ & $3.2386$ & $0.1421$ & $2804$ \\
        $75742$ & $3.2343$ & $0.1418$ & $2884$ \\
        $78242$ & $3.2320$ & $0.1416$ & $2729$ \\
        $80742$ & $3.2235$ & $0.1417$ & \textbf{3257} \\
        $87949$ & $3.2286$ & $0.1413$ & $2729$ \\
        $92949$ & $3.2244$ & $0.1416$ & $2766$ \\
        $100156$ & $3.2258$ & $0.1412$ & $3205$ \\
        $107656$ & $3.2220$ & $0.1411$ & $2884$ \\
        $124570$ & $3.2202$ & $0.1411$ & $2692$ \\
        $139277$ & $3.2213$ & $0.1410$ & $2884$ \\
        $153984$ & $3.2106$ & $0.1410$ & $2884$ \\
        $163691$ & $3.2150$ & $0.1412$ & $2729$ \\
        $168691$ & $3.2133$ & $0.1410$ & $2589$ \\
        $175898$ & $3.2116$ & $0.1410$ & $3014$ \\
        $178398$ & $3.2074$ & $0.1411$ & $3014$ \\
        $180898$ & $3.2106$ & $0.1412$ & $2884$ \\
        $188105$ & $3.2096$ & $0.1412$ & $3014$ \\
        $193105$ & $3.2072$ & \textbf{0.1409} & $2729$ \\
        $200312$ & $3.2088$ & $0.1411$ & $3106$ \\
        $205312$ & $3.2100$ & $0.1410$ & $3059$ \\
        $210019$ & $3.2100$ & $0.1411$ & $3155$ \\
        $212519$ & $3.2119$ & $0.1412$ & $2969$ \\
        $222226$ & $3.2125$ & $0.1412$ & \textbf{3257} \\
        $227226$ & \textbf{3.2039} & $0.1411$ & $2766$ \\
        $234433$ & $3.2106$ & $0.1415$ & $3014$ \\
        $239433$ & $3.2081$ & $0.1414$ & $2766$ \\
        $241933$ & $3.2074$ & $0.1413$ & $2729$ \\
        \bottomrule
    \end{tabular}
\end{table}

\section{More Jet Generation Results}
\label{app:othergenresults}

\begin{figure}[t]
    \centering
        \includegraphics[width=0.99\linewidth]{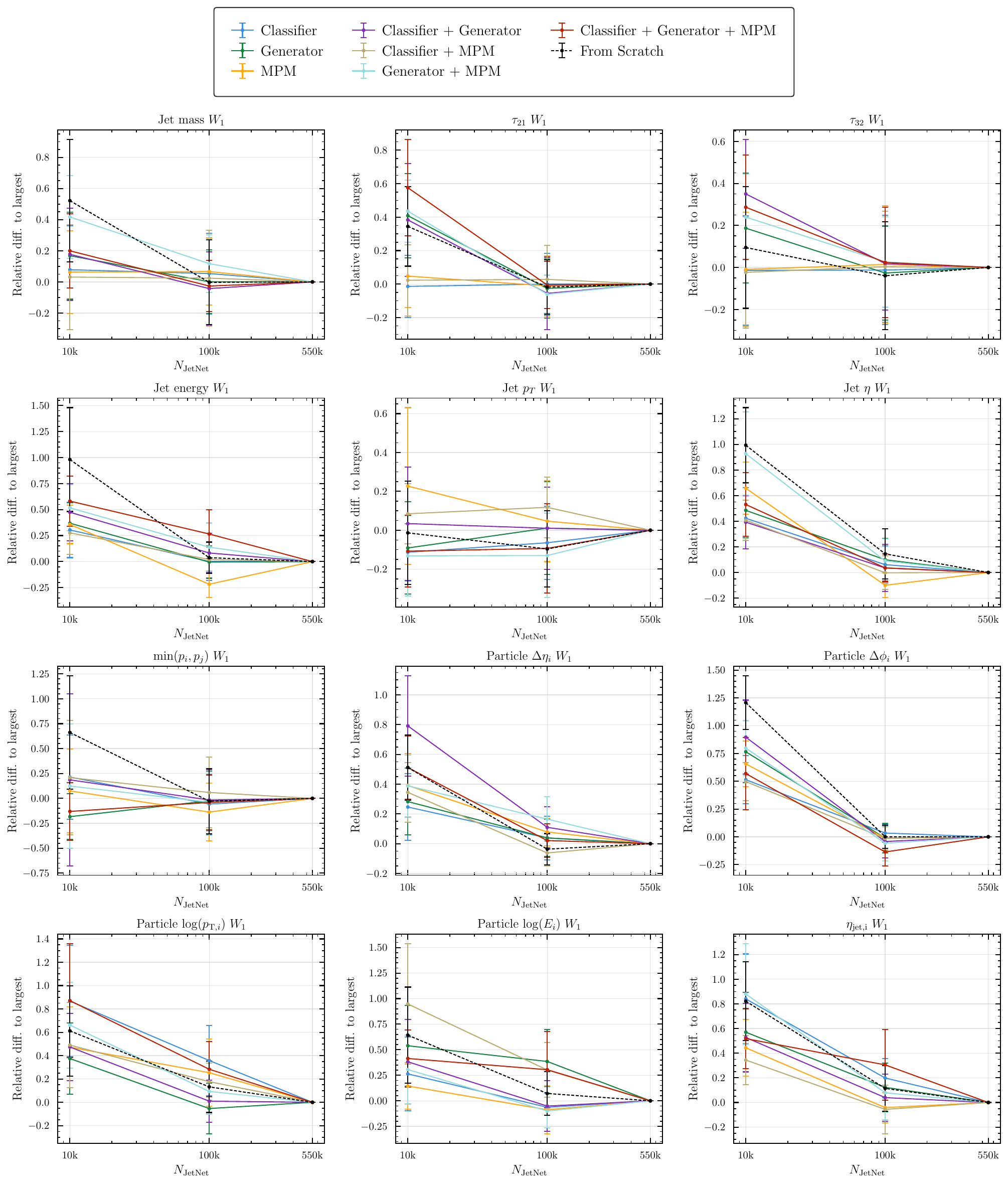}
     \caption{
        Relative performance difference of the generative model as a function of the training
        dataset size compared to the corresponding model obtained with the full
        dataset size ($N_\mathrm{JetNet}=550\mathrm{k}$) for the medium model.
        The relative performance difference is given by 
        $\left(W_1(N_\mathrm{JetNet}) - W_1(550\mathrm{k})\right)/W_1(550\mathrm{k})$
        for the corresponding metric.
    }
    \label{fig:jetnet_metrics_vs_dataset_size_medium_extended}
\end{figure}

\begin{figure}[t]
    \centering
        \includegraphics[width=0.99\linewidth]{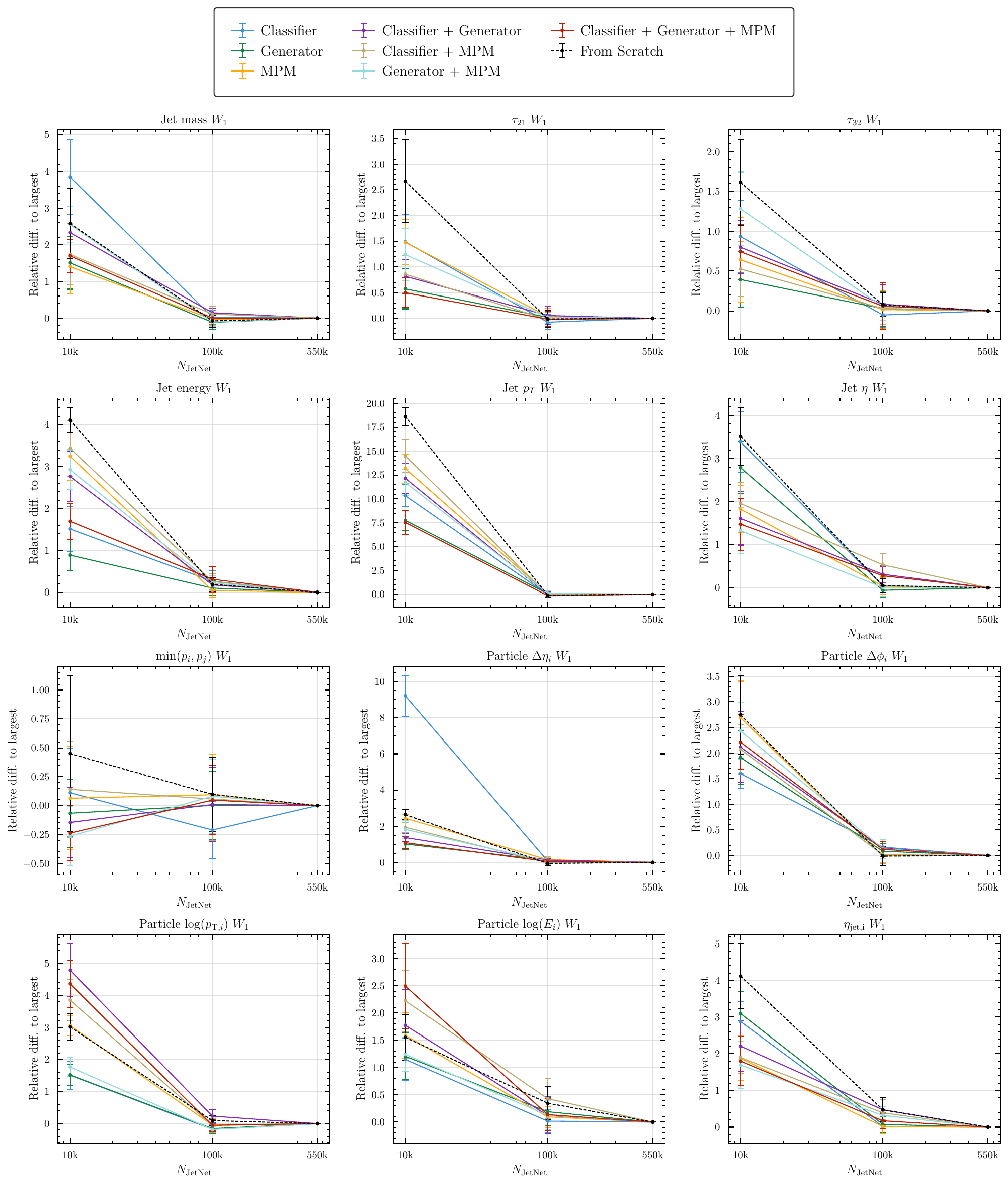}
    \caption{
        Relative performance difference of the generative model as a function of the training
        dataset size compared to the corresponding model obtained with the full
        dataset size ($N_\mathrm{JetNet}=550\mathrm{k}$) for the small model.
        The relative performance difference is given by 
        $\left(W_1(N_\mathrm{JetNet}) - W_1(550\mathrm{k})\right)/W_1(550\mathrm{k})$
        for the corresponding metric.
    }
    \label{fig:jetnet_metrics_vs_dataset_size_small_extended}
\end{figure}

\begin{figure}[t]
    \centering
        \includegraphics[width=0.99\linewidth]{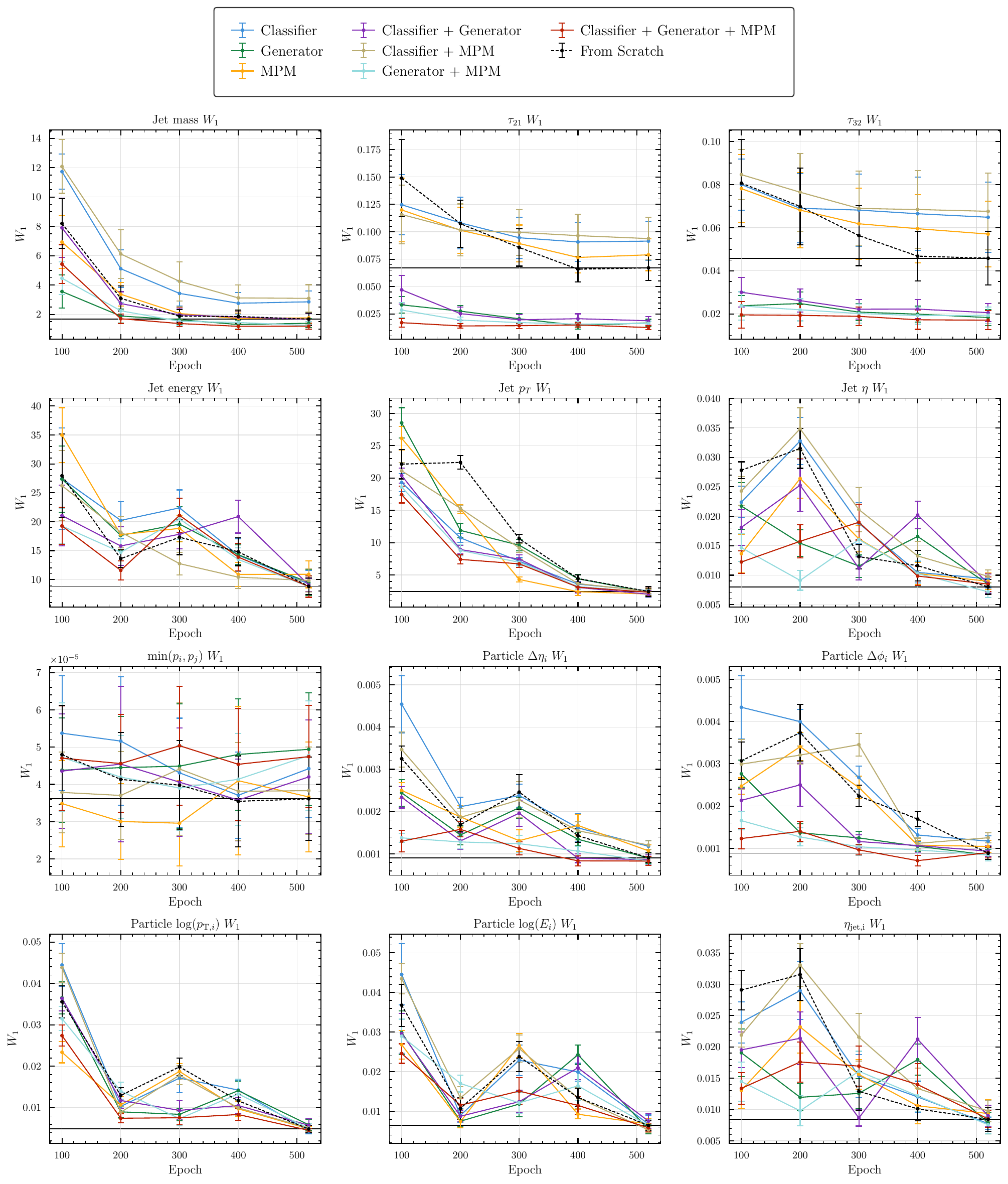}
    \caption{
        Performance of the generative models as a function of the training epoch 
        using a training dataset size $N_\mathrm{JetNet} = 550\mathrm{k}$ for
        the medium model size.
        The black horizontal line marks the value corresponding to the final epoch 
        of the from scratch baseline.
    }
    \label{fig:jetnet_metrics_vs_epoch_medium_extended}
\end{figure}

\begin{figure}[t]
    \centering
        \includegraphics[width=0.99\linewidth]{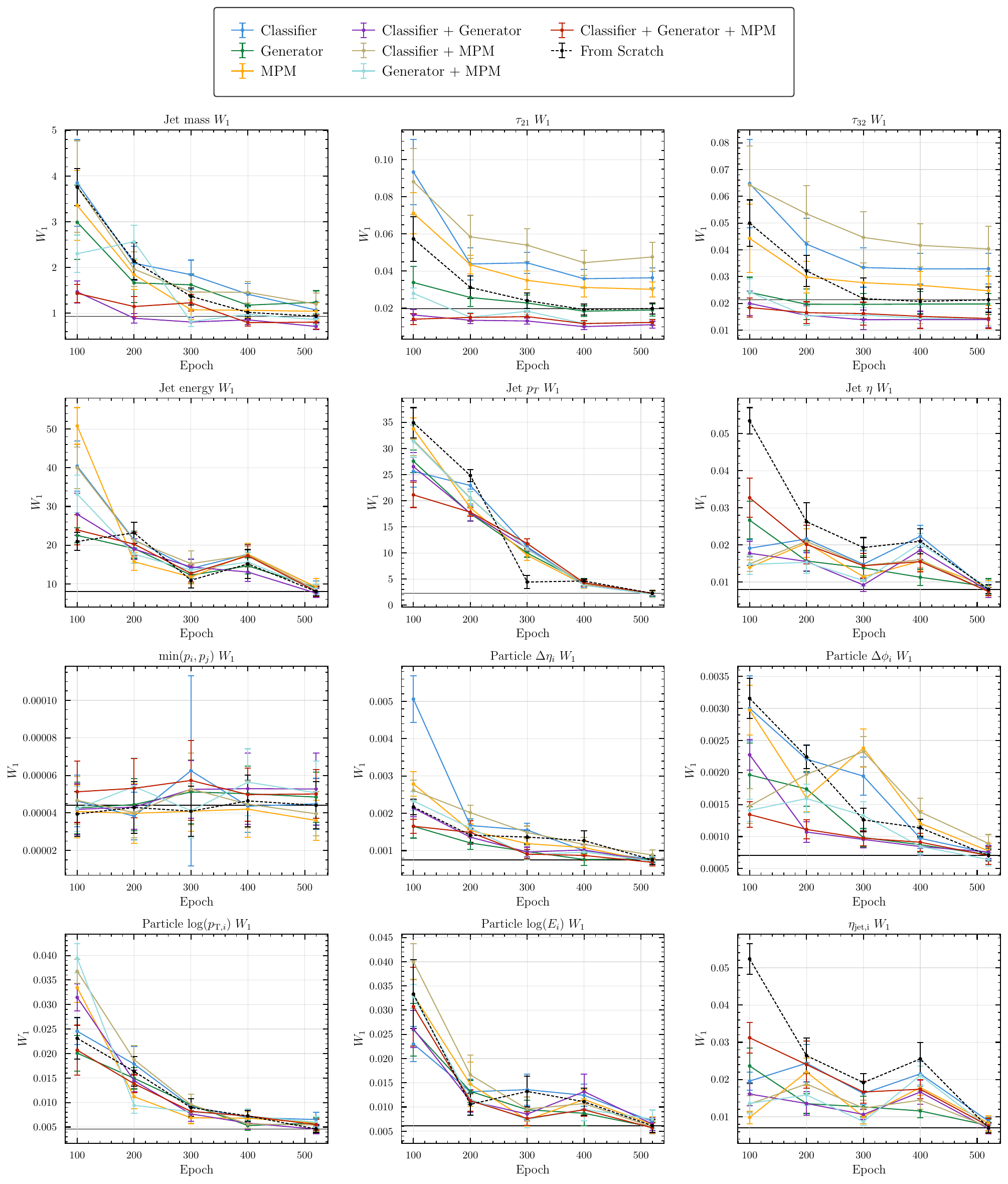}
    \caption{
        Performance of the generative models as a function of the training epoch 
        using a training dataset size $N_\mathrm{JetNet} = 550\mathrm{k}$ for
        the small model size.
        The black horizontal line marks the value corresponding to the final epoch 
        of the from scratch baseline.
    }
    \label{fig:jetnet_metrics_vs_epoch_small_extended}
\end{figure}

This section provides additional results related to the generative downstream task.
Table~\ref{tab:jetnet_metrics_full_dataset_last_epoch_additional_metrics} shows additional
metrics calculated for the evaluation of the generative models.
Those metrics complement the ones shown in Table~\ref{tab:jetnet_metrics_full_dataset_last_epoch}
in Section~\ref{sec:jetnetres}, but were found to be less informative in terms of
effects that different pre-training methods have on the downstream performance.
The corresponding performance difference seen between different dataset sizes
is shown for the extended set of metrics in Figures~\ref{fig:jetnet_metrics_vs_dataset_size_medium_extended} and
\ref{fig:jetnet_metrics_vs_dataset_size_small_extended}, and the corresponding
performance vs. training epoch trajectories are shown in
Figures~\ref{fig:jetnet_metrics_vs_epoch_medium_extended} and
\ref{fig:jetnet_metrics_vs_epoch_small_extended}.

\begin{table}[ht]
    \centering
    \caption{
        Performance of the different generative models averaged over all jet types.
        Each model is trained for 520 epochs on the full dataset
        ($N_\mathrm{JetNet}=550\mathrm{k}$).
    }
    \label{tab:jetnet_metrics_full_dataset_last_epoch_additional_metrics}

\vspace{1.5em} (a) Medium \\[0.5em]

\begin{tabular}{lccccccc}
\toprule
 & Jet $p_\mathrm{T}$ & Jet $\eta$ & $\min(p_i \, p_j)$ & $\Delta\eta_i$ & $\Delta\phi_i$ & $\log p_{\mathrm{T},i}$ & $\log E_i$ \\
 & (GeV) & $\times 10^{-3}$ & $\times 10^{-5}$ & $\times 10^{-4}$ & $\times 10^{-4}$ & $\times 10^{-3}$ & $\times 10^{-3}$ \\
\midrule
From Scratch & $2.4 \pm 0.8$ & $8.0 \pm 1.3$ & $\mathbf{3.6 \pm 1.1}$ & $9.1 \pm 1.1$ & $8.8 \pm 1.5$ & $4.8 \pm 0.9$ & $6.4 \pm 1.2$ \\
Generator & $2.4 \pm 0.5$ & $8.5 \pm 1.2$ & $4.9 \pm 1.5$ & $9.0 \pm 1.3$ & $\mathbf{8.4 \pm 1.4}$ & $5.9 \pm 1.4$ & $5.4 \pm 1.1$ \\
Classifier + Generator + MPM & $2.3 \pm 0.6$ & $8.5 \pm 1.2$ & $4.7 \pm 1.4$ & $\mathbf{8.3 \pm 1.1}$ & $9.0 \pm 1.3$ & $4.5 \pm 0.5$ & $5.8 \pm 1.0$ \\
Classifier + Generator & $\mathbf{2.0 \pm 0.5}$ & $8.6 \pm 1.6$ & $4.2 \pm 1.5$ & $8.6 \pm 1.2$ & $9.3 \pm 1.2$ & $5.7 \pm 1.6$ & $7.5 \pm 1.9$ \\
Generator + MPM & $2.3 \pm 0.8$ & $\mathbf{7.2 \pm 1.0}$ & $4.8 \pm 1.5$ & $8.3 \pm 1.0$ & $8.5 \pm 1.5$ & $5.0 \pm 1.1$ & $6.4 \pm 1.5$ \\
Classifier & $2.6 \pm 0.5$ & $9.4 \pm 1.5$ & $4.4 \pm 1.3$ & $11.8 \pm 1.4$ & $11.7 \pm 1.3$ & $\mathbf{4.3 \pm 0.7}$ & $6.8 \pm 2.3$ \\
MPM & $2.1 \pm 0.5$ & $9.0 \pm 1.4$ & $3.7 \pm 1.5$ & $10.7 \pm 1.4$ & $10.4 \pm 1.4$ & $4.9 \pm 0.7$ & $6.8 \pm 1.5$ \\
Classifier + MPM & $2.5 \pm 0.3$ & $9.8 \pm 1.1$ & $3.8 \pm 1.2$ & $12.1 \pm 1.0$ & $12.5 \pm 1.2$ & $5.0 \pm 0.8$ & $\mathbf{5.3 \pm 0.8}$ \\
\bottomrule
\end{tabular}

\vspace{1.5em} (b) Small \\[0.5em]

\begin{tabular}{lccccccc}
\toprule
 & Jet $p_\mathrm{T}$ & Jet $\eta$ & $\min(p_i \, p_j)$ & $\Delta\eta_i$ & $\Delta\phi_i$ & $\log p_{\mathrm{T},i}$ & $\log E_i$ \\
 & (GeV) & $\times 10^{-3}$ & $\times 10^{-5}$ & $\times 10^{-4}$ & $\times 10^{-4}$ & $\times 10^{-3}$ & $\times 10^{-3}$ \\
\midrule
From Scratch & $2.3 \pm 0.5$ & $8.0 \pm 1.1$ & $4.4 \pm 1.3$ & $7.5 \pm 1.2$ & $7.1 \pm 0.8$ & $4.6 \pm 0.7$ & $6.1 \pm 1.3$ \\
Generator & $2.2 \pm 0.7$ & $8.7 \pm 2.1$ & $4.8 \pm 1.3$ & $7.8 \pm 0.9$ & $7.2 \pm 1.0$ & $5.8 \pm 1.1$ & $5.8 \pm 1.0$ \\
Classifier + Generator + MPM & $2.3 \pm 0.5$ & $7.4 \pm 1.1$ & $5.0 \pm 1.3$ & $\mathbf{6.9 \pm 0.9}$ & $7.0 \pm 1.4$ & $5.5 \pm 1.1$ & $\mathbf{5.7 \pm 1.0}$ \\
Classifier + Generator & $2.2 \pm 0.6$ & $7.6 \pm 1.8$ & $5.3 \pm 1.9$ & $7.2 \pm 0.9$ & $7.6 \pm 1.0$ & $\mathbf{4.6 \pm 1.0}$ & $6.5 \pm 1.3$ \\
Generator + MPM & $\mathbf{2.1 \pm 0.5}$ & $8.3 \pm 1.6$ & $5.1 \pm 1.7$ & $7.2 \pm 1.1$ & $\mathbf{6.4 \pm 1.0}$ & $6.0 \pm 1.2$ & $7.5 \pm 1.9$ \\
Classifier & $2.3 \pm 0.7$ & $7.5 \pm 1.0$ & $4.5 \pm 1.4$ & $7.7 \pm 0.9$ & $7.4 \pm 1.0$ & $6.5 \pm 1.5$ & $7.0 \pm 2.3$ \\
MPM & $2.2 \pm 0.5$ & $8.6 \pm 1.8$ & $\mathbf{3.6 \pm 1.1}$ & $7.1 \pm 1.0$ & $7.8 \pm 1.2$ & $5.9 \pm 1.4$ & $6.4 \pm 1.6$ \\
Classifier + MPM & $2.2 \pm 0.7$ & $\mathbf{7.3 \pm 1.1}$ & $4.0 \pm 1.2$ & $8.8 \pm 1.3$ & $8.9 \pm 1.4$ & $5.4 \pm 1.2$ & $5.8 \pm 1.4$ \\
\bottomrule
\end{tabular}

\vspace{1.5em} (c) Micro \\[0.5em]

\begin{tabular}{lccccccc}
\toprule
 & Jet $p_\mathrm{T}$ & Jet $\eta$ & $\min(p_i \, p_j)$ & $\Delta\eta_i$ & $\Delta\phi_i$ & $\log p_{\mathrm{T},i}$ & $\log E_i$ \\
 & (GeV) & $\times 10^{-3}$ & $\times 10^{-5}$ & $\times 10^{-4}$ & $\times 10^{-4}$ & $\times 10^{-3}$ & $\times 10^{-3}$ \\
\midrule
From Scratch & $\mathbf{2.0 \pm 0.3}$ & $10.3 \pm 1.2$ & $3.3 \pm 0.8$ & $\mathbf{14.7 \pm 1.7}$ & $\mathbf{20.2 \pm 2.1}$ & $5.5 \pm 1.2$ & $7.4 \pm 1.3$ \\
Generator & $3.2 \pm 0.4$ & $9.7 \pm 1.7$ & $3.3 \pm 1.3$ & $36.8 \pm 3.5$ & $36.9 \pm 2.7$ & $6.9 \pm 1.4$ & $7.2 \pm 1.3$ \\
Classifier + Generator + MPM & $3.1 \pm 0.5$ & $12.8 \pm 2.1$ & $6.1 \pm 1.8$ & $43.0 \pm 6.6$ & $53.7 \pm 16.1$ & $6.2 \pm 0.9$ & $6.9 \pm 2.1$ \\
Classifier + Generator & $9.6 \pm 0.9$ & $13.4 \pm 2.1$ & $5.8 \pm 1.7$ & $22.4 \pm 2.7$ & $21.5 \pm 2.9$ & $9.3 \pm 2.2$ & $8.4 \pm 2.3$ \\
Generator + MPM & $3.1 \pm 0.5$ & $12.0 \pm 2.0$ & $\mathbf{3.3 \pm 0.9}$ & $33.8 \pm 1.9$ & $34.9 \pm 2.3$ & $6.1 \pm 1.2$ & $7.1 \pm 1.6$ \\
Classifier & $3.0 \pm 0.5$ & $11.6 \pm 1.3$ & $5.9 \pm 1.4$ & $31.7 \pm 4.4$ & $33.5 \pm 5.1$ & $5.7 \pm 0.7$ & $6.8 \pm 1.4$ \\
MPM & $2.1 \pm 0.3$ & $9.4 \pm 1.5$ & $3.8 \pm 1.0$ & $22.2 \pm 2.8$ & $23.8 \pm 2.4$ & $5.3 \pm 0.9$ & $6.5 \pm 1.6$ \\
Classifier + MPM & $2.5 \pm 0.7$ & $\mathbf{8.5 \pm 1.4}$ & $3.6 \pm 0.9$ & $17.5 \pm 2.4$ & $20.3 \pm 3.7$ & $\mathbf{5.1 \pm 0.6}$ & $\mathbf{5.8 \pm 1.4}$ \\
\bottomrule
\end{tabular}

\vspace{1em}
\end{table}

\begin{figure}[h!]
    \centering
    \subfloat{
        \includegraphics[width=0.98\linewidth]{images/jetnet_generation/feature_distributions/feature_distributions_small_vs_downstream_dataset_size_scratch_vs_class-gen_light_quark.pdf}
    }\\[-1.5em]
    (a) light quark jets \\
    \subfloat{
        \includegraphics[width=0.98\linewidth]{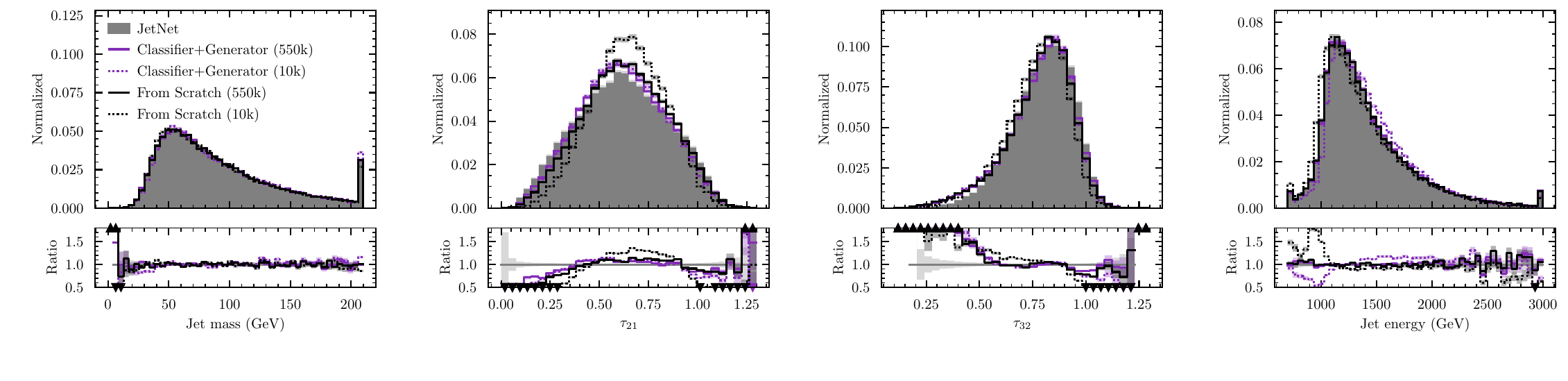}
    }\\[-1.5em]
    (b) gluon jets \\
    \subfloat{
        \includegraphics[width=0.98\linewidth]{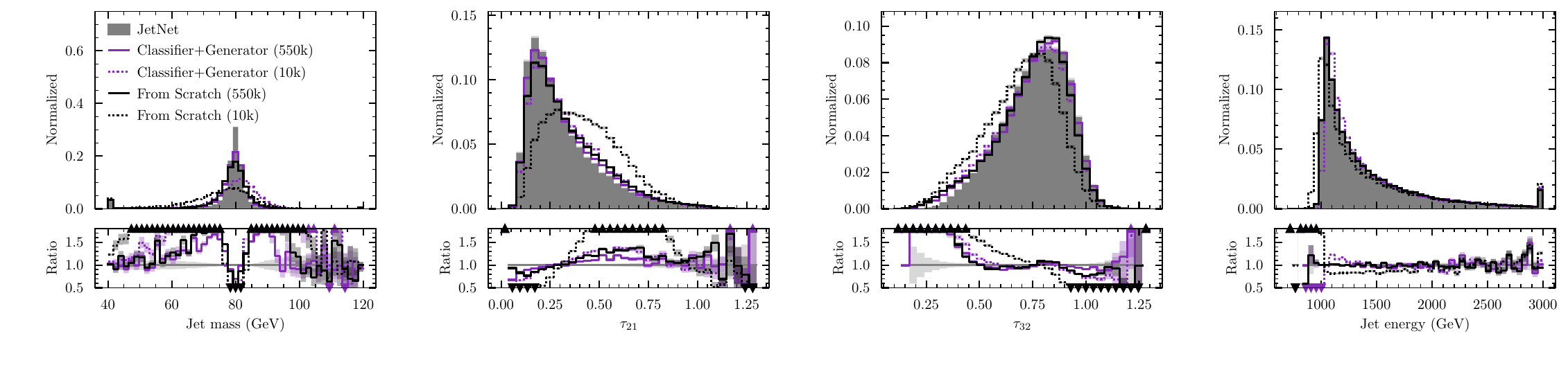}
    }\\[-1.5em]
    (c) $W$ boson jets \\
    \subfloat{
        \includegraphics[width=0.98\linewidth]{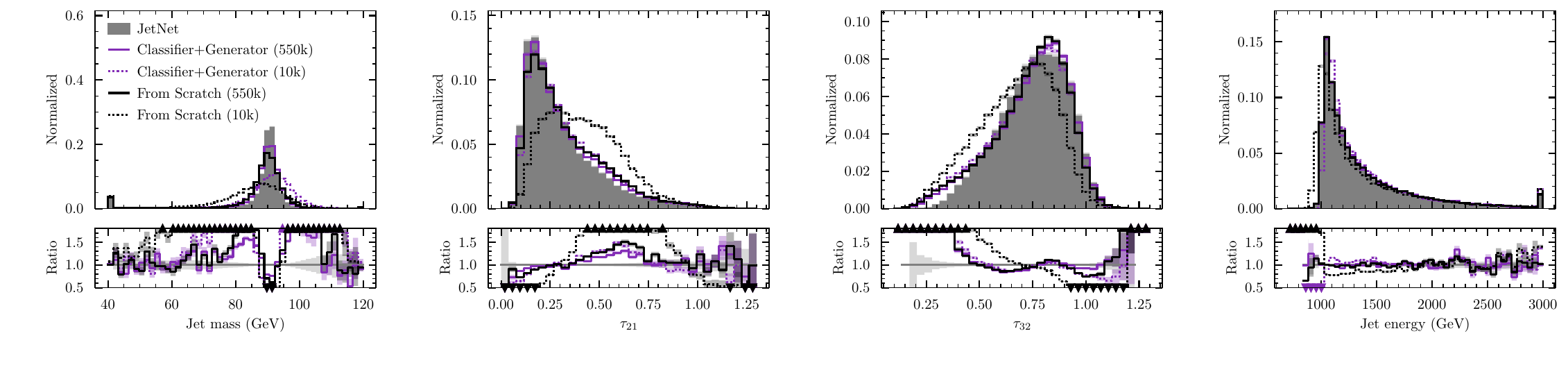}
    }\\[-1.5em]
    (d) $Z$ boson jets \\
    \caption{
        Jet-level distributions of jets initiated by (a) light quarks, (b)
        gluons, (c) $W$ bosons and (d) $Z$ bosons. 
        Generated jets are obtained from the small model with the
        Classifier+Generator pre-training method compared with the from scratch
        baseline.
        Two downstream dataset sizes are shown in dotted
        ($N_\mathrm{JetNet}=10\mathrm{k}$) and solid
        ($N_\mathrm{JetNet}=550\mathrm{k}$) lines.
        The gray histogram shows jets from the JetNet dataset.
    }
    \label{fig:jetnet_distributions_vs_dataset_size_class_gen_W-Z-light-gluon}
\end{figure}

\begin{figure}[h!]
    \centering
    \subfloat{
        \includegraphics[width=0.98\linewidth]{images/jetnet_generation/feature_distributions/feature_distributions_small_vs_downstream_dataset_size_scratch_vs_class-mpm_light_quark.pdf}
    }\\[-1.5em]
    (a) light quark jets \\
    \subfloat{
        \includegraphics[width=0.98\linewidth]{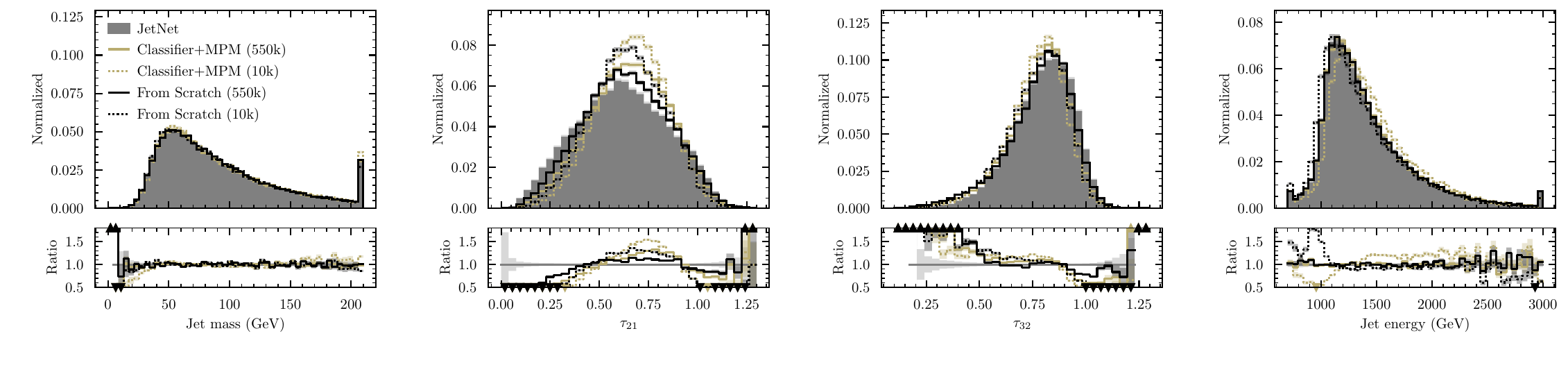}
    }\\[-1.5em]
    (b) gluon jets \\
    \subfloat{
        \includegraphics[width=0.98\linewidth]{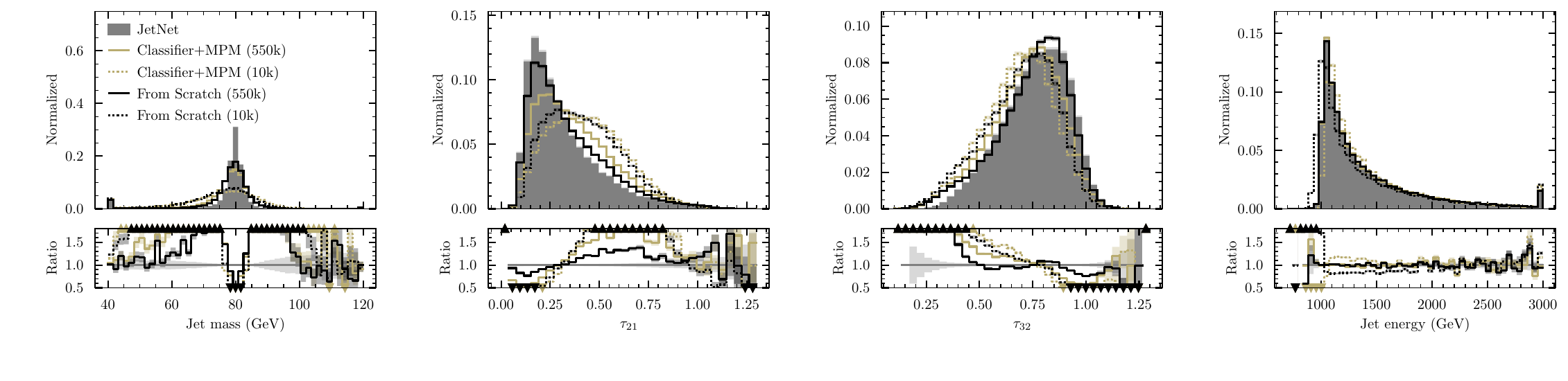}
    }\\[-1.5em]
    (c) $W$ boson jets \\
    \subfloat{
        \includegraphics[width=0.98\linewidth]{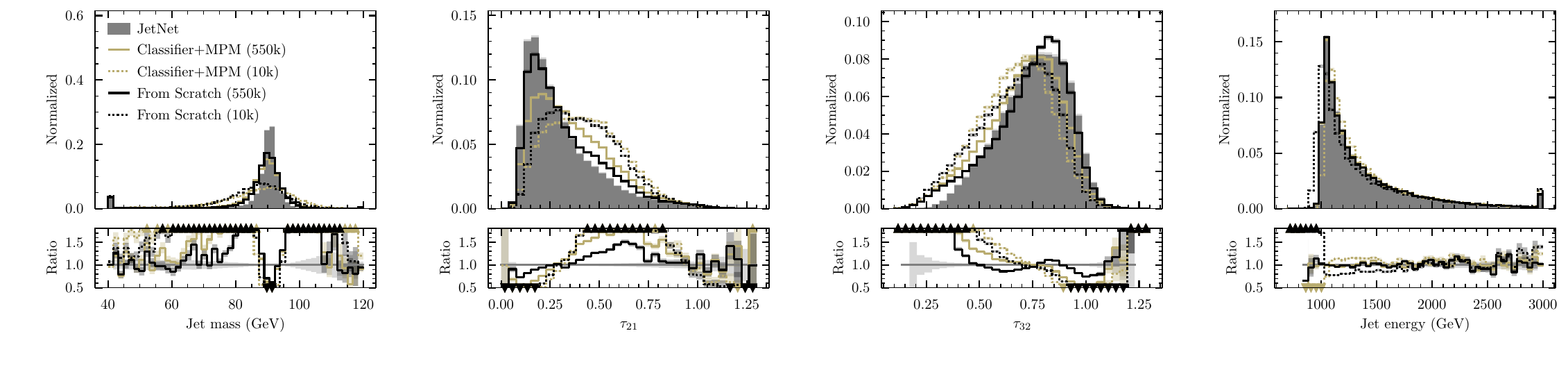}
    }\\[-1.5em]
    (d) $Z$ boson jets \\
    \caption{
        Jet-level distributions of jets initiated by (a) light quarks, (b)
        gluons, (c) $W$ bosons and (d) $Z$ bosons. 
        Generated jets are obtained from the small model with the
        Classifier+MPM pre-training method compared with the from scratch
        baseline.
        Two downstream dataset sizes are shown in dotted
        ($N_\mathrm{JetNet}=10\mathrm{k}$) and solid
        ($N_\mathrm{JetNet}=550\mathrm{k}$) lines.
        The gray histogram shows jets from the JetNet dataset.
    }
    \label{fig:jetnet_distributions_vs_dataset_size_class_mpm_W-Z-light-gluon}
\end{figure}

\begin{figure}[h!]
    \centering
    \subfloat{
        \includegraphics[width=0.98\linewidth]{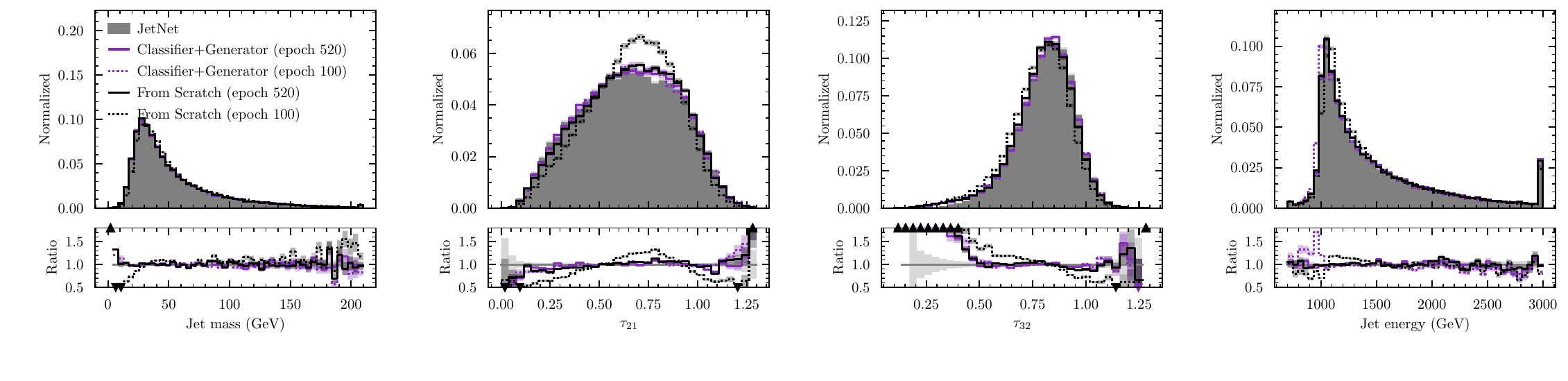}
    }\\[-1.5em]
    (a) light quark jets \\
    \subfloat{
        \includegraphics[width=0.98\linewidth]{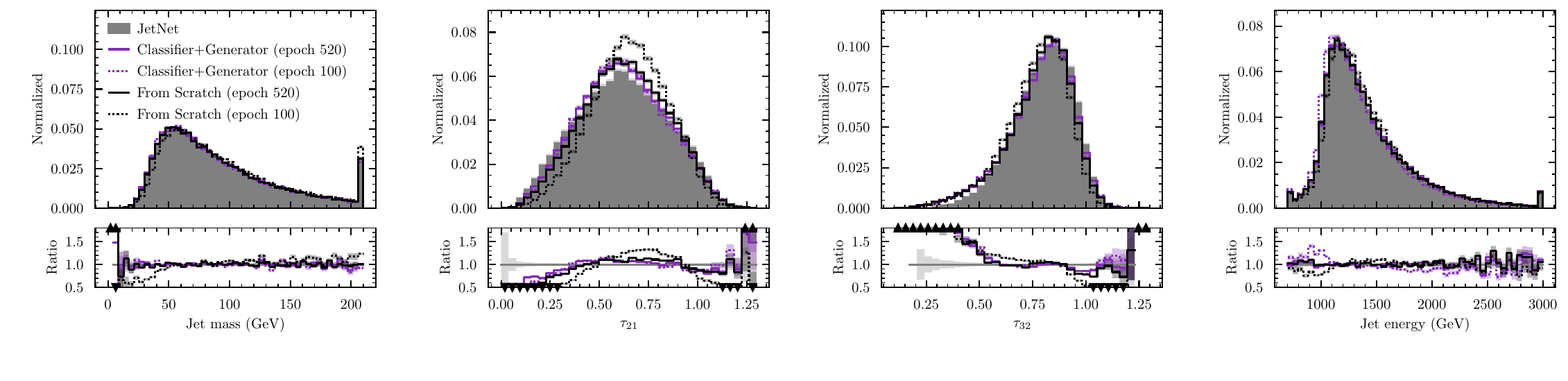}
    }\\[-1.5em]
    (b) gluon jets \\
    \subfloat{
        \includegraphics[width=0.98\linewidth]{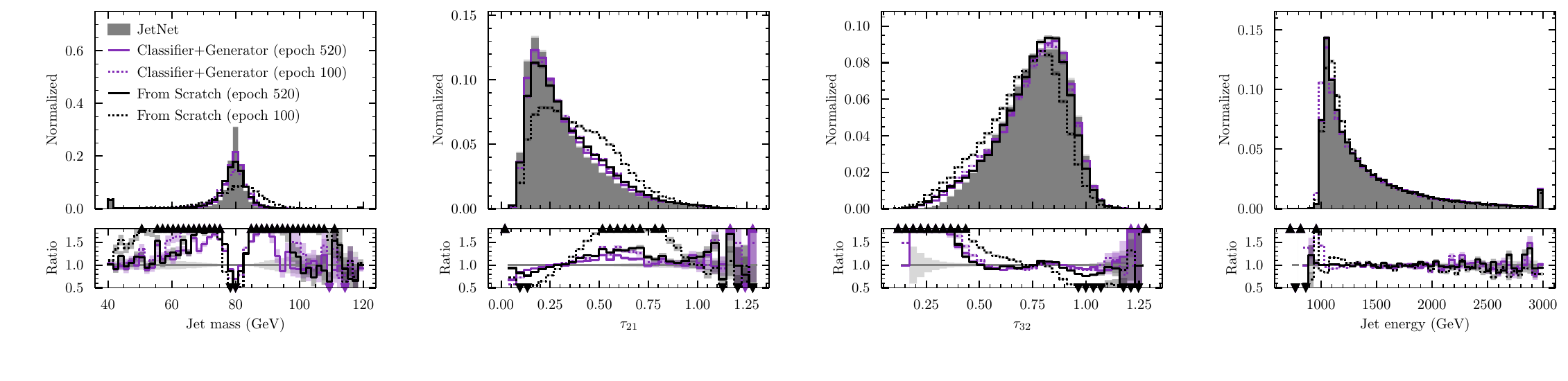}
    }\\[-1.5em]
    (c) $W$ boson jets \\
    \subfloat{
        \includegraphics[width=0.98\linewidth]{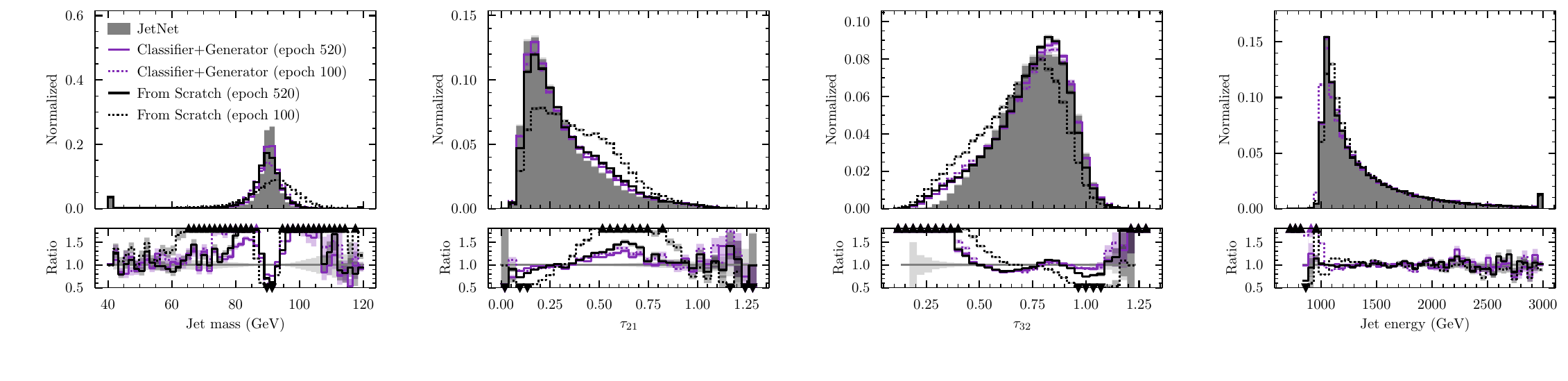}
    }\\[-1.5em]
    (d) $Z$ boson jets \\
    \caption{
        Jet-level distributions of jets initiated by (a) light quarks, (b)
        gluons, (c) $W$ bosons and (d) $Z$ bosons. 
        Generated jets are obtained from the small model with the
        Classifier+Generator pre-training method compared with the from scratch
        baseline.
        Two evaluated epochs are shown in dotted (epoch 100) and solid (epoch
        520) lines. The gray histogram shows jets from the JetNet dataset.
    }
    \label{fig:jetnet_distributions_vs_epoch_class_gen_W-Z-light-gluon}
\end{figure}

\begin{figure}[h!]
    \centering
    \subfloat{
        \includegraphics[width=0.98\linewidth]{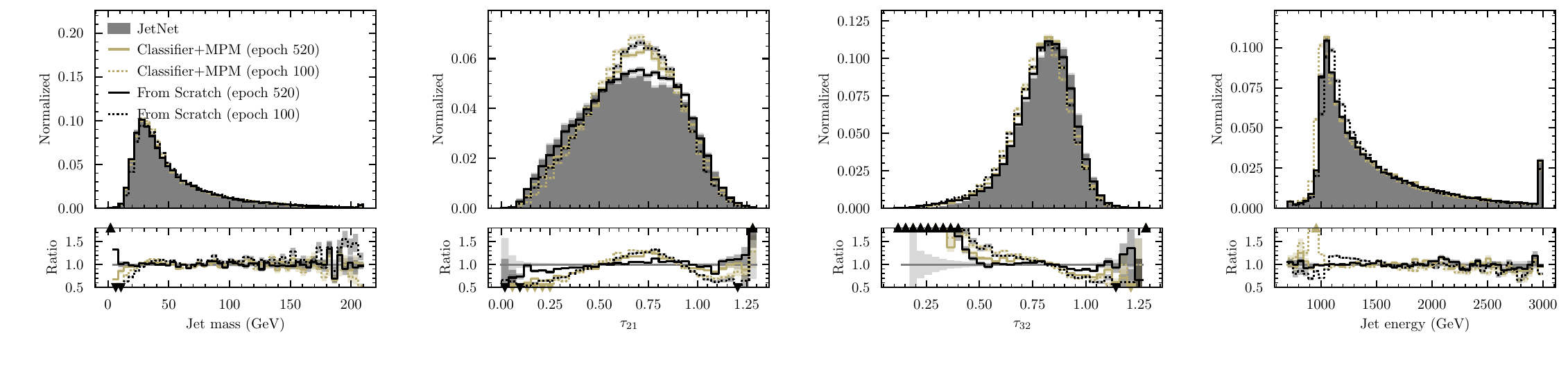}
    }\\[-1.5em]
    (a) light quark jets \\
    \subfloat{
        \includegraphics[width=0.98\linewidth]{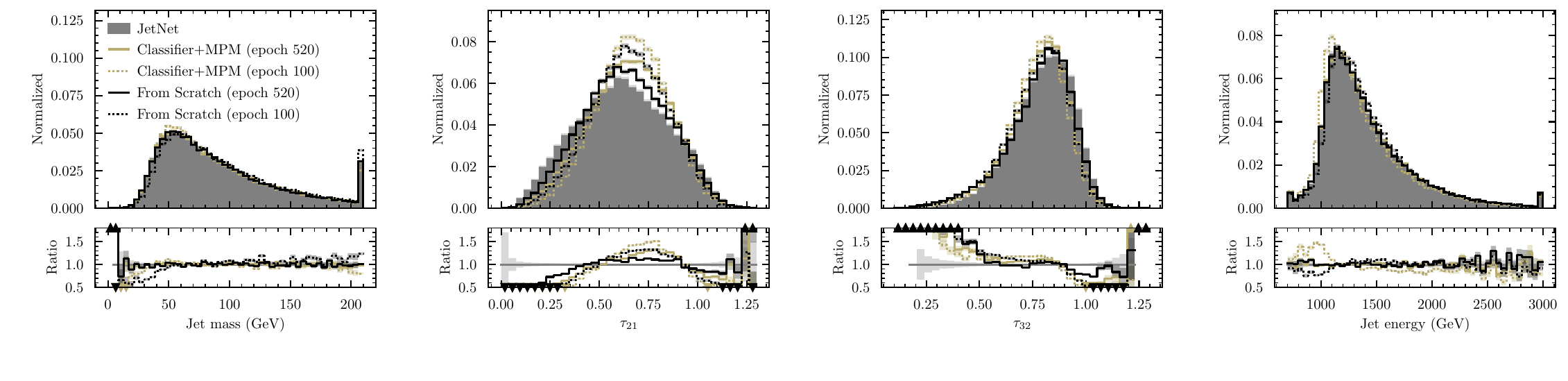}
    }\\[-1.5em]
    (b) gluon jets \\
    \subfloat{
        \includegraphics[width=0.98\linewidth]{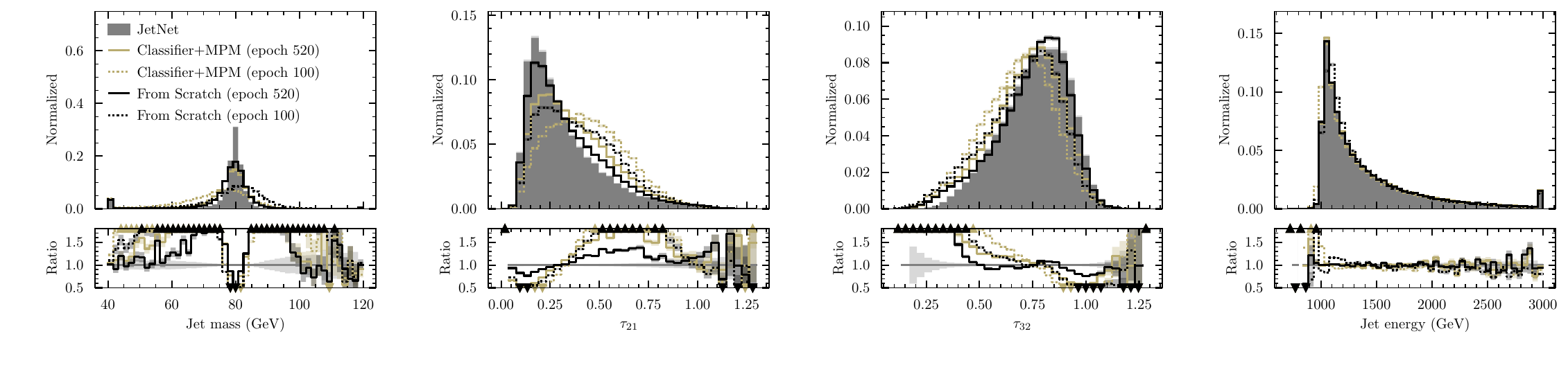}
    }\\[-1.5em]
    (c) $W$ boson jets \\
    \subfloat{
        \includegraphics[width=0.98\linewidth]{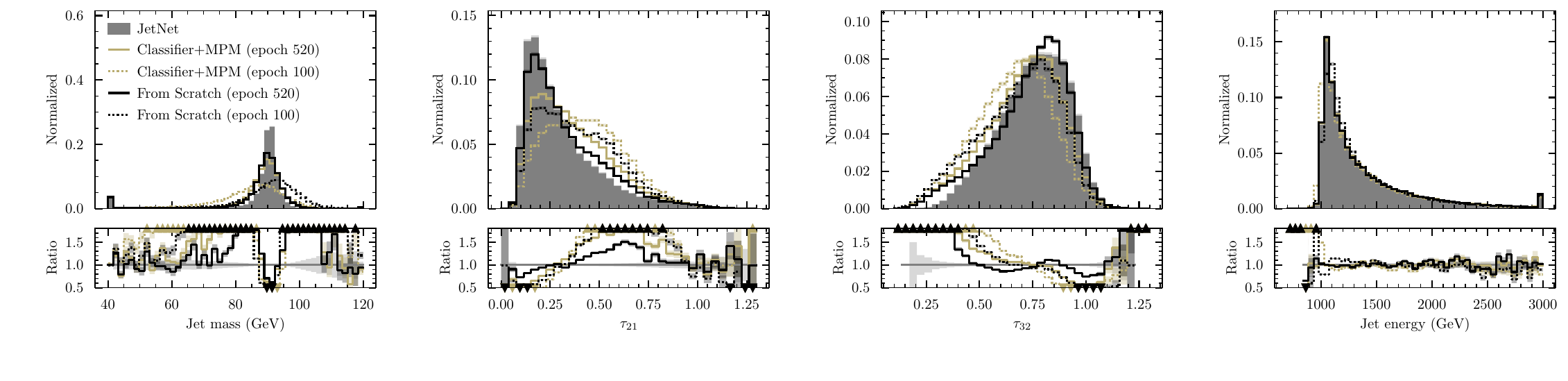}
    }\\[-1.5em]
    (d) $Z$ boson jets \\
    \caption{
        Jet-level distributions of jets initiated by (a) light quarks, (b)
        gluons, (c) $W$ bosons and (d) $Z$ bosons. 
        Generated jets are obtained from the small model with the
        Classifier+MPM pre-training method compared with the from scratch
        baseline.
        Two evaluated epochs are shown in dotted (epoch 100) and solid (epoch
        520) lines. The gray histogram shows jets from the JetNet dataset.
    }
    \label{fig:jetnet_distributions_vs_epoch_class_mpm_W-Z-light-gluon}
\end{figure}

\end{document}